\documentclass[acmsmall,screen,nonacm,dvipsnames]{acmart}

\AtBeginDocument{%
  }

\setcopyright{none}
\copyrightyear{2027}
\acmYear{2027}
\acmJournal{PACMPL}
\acmNumber{PLDI}

\usepackage{lipsum}

\usepackage{url}
\usepackage{fancyvrb}
\usepackage{mdwlist}  %
\usepackage{bookmark}
\usepackage{xspace}
\usepackage{xstring}
\usepackage{comment}
\usepackage{booktabs} %
\usepackage{xr}

\usepackage{enumitem}

\setlist[enumerate]{leftmargin=5.5mm}
\setlist[itemize]{leftmargin=4.5mm}

\usepackage{pifont} %

\usepackage[euler]{textgreek}

\usepackage{amsmath}
\usepackage{mathpartir}

\usepackage{amssymb}
\usepackage{bm}       %
\usepackage{thmtools, thm-restate}
\usepackage{annotate-equations}

\usepackage[normalem]{ulem} %

\usepackage{tikz}
\usetikzlibrary{matrix}
\usetikzlibrary{arrows.meta,automata}
\usetikzlibrary{positioning}
\usetikzlibrary{shapes.geometric}
\usetikzlibrary{petri,topaths}
\usetikzlibrary{calc}
\usetikzlibrary{patterns}
\usepackage{wrapfig}

\pgfdeclarepatternformonly{soft horizontal lines}{\pgfpointorigin}{\pgfqpoint{100pt}{1.5pt}}{\pgfqpoint{100pt}{1.5pt}}%
{
  \pgfsetstrokeopacity{0.3}
  \pgfsetlinewidth{0.2pt}
  \pgfpathmoveto{\pgfqpoint{0pt}{0.5pt}}
  \pgfpathlineto{\pgfqpoint{100pt}{0.5pt}}
  \pgfusepath{stroke}
}

\pgfdeclarepatternformonly{soft vertical lines}{\pgfpointorigin}{\pgfqpoint{1pt}{100pt}}{\pgfqpoint{1.5pt}{100pt}}%
{
  \pgfsetstrokeopacity{0.3}
  \pgfsetlinewidth{0.2pt}
  \pgfpathmoveto{\pgfqpoint{0.5pt}{0pt}}
  \pgfpathlineto{\pgfqpoint{0.5pt}{100pt}}
  \pgfusepath{stroke}
}
\pgfdeclarepatternformonly{soft crosshatch}{\pgfqpoint{-1pt}{-1pt}}{\pgfqpoint{4pt}{4pt}}{\pgfqpoint{3pt}{3pt}}%
{
  \pgfsetstrokeopacity{0.3}
  \pgfsetlinewidth{0.2pt}
  \pgfpathmoveto{\pgfqpoint{3.1pt}{0pt}}
  \pgfpathlineto{\pgfqpoint{0pt}{3.1pt}}
  \pgfpathmoveto{\pgfqpoint{0pt}{0pt}}
  \pgfpathlineto{\pgfqpoint{3.1pt}{3.1pt}}
  \pgfusepath{stroke}
}

\usepackage{stmaryrd}
\usepackage{mathtools}  %

\usepackage{savesym}
\savesymbol{checkmark}
\usepackage{dingbat}

\usepackage{longtable}
\usepackage{subcaption}
\usepackage{xcolor}
\usepackage{array}

\theoremstyle{plain}
\newtheorem{theorem}{Theorem}[section]

\newtheorem{lemma}[theorem]{Lemma}

\usepackage{longtable}
\usepackage{cleveref}

\usepackage{multirow}

\usepackage{scalerel} %

\usepackage{listings, listings-rust}
\usepackage[scaled=0.85]{beramono}

\definecolor{listing-quote}{RGB}{211,54,130}
\definecolor{listing-keyword}{RGB}{24,78,150}
\definecolor{listing-comment}{RGB}{110,110,110}
\definecolor{listing-string}{RGB}{150,45,45}
\lstdefinelanguage{haskell}{
  keywords={self, Void, Class, All, extends, this, trait, inherits, super, type, Trait, override, new, if, then, else, let, macro, in, letrec, rec, fun, module, struct, end, functor, where, import, String, Quote, Show, Num, Code, instance, class, data, sig, val},
  identifierstyle=\color{black},
  morecomment=[l]{--},
  morecomment=[l]{//},
  morecomment=[s]{(*}{*)},
  morestring=[b]",
  xleftmargin  = 2mm,
}

\lstnewenvironment{lstlistingnb}[1][]%
  {\noindent\minipage{\linewidth}\medskip 
   \lstset{#1}}
  {\endminipage}

\theoremstyle{remark}

\newcounter{mycnumber}

\definecolor{sidecolor}{HTML}{C32148}

\newcounter{mytsnumber}

\newcounter{myanumber}

\newcounter{mysnumber}

\Crefformat{figure}{#2Fig.~#1#3}
\Crefformat{section}{\S#2#1#3}
\Crefformat{subsection}{\S#2#1#3}
\Crefformat{subsubsection}{\S#2#1#3}
\Crefformat{appendix}{Appendix~#2#1#3}
\Crefname{theorem}{Thm.}{Thms.}
\usepackage[utf8]{inputenc}
\usepackage[T1]{fontenc} %

\usepackage{tabularx}
\newcommand\doubleplus{+\kern-0.6ex+}
\newcommand\Lleftrightarrow{\Lleftarrow\kern-1.7ex\Rrightarrow}

\definecolor{ibmblue}{RGB}{100, 143, 255}
\definecolor{ibmpurple}{RGB}{120, 94, 240}
\definecolor{ibmred}{RGB}{220, 38, 127}
\definecolor{ibmorange}{RGB}{254, 97, 0}
\definecolor{ibmyellow}{RGB}{255, 176, 0}

\colorlet{target}{ibmblue}

\ExplSyntaxOn
\cs_new:Npn \expandableinput_ #1
  { \use:c { @@input } { \file_full_name:n {#1} } }
\AddToHook{env/tabular/begin}
  { \cs_set_eq:NN \input \expandableinput_ }
\AddToHook{env/tabularx/begin}
  { \cs_set_eq:NN \input \expandableinput_ }
\AddToHook{env/tabulary/begin}
  { \cs_set_eq:NN \input \expandableinput_ }
\ExplSyntaxOff

\renewcommand\MathparLineskip{\lineskip=0.1cm}

\makeatletter
\def\@parfont{\normalfont\bfseries\itshape}
\makeatother

\newcommand{\Dom}{\mathrm{dom}\,}
\newcommand{\Erase}[1]{\lceil #1 \rceil}
\newcommand{\Rst}[1]{{\downarrow}#1}
\newcommand{\Ends}[1]{\leadsto^{\dagger #1}}
\newcommand{\LftEnd}[1]{\overset{\dagger #1}{\leadsto}}
\newcommand{\Arr}[4]{#3 \mathbin{\multimap^{#1}_{#2}} #4}
\newcommand{\ArrT}[6]{\forall \bar{#1}.\,\Arr{#2 \Ends{#3}}{#4}{#5}{#6}}
\newcommand{\Qual}[1]{\lvert #1 \rvert}
\newcommand{\Lin}{\mathsf{lin}}
\newcommand{\Aff}{\mathsf{aff}}
\newcommand{\Un}{\mathsf{un}}
\newcommand{\Rview}[2]{\mathsf{read}_{#1}(#2)}
\newcommand{\Gpre}[1]{{\color{gray}#1}}
\newcommand{\FL}{\mathrm{FL}}
\newcommand{\Head}{\mathrm{hd}}
\newcommand{\Heads}{\mathrm{heads}}
\newcommand{\Subs}{\mathrm{subs}}
\newcommand{\Vars}{\mathrm{vars}}
\newcommand{\Mut}{\mathsf{mut}}
\newcommand{\RefT}{\mathsf{ref}}
\newcommand{\Kw}[1]{\mathsf{#1}}
\providecommand{\Dashv}{\mathrel{\text{\reflectbox{$\vDash$}}}}
\newcommand{\rdef}[1]{\hypertarget{rule:#1}{\textsc{#1}}}
\newcommand{\rref}[1]{\hyperlink{rule:#1}{\textsc{#1}}}
\usepackage{xspace}
\newcommand{\rocq}{\;\raisebox{-0.25ex}{\includegraphics[height=1.8ex]{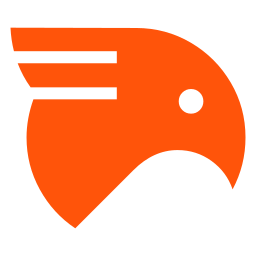}}}
\newcommand{\LBorrow}{$\lambda^{bo\rho}$\xspace}

\newlength{\lstcellw}
\AtBeginDocument{\settowidth{\lstcellw}{\fontsize{8.5}{8.5}\selectfont\ttfamily x}}
\newcommand{\lstann}[2]{\makebox[#1\lstcellw][l]{\normalcolor$#2$}}

\lstdefinelanguage{calc}{
  morekeywords={let, in, ref, read, swap, drop, if, then, else, loop, letlft, endlft, true, false},
  morecomment=[l]{--},
  sensitive=true,
}

\usepackage[skins,breakable]{tcolorbox}
\newtcolorbox{taggedbox}[2][]{%
  enhanced, breakable,
  colback=white, colframe=black!60, boxrule=0.4pt, arc=2pt,
  left=8pt, right=8pt, top=5pt, bottom=5pt, boxsep=0pt,
  before skip=6pt plus 2pt minus 2pt, after skip=6pt plus 2pt minus 2pt,
  overlay unbroken and first={%
    \node[anchor=north east, inner xsep=3pt, inner ysep=1.5pt, fill=white,
          font=\scriptsize\sffamily\color{black!60}]
      at ([xshift=-5pt]frame.north east) {#2};},
  #1
}

\begin{document}

\title{A Calculus for Borrowability}

\author{Yanning Chen}
\affiliation{%
  \institution{University of Toronto}
  \city{Toronto}
  \country{Canada}}
\email{yanning@cs.toronto.edu}
\author{Fan Long}
\affiliation{%
  \institution{University of Toronto}
  \city{Toronto}
  \country{Canada}}
\email{fanl@cs.toronto.edu}
\author{Ningning Xie}
\affiliation{%
  \institution{University of Toronto}
  \city{Toronto}
  \country{Canada}}
\email{ningningxie@cs.toronto.edu}

\begin{abstract}
  Borrowing grants temporary shared or exclusive access to a linear resource
  without transferring ownership. However, capturing borrowing in a type system
  poses two challenges: a lender, the reference a borrow was taken from,
  should remain a first-class value that can be moved, and regain
  its permission when a non-lexical borrow ends. We propose
  \textit{borrowability}, an index on reference types that
  tracks how access permissions are lent across lifetimes while separating them
  from the owner's obligation to release the resource. A lender is then
  an ordinary value whose borrowability records its remaining access permissions
  and the lifetimes of its active borrows, recovering borrowed permissions in
  place when those lifetimes end. We formalize borrowability in the
  calculus \LBorrow{}, mechanize its semantic type
  soundness and memory safety proofs in Rocq using Iris, and show that our design rules
  out all 24 in-scope soundness issues reported against the Rust compiler.
\end{abstract}

\maketitle

\section{Introduction}
\label{sec:introduction}

\textit{Linear types}~\cite{wadlerLinearTypesCan1990} give a type system precise
control over resources: a file handle or memory cell is released exactly
once. However, every operation on a resource must consume its reference and hand
one back, threading the resource through the program in a cumbersome
\emph{capability-passing style}. Moreover, linearity forbids aliasing
altogether, disallowing even harmless read-only ones.

\emph{Borrowing} lifts both restrictions. A borrow grants temporary shared or
exclusive access to a resource without transferring ownership, so a read no
longer has to consume and return the reference, and several readers may
coexist, while the owner keeps the obligation to release the resource and
regains full access when the borrow ends. The idea goes back to aliasing
control in object-oriented languages~\cite{hogg1991islands} and to observers
for linear types~\cite{oderskyObserversLinearTypes1992}, and has been refined
by a long line of
work~\cite{fahndrichAdoptionFocusPractical2002,swamySafeManualMemory2006,nadenTypeSystemBorrowing2012,chargueraudTemporaryReadOnlyPermissions2017}.
Rust~\cite{matsakisRustLanguage2014} popularized borrowing in mainstream systems
programming. Its borrow checker rules out use-after-free and data races at
compile time without a garbage collector, a major reason for its
adoption, and its ownership model now shapes memory management in other
languages as well~\cite{lorenzenOxidizingOCamlModal2024,georgesDataRaceFreedom2025}.
Borrowing has been extensively studied, from semantic
models of Rust's guarantees~\cite{jungRustBeltSecuringFoundations2017} and
aliasing models for its unsafe
code~\cite{jungStackedBorrowsAliasing2019,villaniTreeBorrows2025} to core
calculi~\cite{weissOxideEssenceRust2021} and type systems for borrowing in
functional
languages~\cite{radanneKindlyBentFree2020,wagnerLinearityBorrowing2025,marshallFunctionalOwnershipFractional2024,matsushitaPureBorrowLinear2026}.

Despite this progress, two challenges remain in typing borrowing, and no
existing system meets both. First, a reference from which a borrow has been
taken, a \emph{lender}, should remain a first-class value that can, for example,
be captured by a closure that will later release the resource. However, Rust
forbids moving a borrowed reference because its type system does not record
outstanding borrows in the lender's type. Several existing type systems for
borrowing~\cite{wagnerLinearityBorrowing2025,radanneKindlyBentFree2020}
likewise withhold the lender from the borrower's context while a borrow is
active. Second, borrows may be non-lexical, ending along control-flow paths
without following lexical scope, and a lender must regain its permission when a
borrow ends. Existing type systems, in turn, either tie the end of a borrow to a
lexical
scope~\cite{wagnerLinearityBorrowing2025,radanneKindlyBentFree2020} or
require explicitly returning the borrow or reclaiming
the lender~\cite{marshallFunctionalOwnershipFractional2024,matsushitaPureBorrowLinear2026}.

We propose \emph{borrowability}, an index on reference types
that tracks how \emph{access permissions} are lent across lifetimes while separating
them from the owner's linear \emph{obligation} to release the resource. A lender is
then an ordinary value whose borrowability records its remaining access
permissions and the lifetimes of its active borrows, and ending a lifetime
restores those permissions in place with no return or join operation. We develop
this idea in a calculus \LBorrow{}.

\paragraph{Contributions}
\Cref{sec:running-example} gives background on linear types and borrowing,
and \Cref{sec:challenges} illustrates the two challenges with examples. We then
make the following contributions:
\begin{itemize}
\item We introduce borrowability and show that it models borrowing as
  reference type splitting across typing contexts and enables in-place
  permission restoration when lifetimes end, supporting both first-class lenders
  and non-lexical lifetimes
  (\Cref{sec:borrowability}--\Cref{sec:overview-mem-ops}). We further
  introduce \emph{lifetime effects}, which let closures end lifetimes safely
  (\Cref{sec:overview-functions}).
\item We formalize \LBorrow{}, presenting its syntax and type system
  (\Cref{sec:calculus}), in which borrowing, restoration, and lifetime
  effects are all built into the typing rules.
\item We present the semantics of \LBorrow{} (\Cref{sec:semantics}), and prove
  its semantic type soundness and memory safety.
  The calculus and its metatheory are all mechanized in
  Rocq~\cite{therocqdevelopmentteamRocqProver2025}.
\item We survey the soundness issues reported against the Rust compiler and
  show that \LBorrow{} rules out all 24 issues within its
  scope~(\Cref{sec:case-study}).
\end{itemize}

\noindent
We discuss related work and future work in \Cref{sec:related}.

\section{Overview}
\label{sec:overview}

This section introduces \LBorrow{} informally, using a file logger as a running
example. \Cref{sec:running-example} motivates borrowing with the logger, and
\Cref{sec:challenges} identifies the challenges of capturing borrowing in a type
system. \Cref{sec:borrowability} to \Cref{sec:overview-functions} then introduce
and develop the idea of \emph{borrowability}.

\subsection{From Linear Types to Borrowing}
\label{sec:running-example}

\paragraph{Linear types}

Consider a logger with a primitive type $\Kw{File}$ for open files. The
operation $\Kw{open}: \Kw{string} \to \RefT\,\Kw{File}$ returns a linear
reference~\cite{wadlerLinearTypesCan1990}, and $\Kw{close}: \RefT\,\Kw{File}
\multimap ()$ consumes it and closes the file. Reading and writing preserve the
resource by returning its handle alongside the result: $\Kw{fread}:
\RefT\,\Kw{File} \multimap \Kw{string} \otimes \RefT\,\Kw{File}$ and
$\Kw{fwrite}: \RefT\,\Kw{File} \multimap \Kw{string} \multimap
\RefT\,\Kw{File}$.

With linear types, a file can neither be closed twice nor be left open. For
example, given \verb|f| returned by \verb|open|, only the first of these
programs is well-typed:

\begin{taggedbox}{linear types}
\begin{lstlisting}[language=calc, numbers=none, xleftmargin=2mm, aboveskip=0pt, belowskip=0pt]
let (s, f1) = fread f in close f1; s             -- well-typed
let (s, f1) = fread f in s                       -- ill-typed! f1 is never closed
let (s, f1) = fread f in close f1; close f1; s   -- ill-typed! f1 is closed twice
\end{lstlisting}
\end{taggedbox}

However, passing the handle to an operation transfers ownership, so the
operation must return it for the next one to use. Two writes and a close
therefore thread the handle through each step:

\begin{taggedbox}{linear types}
\begin{lstlisting}[language=calc, numbers=none, xleftmargin=2mm, aboveskip=0pt, belowskip=0pt]
let f1 = fwrite f "started" in let f2 = fwrite f1 "done" in close f2
\end{lstlisting}
\end{taggedbox}

\noindent
This \emph{capability-passing style} burdens the caller even when an operation
only needs temporary access to the file.

Moreover, a linear reference cannot be shared, even for reading. Suppose a
function \verb|cat| concatenates the contents of two files:

\begin{taggedbox}{linear types}
\begin{lstlisting}[language=calc, numbers=none, xleftmargin=2mm, aboveskip=0pt, belowskip=0pt]
cat f f   -- ill-typed! f is used twice
\end{lstlisting}
\end{taggedbox}

\noindent
The only fix is a second reference to the file, which is unnecessary and, if
all we have is \verb|f|, impossible. These two examples motivate borrowing: we
want to share access temporarily without transferring ownership through every
call.

\paragraph{Borrowing in Rust}
Rust supports these patterns through immutable and mutable borrows. An immutable borrow
\verb|&f| provides read-only access and can coexist with other immutable borrows:%
\footnote{Our Rust examples use an illustrative API rather than
\texttt{std::fs::File}: \texttt{open} returns a \texttt{Box<File>},
\texttt{close} consumes that box, and the other operations take borrows of
\texttt{File}. Each example starts with its own open handle \texttt{f}.}

\begin{taggedbox}{Rust}
\begin{lstlisting}[language=rust, numbers=none, xleftmargin=2mm, aboveskip=0pt, belowskip=0pt]
let s = cat(&f, &f);  // cat: fn(&File, &File) -> String
\end{lstlisting}
\end{taggedbox}

A mutable borrow \verb|&mut f| provides exclusive access. The two writes now
reuse the same handle:

\begin{taggedbox}{Rust}
\begin{lstlisting}[language=rust, numbers=none, xleftmargin=2mm, aboveskip=0pt, belowskip=0pt]
fwrite(&mut f, "started"); fwrite(&mut f, "done"); close(f)
\end{lstlisting}
\end{taggedbox}

\noindent
Each call temporarily borrows \verb|f|, and ownership stays with the caller. The
first borrow ends after the first call, so \verb|f| regains its permission and
can be borrowed again by the second.

Rust permits multiple immutable borrows or a single mutable borrow of the same
\emph{place} (a memory location such as a variable or a field) at any time, a
rule often called \emph{aliasing XOR mutability}. For example, Rust rejects
copying the file onto itself, which would read and overwrite it at once:

\begin{taggedbox}{Rust}
\begin{lstlisting}[language=rust, numbers=none, xleftmargin=2mm, aboveskip=0pt, belowskip=0pt]
copy(&f, &mut f);  // rejected! copy: fn(&File, &mut File)
\end{lstlisting}
\end{taggedbox}

Borrowing thus lifts both restrictions of linearity while still excluding
conflicting accesses. The question is how to capture it in a type system.

\subsection{Challenges in Typing Borrowing}
\label{sec:challenges}

We call a reference from which a borrow has been taken a \emph{lender}. (A
lender may be an owner, as in the logger, or itself a borrow when we
\textit{reborrow}.) Formalizing borrowing in a type system requires deciding
how to type a lender while a borrow is active and how to restore its permissions
when the borrow ends. In practice, this poses two challenges.

\paragraph{Challenge 1: Typing first-class lenders}

Suppose we want to package the eventual closing of the log into a closure while
a reader still uses the file:

\begin{taggedbox}{Rust}
\begin{lstlisting}[language=Rust, escapeinside={(*@}{@*)}, aboveskip=0pt, belowskip=0pt]
let r: &File = &f;         // f: Box<File>
let finish = (*@\normalcolor\texttt{||}@*) close(f);  // rejected! f is captured while borrowed
show(fread(r));
finish();
\end{lstlisting}
\end{taggedbox}

\noindent
The closure \verb|finish| (\verb+||+ is an empty parameter list) takes
ownership of \verb|f| because \verb|close| consumes it. It runs only after the
last use of \verb|r|, so the program is safe. Nevertheless, Rust rejects
capturing \verb|f| while it is borrowed, making lenders second-class values that
cannot be moved or captured like other references. Ideally, lenders should be
first-class: the type system should permit moving \verb|f| into the closure
while ensuring that the closure cannot close the file before the borrow ends.

Supporting this transfer requires encoding what \verb|f| can do while borrowed
in its type. However, tracking only the lender's \emph{current} access
permission is insufficient. Although an immutably borrowed lender only has read
access, it cannot be freely duplicated like an ordinary read-only reference
because each copy would regain full ownership when the borrow ends. Similarly, a
mutably borrowed owner has no current access to the referent, yet it cannot be
discarded because it still carries the obligation to close the file.
In Rust, the type of \verb|f| remains \verb|Box<File>| even while borrowed, and
borrowing restrictions are instead enforced by a separate \emph{borrow
  checker}~\cite{rustMIRBorrowChecker} over memory places, which forbids moving
\verb|f| into \verb|finish|.

\paragraph{Challenge 2: Restoring permissions non-lexically}

Once a borrow ends, the lender regains its permission. The simplest design ties
each borrow to a lexical scope and restores the lender when the scope ends. In
practice, however, this is often too restrictive. Suppose the logger reads an
existing log, but writes a header if the log is empty:

\begin{taggedbox}[unbreakable]{Rust}
\begin{lstlisting}[language=Rust, deletekeywords={[3]size}, aboveskip=0pt, belowskip=0pt]
let mut f = open("app.log");
let r = &f;
if size(r) > 0 { show(fread(r)) }    // borrow r ends after fread(r)
else { fwrite(&mut f, "header") }    // .. but before fwrite in this branch
\end{lstlisting}
\end{taggedbox}

\noindent
The borrow \verb|r| ends after \verb|fread(r)| in the \verb|if| branch but
before \verb|fwrite| in the \verb|else| branch, where \verb|fwrite| needs a
mutable borrow, so the span of \verb|r| does not align with any lexical scope.
Rust accepts this program by assigning each borrow a \emph{lifetime}, the set of
program points during which it is active. Notably, a borrow's lifetime is
determined by where the borrow is used in the control flow rather than by
lexical scope, and this feature is known as \emph{non-lexical lifetimes} (NLL).

To support non-lexical lifetimes, a type system for borrowing must restore the
lender's permissions when a borrow ends, at potentially different points across
control-flow branches, while ruling out any later uses of the expired borrow.
Moreover, restoration must happen in place without requiring the programmer to
explicitly return the borrow or reclaim the lender, which would reintroduce
capability passing.

\paragraph{Prior approaches: Two perspectives on borrowing}

Existing type systems for borrowing struggle to meet both challenges because of
how they view borrowing.
One perspective views borrowing as a \emph{temporary transfer of
  permission}, where the lender is made unavailable in the typing context during
the borrow and reinstated afterwards.
BoCa~\cite{wagnerLinearityBorrowing2025} keeps the lender outside the
borrower's context, while Affe~\cite{radanneKindlyBentFree2020} suspends its
binding within a lexical scope. Neither needs the borrower to return the borrow
when the scope ends, which addresses the lexical part of Challenge~2.
Oxide~\cite{weissOxideEssenceRust2021}, a formalization of Rust's type system, keeps
loan information in the typing context instead and thereby supports NLL. None
of the three, however, lets a lender be moved while borrowed, so Challenge~1
remains.

Alternatively, one can view a borrow as a \emph{split of permissions}
recorded in types. Fractional
uniqueness~\cite{marshallFunctionalOwnershipFractional2024}, for example,
splits a borrow into fractional references that can be passed around within the borrow's
scope. Recovering full access, however, requires explicitly joining the
fractions and returning the result,
which falls short of Challenge~2 and brings back the capability-passing burden.
\Cref{sec:related} compares these systems in more detail~(\Cref{tbl:borrowing-foundations}).

\paragraph{Our approach: Borrowability}
Different as they may seem, the two perspectives are complementary and can be
combined. Our answer to both challenges rests on separating the \emph{access
  permission} of a reference, from the owner's
linear \emph{obligation} to eventually release the resource. A borrow takes over
some or all of its lender's access permission, while the obligation stays with
the owner. The lender thus remains an ordinary value whose type records both the
access permission it has left and the borrow's lifetime, while the owner keeps
its obligation however it is moved (Challenge~1). When the borrow's lifetime
ends, the lent access permission is restored in the lender's type in place,
requiring no return or join operation (Challenge~2). This leads us to the notion
of \emph{borrowability}, which we develop in the next subsection.

\subsection{Borrowability}
\label{sec:borrowability}

We now present the definition of \textit{borrowability}, the key notion in our calculus.
\[
\begin{array}{rcl@{\qquad}l}
a, b, c, p & & & \text{lifetime variable} \\
\alpha & ::= & a \mid a^{\Mut} & \text{borrowability atom} \\
\rho & ::= & U \mid \alpha \mid \rho - \alpha & \text{borrowability} \\
\end{array}
\]
A borrowability describes the permission a reference holds: $U$ is full ownership, an atom $a$ or
$a^{\Mut}$ is an immutable or mutable borrow with lifetime $a$, and $\rho - \alpha$ is a
reference from which the borrow $\alpha$ has been taken. Owned values and borrows are all
pointers to places in memory that differ only in their permissions, so we give them a
single reference type $\RefT^{\rho}\tau$.

For example, take our log file $f : \RefT^{U}\,\Kw{File}$ and borrow it mutably, naming
the borrow $m$ (how borrows are created is the subject of
\Cref{sec:borrowing-context-splitting}). After borrowing, the two references are typed as
\[
f: \RefT^{U - a^\Mut}\,\Kw{File}, \quad m: \RefT^{a^\Mut}\,\Kw{File}
\]
Here, $U - a^\Mut$ is the residual borrowability of $f$ after giving the borrow $a^\Mut$
to $m$, recording that $f$ originally held $U$ and will recover $U$ once
lifetime $a$ ends. A borrowability thus describes both the permission currently
available through a reference and how it changes over time; that is, it
captures both the \emph{span} and the \emph{permission} of a borrow.

For the span of a borrow, at each program point,
we maintain a set of live lifetimes. Lifetimes are introduced and ended by
explicit operations: $\Kw{letlft}\ a$ creates a fresh live lifetime $a$,
optionally with \emph{outlives constraints} $a :> b$ stating that $b$ outlives
$a$, and $\Kw{endlft}\ a$ ends it. A reference with lifetime $a$ in its type,
e.g. $\RefT^{a}\tau$ or $\RefT^{a^\Mut}\tau$, is only well-formed when $a$ is
alive. We now turn to permissions.

A \textit{permission} $\pi = (\beta, d)$ pairs an \textit{access permission}
$\beta \in \{\Kw{w}, \Kw{r}, \emptyset\}$ (write, read, or no access) with a \textit{drop
permission} $d \in \{\bullet, \circ\}$ (an obligation to drop, or no obligation). The
obligation is linear, preventing double-free and resource leaks, and stays with the
owner, as otherwise transferring it to a borrower would allow the borrower to close the owner's file.

\textit{Permission extraction} $\Erase{\rho}$ reads the permission $\pi$ from a borrowability
$\rho$. For the three atomic~forms,
\begin{align*}
\Erase{U} &= (\Kw{w}, \bullet) &
\Erase{a} &= (\Kw{r}, \circ) &
\Erase{a^{\Mut}} &= (\Kw{w}, \circ)
\end{align*}
An owned reference carries both write access permission and the obligation to drop. Borrows
carry no obligation: an immutable borrow may only read, while a mutable borrow may write.

The $\Kw{File}$ interface from \Cref{sec:running-example} is now typed as:
\[
  \begin{array}{l@{\qquad\qquad}l}
    \Kw{open}: \Kw{string} \to \RefT^{U}\,\Kw{File} & \Kw{fread}: \RefT^{a}\,\Kw{File} \multimap \Kw{string} \\
    \Kw{close}: \RefT^{U}\,\Kw{File} \multimap ()    & \Kw{fwrite}: \RefT^{a^\Mut}\,\Kw{File} \multimap \Kw{string} \multimap ()
  \end{array}
\]
$\Kw{open}$ returns a writable owned handle that $\Kw{close}$ must eventually
consume. $\Kw{fread}$ and $\Kw{fwrite}$ take immutable and mutable borrows,
respectively, for any lifetime $a$ (we discuss lifetime polymorphism in
\Cref{sec:overview-functions}, \nameref{par:lifetime-polymorphism}). Neither
carries the closing obligation or has to return its borrow, avoiding the handle
threading of the linear interface in \Cref{sec:running-example}.

\begin{figure}[tbp]
  \vspace{-10pt}
  \centering
  \small
  \noindent\makebox[\linewidth]{%
    \(\boxed{\beta \leadsto \beta_1 \oplus \beta_2}\)\hfill\textit{(Access Permission Split)}}
  \vspace{-8pt}
  \begin{mathpar}
    \inferrule*[right=\rdef{Q-Write}]{ }{\Kw{w} \leadsto \Kw{r} \oplus \Kw{r}}
    \and
    \inferrule*[right=\rdef{Q-Read}]{ }{\Kw{r} \leadsto \Kw{r} \oplus \Kw{r}}
    \and
    \inferrule*[right=\rdef{Q-Empty}]{ }{\beta \leadsto \beta \oplus \emptyset}
    \and
    \inferrule*[right=\rdef{Q-Sym}]{\beta \leadsto \beta_2 \oplus \beta_1}{\beta \leadsto \beta_1 \oplus \beta_2}
  \end{mathpar}
  \smallskip
  \begin{minipage}{0.55\textwidth}
  \noindent\makebox[\linewidth]{%
    \(\boxed{d \leadsto d_1 \oplus d_2}\)\hfill\textit{(Drop Permission Split)}}
  \vspace{-6pt}
  \begin{mathpar}
    \inferrule*[right=\rdef{D-Keep}]{ }{d \leadsto d \oplus \circ}
    \and
    \inferrule*[right=\rdef{D-Sym}]{d \leadsto d_2 \oplus d_1}{d \leadsto d_1 \oplus d_2}
  \end{mathpar}
  \end{minipage}
  \hfill
  \begin{minipage}{0.4\textwidth}
    \noindent\makebox[\linewidth]{%
      \(\boxed{\pi \leadsto \pi_1 \oplus \pi_2}\)\hfill\textit{(Permission Split)}}
    \vspace{-6pt}
    \begin{mathpar}
      \inferrule*[right=\rdef{P-Split}]{\beta \leadsto \beta_1 \oplus \beta_2 \\ d \leadsto d_1 \oplus d_2}{(\beta, d) \leadsto (\beta_1, d_1) \oplus (\beta_2, d_2)}
    \end{mathpar}
  \end{minipage}
  \caption{Permission Split}
  \label{fig:permission-split}
  \label{fig:access-permission-split}
\end{figure}

The remaining case, $\Erase{\rho - \alpha}$, needs a way to divide a permission between a
borrow and what remains of the lender. \Cref{fig:permission-split} defines
\textit{permission split} $\pi \leadsto \pi_1 \oplus \pi_2$, pointwise on access and drop
permissions. For access permissions, a write splits into two reads (\rref{Q-Write}), a
read splits into two reads (\rref{Q-Read}), and any permission may give away nothing
(\rref{Q-Empty}). Notice that \rref{Q-Read} allows for duplicability of read-only
permission, which corresponds to the intuition that immutable references admit
contraction. For drop permissions, \rref{D-Keep} says that the obligation (if present) stays with
exactly one side, so borrowing never duplicates it.

With permission split in hand, the subtracted case for permission extraction is defined as
\begin{align*}
\Erase{\rho - \alpha} &= \max\{\, \pi \mid \Erase{\rho} \leadsto \pi \oplus \Erase{\alpha} \,\}
  && \emptyset < \Kw{r} < \Kw{w}
\end{align*}

\begin{wraptable}{r}{0.5\textwidth}
\centering
\small
\setlength{\tabcolsep}{4pt}
\begin{tabular}{@{}l@{\hspace{8pt}}lc@{}}
\toprule
Rust type & \LBorrow type & Permission \\
\midrule
\lstinline[language=Rust]|Box<T>| & \(\RefT^U\, T\) & \((\Kw{w}, \bullet)\) \\
\lstinline[language=Rust]|&'a T| & \(\RefT^a\, T\) & \((\Kw{r}, \circ)\) \\
\lstinline[language=Rust]|&'a mut T| & \(\RefT^{a^{\Mut}}\, T\) & \((\Kw{w}, \circ)\) \\
\lstinline[language=Rust]|Box<T>| mutably & \(\RefT^{U - a^{\Mut}}\, T\) & \((\emptyset, \bullet)\) \\
\quad borrowed by \lstinline[language=Rust]|&'a mut| & & \\
\lstinline[language=Rust]|Box<T>| immutably & \(\RefT^{U - a}\, T\) & \((\Kw{r}, \bullet)\) \\
\quad borrowed by \lstinline[language=Rust]|&'a| & & \\
\bottomrule
\end{tabular}
\vspace{7pt}
\caption{Comparison between Rust and \LBorrow types}
\label{tbl:rust-lborrow-types-comparison}
\end{wraptable}%
\noindent This gives the largest permission that can remain with the lender after
splitting off $\Erase{\alpha}$. In our example,
$\Erase{U - a^\Mut} = (\emptyset, \bullet)$: $f$ loses all access
while $m$ is active, but retains its closing obligation. An immutable borrow
instead leaves $\Erase{U - a} = (\Kw{r}, \bullet)$, retaining both read access
and the obligation.

\Cref{tbl:rust-lborrow-types-comparison} contrasts borrowability with Rust.
Rust leaves the owner's ordinary type unchanged during borrowing, as discussed
in \Cref{sec:challenges}. \LBorrow{} instead makes ownership and its obligation
explicit in $\RefT^U\tau$, and records outstanding borrows in
$\RefT^{\rho-\alpha}\tau$.

\WFclear
\subsection{Borrowing as Context Splitting}
\label{sec:borrowing-context-splitting}

Recall the mutable borrow $m$ of our log file $f : \RefT^{U}\,\Kw{File}$ from
\Cref{sec:borrowability}.
Where does $m$ come from? In \LBorrow, a borrow is created by an ordinary
binding:
\begin{taggedbox}{\LBorrow}
\begin{lstlisting}[language=calc, escapeinside={(*@}{@*)}, aboveskip=0pt, belowskip=0pt]
let m: (*@\lstann{9}{\RefT^{a^\Mut}\,\Kw{File}}@*) = f in        -- (*@$f : \RefT^{a^\Mut}\,\Kw{File}$ in the bound expression@*)
... m ... f ...                -- (*@$f : \RefT^{U-a^\Mut}\,\Kw{File},\quad m : \RefT^{a^\Mut}\,\Kw{File}$ in the body@*)
\end{lstlisting}
\end{taggedbox}
In a linear type system this is ill-typed: the typing context is split between
subexpressions, each variable goes to exactly one side, while $f$ is needed on
both. In \LBorrow, a single reference can instead be split into two, one for each side
of the context split: the bound expression gets $f : \RefT^{a^\Mut}\,\Kw{File}$, and
the body keeps $f : \RefT^{U - a^\Mut}\,\Kw{File}$. Notably, no borrow syntax like
Rust's \verb|&f| is needed.

Borrowability split lifts permission split (\Cref{fig:permission-split}) through
extraction. Intuitively, $\rho \leadsto \rho_1 \oplus \rho_2$ if
$\Erase{\rho} \leadsto \Erase{\rho_1} \oplus \Erase{\rho_2}$.
Reference split (\hyperlink{rule:S-Ref:overview}{\textsc{S-Ref}}) follows directly: a reference $\RefT^{\rho}\tau$ splits
into $\RefT^{\rho_1}\tau$ and $\RefT^{\rho_2}\tau$ whenever $\rho$ splits into $\rho_1$
and $\rho_2$ and both resulting references are well-formed.
\Cref{fig:ref-split} shows the rule with three representative borrowability
splits and their underlying permission splits.

\begin{figure}[htb]
  \centering
  \small
  \vspace{-3ex}
  \begin{mathpar}
    \inferrule*[right=\hypertarget{rule:S-Ref:overview}{\textsc{S-Ref}}]
    { \Delta \vDash \rho \leadsto \rho_1 \oplus \rho_2 \\
      \Gpre{\Delta \vDash \RefT^{\rho_1}\tau} \\
      \Gpre{\Delta \vDash \RefT^{\rho_2}\tau} }
    { \Delta \vDash \RefT^{\rho}\tau \leadsto \RefT^{\rho_1}\tau \oplus \RefT^{\rho_2}\tau }
  \end{mathpar}
  \[
    \begin{aligned}
      &U      & \leadsto && (U-a)              & \,\oplus\, a                                    && \qquad (\Kw{w}, \bullet) & \leadsto && (\Kw{r}, \bullet)    &\,\oplus\, (\Kw{r}, \circ)     & \textit{(immutable borrow)} \\
      &U      & \leadsto && (U-a^\Mut)         & \,\oplus\, a^\Mut                               && \qquad (\Kw{w}, \bullet) & \leadsto && (\emptyset, \bullet) &\,\oplus\, (\Kw{w}, \circ)  & \textit{(mutable borrow)} \\
      &a^\Mut & \leadsto && (a^\Mut - b^\Mut)  & \,\oplus\, b^\Mut \; \text{if } a <: b \enspace && \qquad (\Kw{w}, \circ)   & \leadsto && (\emptyset, \circ)   &\,\oplus\, (\Kw{w}, \circ)      & \textit{(mutable reborrow)}
    \end{aligned}
  \]
  \caption{Reference split, with representative borrowability \& permission splits}
  \label{fig:ref-split}
  \label{fig:rep-borrow-splits}
\end{figure}

Here we illustrate the first two cases with an example program, and we will explain
the third case in \Cref{sec:overview-mem-ops} (\nameref{par:reborrowing}).
Consider the following program that opens the log file and writes to it through a
mutable borrow:

\begin{taggedbox}[unbreakable]{\LBorrow}
\begin{lstlisting}[language=calc, escapeinside={(*@}{@*)}, aboveskip=0pt, belowskip=0pt]
let f = open "app.log" in      -- (*@$f : \RefT^{U}\,\Kw{File}$@*)
letlft a;                      -- (*@introduces a fresh lifetime $a$@*)
let m: (*@\lstann{9}{\RefT^{a^\Mut}\,\Kw{File}}@*) = f in        -- (*@$f : \RefT^{U-a^\Mut}\,\Kw{File},\quad m : \RefT^{a^\Mut}\,\Kw{File}$@*)
fwrite m "started"; ...
\end{lstlisting}
\end{taggedbox}

Line 1 creates the owned handle. Line 2 introduces a fresh lifetime $a$, which
remains live until $\Kw{endlft}$ ends it (\Cref{sec:lifetime-ending}). Line 3 applies
$U \leadsto (U-a^\Mut) \oplus a^\Mut$ from \Cref{fig:ref-split}: $m$ receives
mutable access for $\Kw{fwrite}$, while $f$ retains only its closing obligation.
No further borrow of $f$ is possible while $a$ is live, since there is no $\alpha'$
such that $\Erase{U-a^\Mut} \leadsto \Erase{U-a^\Mut-\alpha'} \oplus \Erase{\alpha'}$.

Had line 3 taken an immutable borrow $r : \RefT^{a}\,\Kw{File}$ instead, $f$ would keep
$\RefT^{U-a}\,\Kw{File}$, permitting reads through $r$ and $f$ but no writes or closing
until $a$ ends. Since $\Kw{r} \leadsto \Kw{r} \oplus \Kw{r}$, further immutable borrows
are allowed, supporting \verb|cat f f| from \Cref{sec:running-example}. No sequence of
splits, however, yields an immutable and a mutable borrow of $f$ at once, so the analogue
of \verb|copy(&f, &mut f)| is rejected.

In each case, borrowing is implicit in context splitting, and an algorithmic
implementation would infer where to split. Moreover, the lender remains an ordinary
reference whose type records its remaining permissions.

\subsection{Lifetime Ending}
\label{sec:lifetime-ending}

A borrow ends when its lifetime ends, which \LBorrow marks explicitly with
$\Kw{endlft}$. Recall from \Cref{sec:borrowability} that every
borrowability has the form $\rho_0 - \overline{\alpha}$, consisting of a
\emph{head} $\rho_0 \in \{U, a, a^\Mut\}$ and a sequence of
\emph{subtractions} $\overline{\alpha}$. The head records the reference's initial borrowability, while each
subtraction in $\overline{\alpha}$ records an active borrow currently taken from
it.

Ending lifetime $a$ requires that no reference in the context still has $a$ at
its head; namely, all borrows of lifetime $a$ must be gone. Since no one then
holds the permission that was split off under $a$, it is safe to hand it back: \textit{restoration} $\rho\Rst{a}$ removes every subtraction $-a$ and
$-a^\Mut$ from $\rho$, and is lifted pointwise to types and contexts
(\Cref{fig:restoration}). Restoration is a type-only strong update: it
changes the type of the lender in the context, with no runtime effect. Indeed, all
lifetime operations, $\Kw{letlft}$ and $\Kw{endlft}$ included, are no-ops at runtime
(\Cref{sec:opsem}).

\begin{figure}[htb]
  \centering
  \small
  \vspace{-2.5ex}
  \[
  \begin{array}{l@{\qquad}r@{\;}c@{\;}l@{\quad}l@{\hspace{3em}}r@{\;}c@{\;}l}
    U\Rst{\bar a} = U
      & \alpha\Rst{\bar a} & = & \alpha & \alpha \in \{b, b^{\Mut}\},\ b \notin \bar a
      & (\RefT^{\rho}\tau)\Rst{\bar a} & = & \RefT^{\rho\Rst{\bar a}}(\tau\Rst{\bar a}) \\
      & (\rho - \alpha)\Rst{\bar a} & = & \rho\Rst{\bar a} & \alpha \in \{b, b^{\Mut}\},\ b \in \bar a
      & (\Gamma, x : \tau)\Rst{\bar a} & = & \Gamma\Rst{\bar a}, x : \tau\Rst{\bar a} \\
      & (\rho - \alpha)\Rst{\bar a} & = & (\rho\Rst{\bar a}) - \alpha & \alpha \in \{b, b^{\Mut}\},\ b \notin \bar a
      & & & \\
  \end{array}
  \]
  \caption{Restoration}
  \label{fig:restoration}
  \vspace{-3pt}
\end{figure}

\begin{wrapfigure}{r}{0.5\textwidth}
\vspace{-\baselineskip}
\begin{taggedbox}[unbreakable]{\LBorrow}
\begin{lstlisting}[language=calc, escapeinside={(*@}{@*)}, aboveskip=0pt, belowskip=0pt, firstnumber=4]
fwrite m "started";  -- (*@consumes $m$@*)
endlft a;            -- (*@$f : \RefT^{U}\,\Kw{File}$@*)
close f
\end{lstlisting}
\end{taggedbox}
\vspace{-\baselineskip}
\end{wrapfigure}
Let us continue the program of \Cref{sec:borrowing-context-splitting}, whose first three
lines leave the context $f : \RefT^{U-a^\Mut}\,\Kw{File}, m : \RefT^{a^\Mut}\,\Kw{File}$.
We then end the mutable borrow before finally closing the file.
Line 4 consumes $m$. In line 5, ending $a$ removes $-a^\Mut$ from $f$'s type and
restores $f : \RefT^{U}\,\Kw{File}$, allowing line 6 to close it.

We give another example with non-lexical lifetimes.
The following program mirrors the Rust program
in \Cref{sec:challenges}. Like $\Kw{fread}$, the size test
$\Kw{size}: \RefT^{a}\,\Kw{File} \multimap \Kw{int}$ only needs an immutable borrow, and
since $\RefT^{a}\,\Kw{File}$ is unrestricted, both can use the same borrow $r$:

\begin{taggedbox}{\LBorrow}
\begin{lstlisting}[language=calc, escapeinside={(*@}{@*)}, aboveskip=0pt, belowskip=0pt]
let f = open "app.log" in
letlft a;
let r: (*@\lstann{7}{\RefT^{a}\,\Kw{File}}@*) = f in          -- (*@$f : \RefT^{U-a}\,\Kw{File},\quad r : \RefT^{a}\,\Kw{File}$@*)
if size r > 0 then (let s = fread r in endlft a; close f; show s)
     else (endlft a; letlft b; fwrite f "header"; endlft b; close f)
\end{lstlisting}
\end{taggedbox}

\noindent
After line 3, the typing context is
$f : \RefT^{U-a}\,\Kw{File}, r : \RefT^{a}\,\Kw{File}$.
The two branches can end lifetime $a$ at different program points, as long as
they agree on the set of live lifetimes when they join. The \texttt{then} branch
reads through $r$ first and ends $a$ afterwards. The \texttt{else} branch never
uses $r$, so it discards $r$ (weakening, allowed since a borrow carries no
obligation) and ends $a$ right away. This restores $f$ to
$\RefT^{U}\,\Kw{File}$, so that it can be written to, through a fresh borrow at
lifetime $b$, and then closed.

\subsection{Memory Operations}
\label{sec:overview-mem-ops}

Examples so far have used the primitive $\Kw{File}$ interface in
\Cref{sec:borrowability}. We now turn to \LBorrow{}'s own memory operations:
$\RefT$, $\Kw{swap}$, $\Kw{read}$, and $\Kw{drop}$.
Consider a logging service that periodically replaces the active file in
$log: \RefT^U (\RefT^U\,\Kw{File})$. The inner reference is a file handle
managed by $\Kw{open}$ and $\Kw{close}$; the outer one is a memory cell allocated
and freed by \LBorrow{}'s $\RefT$ and $\Kw{drop}$.

\paragraph{Swapping and dropping}
Switching the log file replaces the old handle. A cell's contents may carry a
drop obligation, as the handle does, and must not be discarded. A naive
\(\Kw{store}: \RefT^{a^\Mut}\tau \multimap \tau \multimap ()\)
returns nothing and would overwrite and leak the old handle. Instead,
\LBorrow{} provides
\[\Kw{swap}: \RefT^{a^\Mut} \tau \multimap \tau \multimap \RefT^{a^\Mut} \tau \otimes \tau\]
A \textit{swap} writes the new value and returns both the mutable reference and
the old value,
allowing us to close the old file in the following example:

\begin{taggedbox}{\LBorrow}
\begin{lstlisting}[language=calc, escapeinside={(*@}{@*)}, aboveskip=0pt, belowskip=0pt]
let log = ref (open "app.log") in  -- (*@$log : \RefT^{U}(\RefT^{U}\,\Kw{File})$@*)
letlft a;
let m: (*@\lstann{14}{\RefT^{a^\Mut}(\RefT^{U}\,\Kw{File})}@*) = log in     -- (*@$log : \RefT^{U - a^{\Mut}}(\RefT^{U}\,\Kw{File}),\quad m : \RefT^{a^{\Mut}}(\RefT^{U}\,\Kw{File})$@*)
let (m', old_f) = swap m (open "new.log") in  -- (*@$m' : \RefT^{a^{\Mut}}(\RefT^{U}\,\Kw{File}),\quad old\_f : \RefT^{U}\,\Kw{File}$@*)
close old_f;
endlft a;                          -- (*@$log : \RefT^{U}(\RefT^{U}\,\Kw{File})$@*)
let f = drop log in close f        -- (*@frees the cell; $f : \RefT^{U}\,\Kw{File}$@*)
\end{lstlisting}
\end{taggedbox}

\noindent
Deallocation follows the same principle: $\Kw{drop}: \RefT^{U}\,\tau \multimap \tau$
frees an owned cell and returns its contents, whose obligation, if any, then
falls to the caller. Line 7 thus frees $log$ and closes its file.

\paragraph{Reading}
\LBorrow{} inspects the value behind a reference with \textit{read}. Its typing is
subtle because the value being read may itself contain references, and a rule that
simply returns the referent type is unsound.
For example, suppose we take an immutable borrow
$r : \RefT^{a}(\RefT^{U}\,\Kw{File})$ of $log$ to inspect the
active log file. If reading through $r$ returned $\RefT^U\,\Kw{File}$, we
would obtain full ownership of the inner file handle and could close the active log
file while $log$ still points to it:

\begin{taggedbox}{\LBorrow}
\begin{lstlisting}[language=calc, escapeinside={(*@}{@*)}, aboveskip=0pt, belowskip=0pt]
let r: (*@\lstann{12}{\RefT^{a}(\RefT^{U}\,\Kw{File})}@*) = log in  -- (*@$log : \RefT^{U - a}(\RefT^{U}\,\Kw{File}),\quad r : \RefT^{a}(\RefT^{U}\,\Kw{File})$@*)
let f: (*@\lstann{7}{\RefT^{U}\,\Kw{File}}@*) = read r in
close f;                      -- (*@we accidentally close the active log file@*)
endlft a;                     -- (*@unsound: file closed but $log: \RefT^{U}(\RefT^{U}\,\Kw{File})$@*)
\end{lstlisting}
\end{taggedbox}

Instead, reading through a reference of borrowability $\alpha$ produces a restricted
\textit{read view} $\Rview{\alpha}{\tau}$ of the referent type $\tau$:
\[
\Kw{read}: \RefT^\alpha\,\tau \multimap \Rview{\alpha}{\tau}
\]
\noindent
If the referent type is duplicable, such as a base type or an immutable reference,
the read view is the type itself, since its contents can be copied safely.
If it is itself an owned or mutable reference, call it the inner reference, the read
view is a borrow of the inner reference whose lifetime is shortened to the outer
reference's and whose access permission is the meet of the two.
No more permission thus comes out than the outer reference grants,
and no borrow outlives the reference it was read through. The problematic
example is now safe: reading $\RefT^a (\RefT^U\,\Kw{File})$ returns
$\Rview{a}{\RefT^U\,\Kw{File}}=\RefT^a\,\Kw{File}$, an immutable borrow that can
be inspected with $\Kw{size}$ but cannot be passed to $\Kw{close}$:

\begin{taggedbox}{\LBorrow}
\begin{lstlisting}[language=calc, escapeinside={(*@}{@*)}, aboveskip=0pt, belowskip=0pt]
let r: (*@\lstann{12}{\RefT^{a}(\RefT^{U}\,\Kw{File})}@*) = log in  -- (*@$log : \RefT^{U - a}(\RefT^{U}\,\Kw{File}),\quad r : \RefT^{a}(\RefT^{U}\,\Kw{File})$@*)
let f: (*@\lstann{7}{\RefT^{a}\,\Kw{File}}@*) = read r in
let sz = size f in            -- (*@inspects $f$; $\Kw{close}\ f$ would be ill-typed@*)
endlft a; ...                 -- (*@$log : \RefT^{U}(\RefT^{U}\,\Kw{File})$@*)
\end{lstlisting}
\end{taggedbox}

Unlike $\Kw{swap}$, $\Kw{read}$ \textit{consumes} its argument:
its type returns only the read view. Returning a mutable borrow
$\RefT^{a^\Mut}(\RefT^U\,\Kw{File})$ alongside its read view
$\RefT^{a^\Mut}\,\Kw{File}$ would leave two mutable references at $a$, allowing
a swap through the outer one to close the file while the view is still in use.
Moreover, consuming an owned reference
without freeing it would discard its drop obligation. Therefore, $\Kw{read}$ accepts
only $\RefT^\alpha\,\tau$ with $\alpha \in \{a, a^\Mut\}$, which is why the
example first borrows $r$ from $log$.

\paragraph{Reborrowing}
\label{par:reborrowing}

Consumption also matters when we want to switch the log file based on its
current size. Suppose we hold a mutable borrow
$m : \RefT^{a^\Mut}(\RefT^U\,\Kw{File})$. Reading $m$ to check
the size would consume it, and we could no longer update it via $\Kw{swap}$. The
remedy, as with $log$ above, is to borrow from $m$ under a
shorter lifetime $b$ outlived by $a$ and read through that borrow instead. A
borrow taken from another borrow is called a \textit{reborrow}:

\begin{taggedbox}{\LBorrow}
\begin{lstlisting}[language=calc, escapeinside={(*@}{@*)}, aboveskip=0pt, belowskip=0pt]
letlft b :> a;                      -- (*@$m : \RefT^{a^{\Mut}}(\RefT^{U}\,\Kw{File})$@*)
let r: (*@\lstann{12}{\RefT^{b}(\RefT^{U}\,\Kw{File})}@*) = m in          -- (*@reborrow; $m : \RefT^{a^{\Mut} - b}(\RefT^{U}\,\Kw{File}),\quad r : \RefT^{b}(\RefT^{U}\,\Kw{File})$@*)
let f: (*@\lstann{7}{\RefT^{b}\,\Kw{File}}@*) = read r in          -- (*@consumes $r$@*)
let sz = size f in
endlft b;                           -- (*@$m : \RefT^{a^{\Mut}}(\RefT^{U}\,\Kw{File})$@*)
if sz > 1000 then let (m', old_f) = swap m (open "new.log") in
                  close old_f; m'
else m
\end{lstlisting}
\end{taggedbox}

Here, $\Kw{letlft}\ b :> a$ introduces a fresh lifetime $b$ shorter than $a$. Splitting
$m$ reborrows it as a temporary immutable reference
$r : \RefT^b(\RefT^U\,\Kw{File})$, which we read, while $m$ keeps
$\RefT^{a^\Mut - b}(\RefT^U\,\Kw{File})$ and so cannot be mutated while the reborrow
is in use. Once $\Kw{endlft}\ b$ ends $b$, $m$ recovers its full mutable
borrowability and can be passed to $\Kw{swap}$. The example reborrows $m$
immutably. A mutable reborrow, the third case of \Cref{fig:ref-split}, likewise
splits $m$ at a shorter lifetime $b$, leaving
$m : \RefT^{a^\Mut - b^\Mut}(\RefT^U\,\Kw{File})$ until $b$ ends. This
lets several mutating calls use $m$ in turn without the first consuming it.

\subsection{Closures}
\label{sec:overview-functions}

\paragraph{Lifetime polymorphism}
\label{par:lifetime-polymorphism}
\begin{wrapfigure}{r}{0.6\textwidth}
\vspace{-\baselineskip}
\begin{taggedbox}[unbreakable]{\LBorrow}
\begin{lstlisting}[language=calc, escapeinside={(*@}{@*)}, aboveskip=0pt, belowskip=0pt]
let update = lam m.
  let (m, old_f) = swap m (open "new.log") in
  close old_f
\end{lstlisting}
\end{taggedbox}
\vspace{-\baselineskip}
\end{wrapfigure}
We can package the file-switching logic from \Cref{sec:overview-mem-ops} into a
reusable function $update$ that takes a mutable borrow of the active log
cell and swaps in a fresh file.
What type should $update$ have? Its parameter is a mutable borrow of
the log cell, $\RefT^{p^\Mut}(\RefT^U\,\Kw{File})$ for some lifetime $p$, and
the body is well-typed only when $p$ is live. Rather than fixing $p$ to one
lifetime, $update$ quantifies over it, so that it accepts log cells
borrowed at any lifetime:
\[
update : \forall p.\, \RefT^{p^\Mut}(\RefT^U\,\Kw{File}) \multimap^{p} ()
\]
The annotation on the arrow records that $p$ must be live when
$update$ is called. For example:

\begin{taggedbox}{\LBorrow}
\begin{lstlisting}[language=calc, escapeinside={(*@}{@*)}, aboveskip=0pt, belowskip=0pt]
let log = ref (open "app.log") in  -- (*@$log : \RefT^{U}(\RefT^{U}\,\Kw{File})$@*)
letlft a; update log; endlft a;    -- (*@borrows $log$ at lifetime $a$ for the call@*)
let f = drop log in close f        -- (*@$log : \RefT^{U}(\RefT^{U}\,\Kw{File})$ restored@*)
\end{lstlisting}
\end{taggedbox}
Line 2 instantiates $p$ with $a$ and implicitly splits $log$ into the
caller's lender $\RefT^{U-a^\Mut}(\RefT^U\,\Kw{File})$ and the argument
$\RefT^{a^\Mut}(\RefT^U\,\Kw{File})$. After the call, $\Kw{endlft}\ a$ restores
$log$ to $\RefT^U(\RefT^U\,\Kw{File})$.

\paragraph{Lifetime effects}
Suppose we want to parse the active log file. If the file is corrupted, we close
$log$ immediately; otherwise, we parse its body and close
$log$ afterwards. In both cases, we record a final summary of the
file before closing. To avoid duplicating the closing steps, we bundle them into a
$shutdown$ closure:
\begin{taggedbox}[unbreakable]{\LBorrow}
\begin{lstlisting}[language=calc, escapeinside={(*@}{@*)}, aboveskip=0pt, belowskip=0pt]
let log = ref (open "app.log") in
letlft a;
let r: (*@\lstann{12}{\RefT^{a}(\RefT^{U}\,\Kw{File})}@*) = log in       -- (*@$log : \RefT^{U - a}(\RefT^U\,\Kw{File}),\quad r : \RefT^a(\RefT^U\,\Kw{File})$@*)
let f: (*@\lstann{7}{\RefT^{a}\,\Kw{File}}@*) = read r in
let shutdown = lam u.
  summarize f;
  endlft a;                        -- (*@ends captured lifetime $a$; $log : \RefT^U(\RefT^U\,\Kw{File})$@*)
  let old_f = drop log in          -- (*@$old\_f : \RefT^U\,\Kw{File}$@*)
  close old_f
in
if is_corrupted f then shutdown () -- (*@ends $a$ and closes $log$@*)
else parse_body f; shutdown ()     -- (*@reads $f$, then ends $a$ and closes $log$@*)
\end{lstlisting}
\end{taggedbox}
Because $f : \RefT^a\,\Kw{File}$ is an immutable borrow, and hence
unrestricted, both the closure and the caller can use it. The closure captures
$f$ together with the first-class lender
$log : \RefT^{U-a}(\RefT^U\,\Kw{File})$, while the caller keeps
reading $f$ until it calls $shutdown$. Rust rejects this capture
because a borrowed owner cannot be moved, just as for \verb|finish| in
\Cref{sec:challenges}; $shutdown$ is its \LBorrow{} counterpart.

Unlike $update$, $shutdown$ captures references at a fixed
lifetime $a$. The type $() \multimap^{a} ()$ alone, however, would hide the
ending of $a$, allowing the caller to keep reading $f$ after it has
dangled. \LBorrow{} therefore records lifetime ending as a \textit{lifetime effect}:
\[
shutdown : () \multimap^{a \Ends{\{a\}}} ()
\]
The effect $a \Ends{\{a\}}$ states that $a$ must be live when
$shutdown$ is called and is ended when it returns.
As a result, for the caller,
$a$ must stay live while $shutdown$ is in scope, and the call
ends $a$ just as $\Kw{endlft}\ a$ would: borrows at $a$ such as $f$
become unusable, and lenders borrowed at $a$ are restored.

\section{A Calculus for Borrowability}
\label{sec:calculus}

This section formalizes \LBorrow{}, presenting its
syntax (\Cref{sec:calculus-syntax}) and type system~(\Cref{sec:calculus-typing}).

\begin{figure}[tb]
\small
\vspace{-2ex}
\[
\begin{array}{rcl@{\quad}l@{\qquad\quad}rcl@{\quad}l}
a, b, c, p & & & \text{lifetime variable}
  & \beta & ::= & \Kw{w} \mid \Kw{r} \mid \emptyset & \text{access permission} \\
\alpha & ::= & a \mid a^{\Mut} & \text{borrowability atom}
  & d & ::= & \bullet \mid \circ & \text{drop permission} \\
\rho & ::= & U \mid \alpha \mid \rho - \alpha & \text{borrowability}
  & \pi & ::= & (\beta, d) & \text{permission} \\
& & & & q & ::= & \Lin \mid \Aff \mid \Un & \text{qualifier} \\[0.3em]
\tau & ::= & \multicolumn{5}{l@{\quad}}{() \mid \Kw{int} \mid \Kw{bool} \mid \tau_1 \otimes \tau_2 \mid \RefT^{\rho}\tau
  \mid \ArrT{p}{\Delta_f}{L}{q}{\tau_1}{\tau_2}} & \text{type} \\[0.3em]
e & ::= & \multicolumn{5}{l@{\quad}}{x \mid \lambda x.\, e \mid e_1\, e_2 \mid \Kw{let}\ x = e_1\ \Kw{in}\ e_2 \mid e_1; e_2 \mid () \mid n \mid \Kw{true} \mid \Kw{false} \mid (e_1, e_2)} & \text{term} \\
& \mid & \multicolumn{5}{l@{\quad}}{\Kw{let}\ (x, y) = e_1\ \Kw{in}\ e_2 \mid \RefT\ e \mid \Kw{read}\ e \mid \Kw{swap}\ e_1\ e_2 \mid \Kw{drop}\ e} \\
& \mid & \multicolumn{5}{l@{\quad}}{\Kw{if}\ e\ \Kw{then}\ e_1\ \Kw{else}\ e_2 \mid \Kw{loop}\ x = e_1\ \Kw{in}\ e_2 \mid \Kw{letlft}\ c :> \bar a;\, e \mid \Kw{endlft}\ a;\, e} \\[0.3em]
\Delta & ::= & \multicolumn{5}{l@{\quad}}{\cdot \mid \Delta, a \mid \Delta, a <: b \mid \Delta, \dagger a} & \text{lifetime context} \\
\Gamma & ::= & \multicolumn{5}{l@{\quad}}{\cdot \mid \Gamma, x : \tau} & \text{typing context}
\end{array}
\]
\vspace{-1ex}
\normalsize
\caption{Syntax of \LBorrow{}.}
\Description{The syntax of lifetimes, borrowabilities, permissions, qualifiers, types, terms, and contexts.}
\label{fig:calculus-grammar}

\smallskip
\small
\noindent\makebox[\linewidth]{%
  \(\boxed{\Qual{\tau} = q}\)\hfill\textit{(Type Qualifier)}}
\vspace{-8pt}
\[
\Lin \sqsubset \Aff \sqsubset \Un \qquad
\begin{array}{r@{\;=\;}l@{\;\;}l@{\qquad}r@{\;=\;}l}
\Qual{\RefT^{\rho}\tau} & \Lin & (\Erase{\Head(\rho)} = (\Kw{w}, \bullet)) &
\Qual{()} = \Qual{\Kw{int}} = \Qual{\Kw{bool}} & \Un \\
\Qual{\RefT^{\rho}\tau} & \Aff & (\Erase{\Head(\rho)} = (\Kw{w}, \circ)) &
\Qual{\tau_1 \otimes \tau_2} & \Qual{\tau_1} \sqcap \Qual{\tau_2} \\
\Qual{\RefT^{\rho}\tau} & \Un & (\Erase{\Head(\rho)} = (\Kw{r}, \circ)) &
\Qual{\ArrT{p}{\Delta_f}{L}{q}{\tau_1}{\tau_2}} & q
\end{array}
\]
\vspace{-1ex}
\normalsize
\caption{Type qualifiers.}
\Description{Definition of type qualifiers.}
\label{fig:calculus-qual}

\smallskip
\small
\noindent\makebox[\linewidth]{%
  \(\boxed{\Delta \vDash \tau}\)\hfill\textit{(Selected Well-Formedness Rules)}}
\vspace{-6pt}
\begin{mathpar}
\inferrule*[right=\rdef{WF-Ref}]
{ \Vars(\rho) \subseteq \Dom\Delta \\
  \Heads(\rho) \cap \Subs(\rho) = \emptyset \\
  \overline{\Delta \vDash b <: a}^{\,a \in \Heads(\rho),\ b \in \Heads(\tau)} \\
  \Delta \vDash \tau }
{ \Delta \vDash \RefT^{\rho}\tau }
\and
\inferrule*[right=\rdef{WF-Arr}]
{ (\Dom\Delta_f \cup \FL(\tau_1) \cup \FL(\tau_2)) \setminus \bar p \subseteq \Dom\Delta \\
  \Vars(\Delta_f) \cup L \subseteq \Dom\Delta_f \\
  L \cap \bar p \subseteq \FL(\tau_1) \\\\
  \Delta_f \vDash \tau_1 \\
  \Delta_f \setminus L \vDash \tau_2 }
{ \Delta \vDash \ArrT{p}{\Delta_f}{L}{q}{\tau_1}{\tau_2} }
\end{mathpar}
\vspace{-1ex}
\normalsize
\caption{Selected type well-formedness rules.}
\Description{Selected well-formedness rules for reference and function types.}
\label{fig:calculus-wf}
\vspace{-4pt}
\end{figure}

\subsection{Syntax}
\label{sec:calculus-syntax}

\Cref{fig:calculus-grammar} presents the syntax of \LBorrow{}. A lifetime variable
$a$ is an abstract name that identifies the spans of borrows.
A borrowability atom $\alpha$ is a lifetime variable, optionally carrying a mutability
marker (${}^{\Mut}$). Borrowabilities $\rho$
consist of full borrowability $U$ for owned references, borrowability atoms
$\alpha$ for borrows, and subtracted borrowabilities $\rho - \alpha$ for lenders
from which $\alpha$ has been split away. Thus every borrowability has the form
$U\,\overline{-\alpha'}$ or $\alpha\,\overline{-\alpha'}$: an initial head
$\Head(\rho) \in \{U, a, a^{\Mut}\}$ followed by a sequence of subtractions
$\overline{-\alpha'}$. We write $\Heads(\rho)$ for the set of lifetime variables in
$\Head(\rho)$, which is empty or a singleton, $\Subs(\rho)$ for the lifetime
variables subtracted in $\rho$, and $\Vars(\rho) = \Heads(\rho) \cup \Subs(\rho)$
for all of them. All three lift structurally to types $\tau$, excluding the bound
lifetime parameters $\bar p$ in function types, and
$\FL(\tau) = \Heads(\tau) \cup \Subs(\tau)$ is the set of free lifetimes of
$\tau$. \Cref{app:calculus} gives the full definitions.

As introduced in \Cref{sec:borrowability}, a permission $\pi = (\beta, d)$ pairs
an access permission $\beta \in \{\Kw{w}, \Kw{r}, \emptyset\}$ with a drop
permission $d \in \{\bullet, \circ\}$, and permission extraction $\Erase{\rho}$
computes the permission currently available through $\rho$. We also write
$\Erase{\rho}_\beta$ for the access component $\beta$ of $\Erase{\rho}$.

Terms $e$ include variables $x$, lambda abstractions $\lambda x.\, e$,
applications $e_1\, e_2$, let bindings $\Kw{let}\ x = e_1\ \Kw{in}\ e_2$,
sequencing $e_1; e_2$, units $()$, integers $n$, pairs $(e_1, e_2)$ and pair elimination
$\Kw{let}\ (x, y) = e_1\ \Kw{in}\ e_2$, conditionals $\Kw{if}\ e\ \Kw{then}\ e_1\ \Kw{else}\ e_2$,
and loops $\Kw{loop}\ x = e_1\ \Kw{in}\ e_2$. Terms also include
memory operations for allocation $\RefT\ e$, inspection $\Kw{read}\ e$, mutation
$\Kw{swap}\ e_1\ e_2$, and deallocation $\Kw{drop}\ e$. Finally,
$\Kw{letlft}\ c :> \bar a;\, e$ introduces a fresh lifetime $c$ outlived by
$\bar a$, and $\Kw{endlft}\ a;\, e$ ends lifetime $a$.

Types $\tau$ include base types $()$, $\Kw{int}$, and $\Kw{bool}$, multiplicative
products $\tau_1 \otimes \tau_2$, references $\RefT^{\rho}\tau$, and
function types $\ArrT{p}{\Delta_f}{L}{q}{\tau_1}{\tau_2}$.

Following substructural type systems~\cite{ahmedStepindexedModelSubstructural2005,walkerSubstructural2005},
we classify types according to their substructural properties by assigning each
type $\tau$ a qualifier $\Qual{\tau} = q$ (\Cref{fig:calculus-qual}), lifted to
typing contexts as $\Qual{\Gamma}$.
In particular, \textit{linear} ($\Lin$) values admit \textit{exchange} only, \textit{affine}
($\Aff$) values admit exchange and \textit{weakening}, and \textit{unrestricted} ($\Un$) values
admit exchange, weakening, and \textit{contraction}, ordered as $\Lin \sqsubset \Aff \sqsubset \Un$.
Base types are unrestricted, and the qualifier of a product $\tau_1 \otimes \tau_2$
is the meet $\Qual{\tau_1} \sqcap \Qual{\tau_2}$ of its components' qualifiers.
Function types carry an explicit qualifier $q$
indicating the meet of the qualifiers of captured variables.
The qualifier of a reference $\RefT^{\rho}\tau$ is determined by the permission
$\Erase{\Head(\rho)}$ of its head rather than by $\Erase{\rho}$. Since borrows subtracted
in $\rho$ are eventually restored when their lifetimes end, $\Erase{\Head(\rho)}$
is the maximum permission the reference can hold over its lifespan, and a structural rule
applied to the reference now must respect the permission the reference will
regain later. Specifically, the drop
obligation $\bullet$ forbids \textit{weakening}, and only an access permission of at
most $\Kw{r}$ admits \textit{contraction}, so an owned reference is linear, a mutable
borrow affine, and an immutable borrow unrestricted.

\LBorrow{} uses two contexts. A lifetime context $\Delta$ records live
lifetime variables $a$, dead lifetimes $\dagger a$, and outlives constraints $a <: b$
meaning that $a$ outlives $b$.
Notably, we track dead lifetimes so that new lifetimes cannot reuse the names of
ended ones, as we explain in \Cref{sec:calculus-typing}.
We write $\Dom\Delta$, $\dagger\Delta$, and $\Vars(\Delta)$ for the set of live lifetimes,
dead lifetimes, and mentioned lifetime variables in $\Delta$, respectively.
Throughout the paper, we assume all lifetime contexts $\Delta$ are well-formed,
in that live and dead lifetimes are disjoint ($\Dom\Delta \cap \dagger\Delta = \emptyset$)
and outlives constraints only mention live lifetimes ($\Vars(\Delta) \subseteq \Dom\Delta$).
We write $\Delta \vDash a <: b$ for the
reflexive-transitive closure of the outlives constraints in $\Delta$ over
$\Dom\Delta$, and $\Delta \vDash \Delta_0$ when $\Dom\Delta_0 \subseteq \Dom\Delta$ and
$\Delta \vDash a <: b$ for every $(a <: b) \in \Delta_0$. A typing context
$\Gamma$ maps term variables to types. We assume term and lifetime variables are always fresh.

\paragraph{Well-formedness}
\Cref{fig:calculus-wf} presents selected rules for type
well-formedness $\Delta \vDash \tau$, which states that $\tau$ is well-formed
under lifetime context $\Delta$. We write $\Delta \vDash \Gamma$ 
when $\Delta \vDash \tau$ for every $(x : \tau) \in \Gamma$. Rule
\rref{WF-Ref} checks three conditions on a reference $\RefT^{\rho}\tau$ in
addition to $\Delta \vDash \tau$. First, every lifetime variable appearing in
$\rho$ must be currently alive. Second,
the head and subtractions of $\rho$ must be disjoint,
so a reference cannot borrow from its own lifetime. Third, every head
lifetime $b \in \Heads(\tau)$ of the referenced type $\tau$ must outlive the head lifetime
$a \in \Heads(\rho)$ of the reference, ensuring that a
borrow never outlives the references stored in the cell. We explain rule \rref{WF-Arr} for function types along with the typing rules for
functions in \Cref{sec:calculus-typing}.

\subsection{Typing}
\label{sec:calculus-typing}

\begin{figure}[tb]
\small
\def\MathparLineskip{\lineskiplimit=0.18em\lineskip=0.18em}
\noindent\makebox[\linewidth]{%
  \(\boxed{\Delta; \Gamma \vdash e : \tau \dashv \Delta'}\)\hfill\textit{(Selected Typing Rules)}}
\vspace{-6pt}
\begin{mathpar}
\inferrule*[right=\rdef{T-LftIntro}]
{ \Delta, c, \bar a <: c; \Gamma \vdash e : \tau \dashv \Delta'' \\
  c \notin \Dom\Delta \cup \dagger\Delta \\ \bar a \subseteq \Dom\Delta }
{ \Delta; \Gamma \vdash \Kw{letlft}\ c :> \bar a;\, e : \tau \dashv \Delta'' }
\and
\inferrule*[right=\rdef{T-LftEnd}]
{ \Delta \LftEnd{a} \Delta' \\
  \Delta' \vDash \Gamma\Rst{a} \\
  \Delta'; \Gamma\Rst{a} \vdash e : \tau \dashv \Delta'' }
{ \Delta; \Gamma \vdash \Kw{endlft}\ a;\, e : \tau \dashv \Delta'' }
\and
\inferrule*[right=\rdef{T-Seq}]
{ \Delta_1; \Gamma_1 \vdash e_1 : () \dashv \Delta_2 \\
  \Gpre{\Delta_2 \vDash \Gamma_2\Rst{(\Delta_1 \setminus \Delta_2)}} \\
  \Delta_2; \Gamma_2\Rst{(\Delta_1 \setminus \Delta_2)} \vdash e_2 : \tau \dashv \Delta_3 }
{ \Delta_1; \Gamma_1 \oplus \Gamma_2 \vdash e_1; e_2 : \tau \dashv \Delta_3 }
\and
\inferrule*[right=\rdef{T-Let}]
{ \Delta_1; \Gamma_1 \vdash e_1 : \tau_1 \dashv \Delta_2 \\
  \Gpre{\Delta_2 \vDash \Gamma_2\Rst{(\Delta_1 \setminus \Delta_2)}} \\
  \Delta_2; \Gamma_2\Rst{(\Delta_1 \setminus \Delta_2)}, x : \tau_1 \vdash e_2 : \tau_2 \dashv \Delta_3 }
{ \Delta_1; \Gamma_1 \oplus \Gamma_2 \vdash \Kw{let}\ x = e_1\ \Kw{in}\ e_2 : \tau_2 \dashv \Delta_3 }
\and
\inferrule*[right=\rdef{T-Pair}]
{ \Delta_1; \Gamma_1 \vdash e_1 : \tau_1 \dashv \Delta_2 \\
  \Gpre{\Delta_2 \vDash \Gamma_2\Rst{(\Delta_1 \setminus \Delta_2)}} \\
  \Delta_2; \Gamma_2\Rst{(\Delta_1 \setminus \Delta_2)} \vdash e_2 : \tau_2 \dashv \Delta_3 \\
  \Delta_3 \vDash \tau_1\Rst{(\Delta_2 \setminus \Delta_3)} }
{ \Delta_1; \Gamma_1 \oplus \Gamma_2 \vdash (e_1, e_2) : \tau_1\Rst{(\Delta_2 \setminus \Delta_3)} \otimes \tau_2 \dashv \Delta_3 }
\and
\inferrule*[right=\rdef{T-If}]
{ \Delta_1; \Gamma_1 \vdash e : \Kw{bool} \dashv \Delta_2 \\
  \Gpre{\Delta_2 \vDash \Gamma_2\Rst{(\Delta_1 \setminus \Delta_2)}} \\
  \Delta_2; \Gamma_2\Rst{(\Delta_1 \setminus \Delta_2)} \vdash e_1 : \tau \dashv \Delta_3, \dagger\bar c_1 \\
  \Delta_2; \Gamma_2\Rst{(\Delta_1 \setminus \Delta_2)} \vdash e_2 : \tau \dashv \Delta_3, \dagger\bar c_2 }
{ \Delta_1; \Gamma_1 \oplus \Gamma_2 \vdash \Kw{if}\ e\ \Kw{then}\ e_1\ \Kw{else}\ e_2 : \tau
  \dashv \Delta_3, \dagger(\bar c_1 \cup \bar c_2) }
\and
\inferrule*[right=\rdef{T-Loop}]
{ \Delta; \Gamma \vdash e_1 : \tau \dashv \Delta_l \\
  \Delta_l; x : \tau \vdash e_2 : \tau \otimes \Kw{bool} \dashv \Delta_l, \dagger\bar c }
{ \Delta; \Gamma \vdash \Kw{loop}\ x = e_1\ \Kw{in}\ e_2 : \tau \dashv \Delta_l, \dagger\bar c }
\and
\inferrule*[right=\rdef{T-RefOwn}]
{ \Delta; \Gamma \vdash e : \tau \dashv \Delta' }
{ \Delta; \Gamma \vdash \RefT\ e : \RefT^{U}\tau \dashv \Delta' }
\and
\inferrule*[right=\rdef{T-Drop}]
{ \Delta; \Gamma \vdash e : \RefT^{U}\tau \dashv \Delta' }
{ \Delta; \Gamma \vdash \Kw{drop}\ e : \tau \dashv \Delta' }
\and
\inferrule*[right=\rdef{T-Read}]
{ \Delta; \Gamma \vdash e : \RefT^{\alpha}\tau \dashv \Delta'}
{ \Delta; \Gamma \vdash \Kw{read}\ e : \Rview{\alpha}{\tau} \dashv \Delta' }
\and
\inferrule*[right=\rdef{T-Swap}]
{ \Delta; \Gamma_1 \vdash e_1 : \RefT^{a^{\Mut}}\tau_0 \dashv \Delta' \\
  \Gpre{\Delta' \vDash \Gamma_2\Rst{(\Delta \setminus \Delta')}} \\
  \Delta'; \Gamma_2\Rst{(\Delta \setminus \Delta')} \vdash e_2 : \tau \dashv \Delta'' \\
  \tau = \tau_0\Rst{(\Delta' \setminus \Delta'')} \\
  \Delta'' \vDash \RefT^{a^{\Mut}}\tau }
{ \Delta; \Gamma_1 \oplus \Gamma_2 \vdash \Kw{swap}\ e_1\ e_2 : \RefT^{a^{\Mut}}\tau \otimes \tau \dashv \Delta'' }
\and
\inferrule*[right=\rdef{T-Lam}]
{ \bar p = \Dom\Delta_f \setminus \Dom\Delta \\
  \Gpre{\Delta_f \vDash \Gamma} \\
  \Delta_f; \Gamma, x : \tau_1 \vdash e : \tau_2 \dashv \Delta_f \setminus L, \dagger\bar c \\
  \Delta \vDash \ArrT{p}{\Delta_f}{L}{q}{\tau_1}{\tau_2} \\
  \Delta \vDash \Delta_f|_{\Dom\Delta} \\
  q \sqsubseteq \Qual{\Gamma} }
{ \Delta; \Gamma \vdash \lambda x.\, e : \ArrT{p}{\Delta_f}{L}{q}{\tau_1}{\tau_2} \dashv \Delta }
\and
\inferrule*[right=\rdef{T-App}]
{ \Delta; \Gamma_1 \vdash e_1 : \ArrT{p}{\Delta_f}{L}{q}{\tau_1}{\tau_2} \dashv \Delta_1 \\
  \Gpre{\Delta_1 \vDash \Gamma_2\Rst{(\Delta \setminus \Delta_1)}} \\
  \Delta_1; \Gamma_2\Rst{(\Delta \setminus \Delta_1)} \vdash e_2 : \theta\tau_1 \dashv \Delta_2 \\
  \theta \Vdash_{\Delta_2} \ArrT{p}{\Delta_f}{L}{q}{\tau_1}{\tau_2} \\
  \Delta_2 \vDash \ArrT{p}{\Delta_f}{L}{q}{\tau_1}{\tau_2} \\
  \Delta_2 \vDash \theta\Delta_f \\
  \theta\Delta_f \vDash \Delta_2|_{\theta L} \\
  \Delta_2 \LftEnd{\theta L} \Delta_3 }
{ \Delta; \Gamma_1 \oplus \Gamma_2 \vdash e_1\, e_2 : \theta\tau_2 \dashv \Delta_3 }
\end{mathpar}
\vspace{-1ex}
\normalsize
\caption{Selected typing rules of \LBorrow{}.}
\Description{Selected typing rules for lifetimes, sequencing, control flow, memory operations, and functions.}
\label{fig:calculus-typing}
\vspace{-4pt}
\end{figure}

\begin{figure}[tb]
\small
\noindent\makebox[\linewidth]{%
  \(\boxed{\Delta \vDash \rho \leadsto \rho_1 \oplus \rho_2}\)\hfill\textit{(Selected Borrowability Splitting Rule)}}
\vspace{-6pt}
\begin{mathpar}
\inferrule*[right=\rdef{S-Borrow}]
{ \alpha \in \{a, a^{\Mut}\} \\
  a \in \Dom\Delta \\
  a \notin \Heads(\rho) \\
  \Erase{\rho} \leadsto \Erase{\rho - \alpha} \oplus \Erase{\alpha} \\
  \overline{\Delta \vDash a_0 <: a}^{\,a_0 \in \Heads(\rho)} }
{ \Delta \vDash \rho \leadsto (\rho - \alpha) \oplus \alpha }
\end{mathpar}
\vspace{-3.5ex}
\noindent\makebox[\linewidth]{%
  \(\boxed{\Delta \vDash \tau \leadsto \tau_1 \oplus \tau_2}\)\hfill\textit{(Selected Type Splitting Rule)}}
\vspace{-6pt}
\begin{mathpar}
\inferrule*[right=\rdef{S-Ref}]
{ \Delta \vDash \rho \leadsto \rho_1 \oplus \rho_2 \\
  \Gpre{\Delta \vDash \RefT^{\rho_1}\tau} \\
  \Gpre{\Delta \vDash \RefT^{\rho_2}\tau} }
{ \Delta \vDash \RefT^{\rho}\tau \leadsto \RefT^{\rho_1}\tau \oplus \RefT^{\rho_2}\tau }
\end{mathpar}
\vspace{-1ex}
\normalsize
\caption{Selected borrowability and type splitting rules.}
\Description{Rules for splitting borrowabilities and types.}
\label{fig:calculus-split}

\smallskip
\small
\begin{minipage}[t]{0.34\textwidth}
\noindent\makebox[\linewidth]{%
  \(\boxed{\Delta \LftEnd{a} \Delta'}\)\hfill\textit{(Lifetime Ending)}}
\vspace{-6pt}
\begin{mathpar}
\inferrule*[right=\rdef{L-End}]
{ a \in \Dom\Delta \\
  \overline{\Delta \nvDash a <: a'}^{\,a' \in \Dom\Delta \setminus \{a\}} }
{ \Delta \LftEnd{a} (\Delta \setminus \{a\}), \dagger a }
\end{mathpar}
\caption{Lifetime ending.}
\Description{Rule for ending a live lifetime.}
\label{fig:calculus-lft-end}
\end{minipage}
\hfill
\begin{minipage}[t]{0.63\textwidth}
\noindent\makebox[\linewidth]{%
  \(\boxed{\Rview{\alpha}{\tau} = \tau'}\)\hfill\textit{(Read View)}}
\vspace{-6pt}
\[
\begin{array}{r@{\;}c@{\;}l}
\Rview{\alpha}{\tau} & = & \tau \quad \Qual{\tau} = \Un \\
\Rview{\alpha}{\tau_1 \otimes \tau_2} & = & \Rview{\alpha}{\tau_1} \otimes \Rview{\alpha}{\tau_2} \\
\Rview{\alpha}{\RefT^{\rho}\tau} & = & \RefT^{\alpha'}\tau \quad \Qual{\RefT^{\rho}\tau} \neq \Un \\
 & & \quad \alpha, \alpha' \in \{a, a^\Mut\},\; \Erase{\alpha'}_\beta = \Erase{\rho}_\beta \sqcap \Erase{\alpha}_\beta \neq \emptyset
\end{array}
\]
\caption{Read view.}
\Description{Computing the read view of a referenced type.}
\label{fig:calculus-rview}
\end{minipage}
\normalsize
\vspace{-4pt}
\end{figure}

\Cref{fig:calculus-typing} presents the selected typing rules of \LBorrow{}. The typing judgment $\Delta; \Gamma
\vdash e : \tau \dashv \Delta'$ states that under input lifetime context $\Delta$ and
typing context $\Gamma$, expression $e$ has type $\tau$ and produces output lifetime
context $\Delta'$. We omit the standard rules for variables, constants, pair
elimination, and the structural rules; the appendix gives the full system.
For readability, routine well-formedness premises are shaded in \Gpre{gray}.

\paragraph{Borrowability and context splitting}
To type expressions with multiple subexpressions, we introduce the
context splitting judgment $\Delta \vDash \Gamma \leadsto \Gamma_1 \oplus \Gamma_2$,
written inline in typing rules as $\Delta; \Gamma_1 \oplus \Gamma_2 \vdash e: \tau \dashv \Delta'$,
to split bindings in $\Gamma$. In addition to distributing
bindings between $\Gamma_1$ and $\Gamma_2$, context splitting can split a single
binding $x : \tau$ into $x : \tau_1$ and $x : \tau_2$ via type splitting
$\Delta \vDash \tau \leadsto \tau_1 \oplus \tau_2$ and borrowability splitting
$\Delta \vDash \rho \leadsto \rho_1 \oplus \rho_2$ (\Cref{fig:calculus-split}).

Rule \rref{S-Borrow} borrows $\alpha$
from $\rho$ by splitting it into a lender $\rho - \alpha$ and a borrow
$\alpha \in \{a, a^{\Mut}\}$, as long as $a$ is alive and different from the
head of $\rho$, the split itself respects permission splitting
(\Cref{fig:access-permission-split,fig:permission-split}), and any head lifetime
of $\rho$ outlives $a$ so that a reborrow cannot outlive the borrow it is taken from.
Rule \rref{S-Ref} further lifts borrowability
splitting to reference types while checking that both resulting references are
well-formed under $\Delta$, ensuring that any references in $\tau$ outlive the
new borrow. We omit standard rules for unrestricted types and products for brevity.

\paragraph{Lifetimes, restoration, and sequential composition}
Rule \rref{T-LftIntro} extends the lifetime context with a fresh lifetime $c$
and its outlives constraints. Notably, $\Kw{letlft}$ does not delimit a scope for
$c$; rather, $c$ stays alive until an explicit $\Kw{endlft}\ c$.
Also,
$c$ must be fresh with respect to both live and dead lifetimes in $\Delta$,
preventing a new lifetime from reusing the name of an ended lifetime and making
it appear alive again in the output context.

Rule \rref{T-LftEnd} ends a live lifetime $a$ and restores the typing
context. First, it updates the lifetime
context via $\Delta \LftEnd{a} \Delta'$ (\Cref{fig:calculus-lft-end}), where
rule \rref{L-End} removes $a$ and its outlives constraints from $\Delta$ and
records $\dagger a$, provided that no other live lifetime in $\Delta$ is
outlived by $a$ so that shorter reborrows end before their parent borrow.
Next,
it restores the typing context via $\Gamma\Rst{a}$, which strips $-a$ and
$-a^{\Mut}$ from borrowabilities recursively (\Cref{fig:restoration}), and
checks $\Delta' \vDash \Gamma\Rst{a}$ to ensure that no active borrow or closure
still references $a$.

In \rref{T-Seq} and \rref{T-Let}, context
splitting divides the initial context $\Gamma$ into $\Gamma_1$ and $\Gamma_2$. Evaluating
$e_1$ under $\Gamma_1$ has the side effect of ending the lifetimes in
$\Dom \Delta_1 \setminus \Dom \Delta_2$, which we abbreviate as
$\Delta_1 \setminus \Delta_2$. Both rules therefore restore $\Gamma_2$ by these
lifetimes, as $\Gamma_2\Rst{(\Delta_1 \setminus \Delta_2)}$, and require the
restored context to be well-formed under $\Delta_2$.
Rule \rref{T-Pair} additionally restores the first component's type
$\tau_1$ by the lifetimes $\Delta_2 \setminus \Delta_3$ ended by $e_2$, since the old
type $\tau_1$ is no longer precise after $e_2$ has ended the relevant lifetimes. Rule
\rref{T-If} requires both branches to agree on the live lifetimes and constraints under
the restored context, while combining their dead lifetimes. Finally, rule \rref{T-Loop}
introduces stateful loop, which is equivalent to a simple form of tail-recursive function \cite{henriksenFutharkPurelyFunctional2017}.
It threads the loop state $x : \tau$ from $e_1$ through the body $e_2$, which produces the
next state and a continuation flag, requiring live lifetimes to be stable across iterations.

\paragraph{Memory operations}

Rule \rref{T-RefOwn} allocates an owned reference $\RefT^{U}\tau$, and dually rule
\rref{T-Drop} deallocates one and returns the stored value of type $\tau$.

Rule \rref{T-Read} inspects a borrowed reference $e : \RefT^{\alpha}\tau$
and returns its read view $\Rview{\alpha}{\tau}$ (\Cref{fig:calculus-rview}).
When $\tau$ is unrestricted, read preserves its type; in particular, a nested
immutable borrow keeps its own lifetime. When the nested
reference $\RefT^{\rho}\tau_0$ is linear or affine instead, that is, owned or mutably
borrowed, reading yields a borrow at the outer borrow's lifetime $a$ whose access
permission is the meet $\Erase{\rho}_\beta \sqcap \Erase{\alpha}_\beta \neq
\emptyset$. Because \rref{WF-Ref} guarantees that the
inner reference outlives $a$, this never extends the inner lifetime and is therefore sound. A nested
reference with empty access permission $\Erase{\rho}_\beta = \emptyset$ cannot be
read at all.

Rule \rref{T-Swap} replaces the contents of a mutable borrow
$e_1 : \RefT^{a^{\Mut}}\tau_0$ with $e_2 : \tau$ and returns the updated borrow
together with the old value. As in \rref{T-Pair}, we restore $\tau_0$ by the
lifetimes $e_2$ ends and check that the result remains well-formed.

\paragraph{Functions}\label{par:functions}

The function type $\ArrT{p}{\Delta_f}{L}{q}{\tau_1}{\tau_2}$ extends the notation
of \Cref{sec:overview-functions} with four components: the lifetime context
$\Delta_f$ that the function body requires; the bound lifetime parameters
$\bar p \subseteq \Dom\Delta_f$, the remaining lifetimes
$\Dom\Delta_f \setminus \bar p$ being captured from the environment; the
\textit{lifetime effect} $L \subseteq \Dom\Delta_f$, i.e., the set of lifetimes
that a call would end; and the qualifier $q$. As defined in \rref{WF-Arr}
(\Cref{fig:calculus-wf}), a function type is \textit{well-formed under $\Delta$} if
the following conditions hold: all free lifetimes of the function type are alive in $\Delta$;
$\Delta_f$ and $L$ mention only lifetimes in $\Dom\Delta_f$; every bound parameter
that the function ends appears in the argument type $\tau_1$, so that the argument
at a call site determines which lifetime is ended; and $\tau_1$ is well-formed under
$\Delta_f$ while $\tau_2$ is well-formed under $\Delta_f \setminus L$ after applying the lifetime effect.

Rule \rref{T-Lam} checks that the function body discharges the lifetime effect $L$,
that the function type is well-formed, and enforces a
set of constraints on lifetimes. Here, the bound parameters
$\bar p$ must not shadow $\Delta$, and $\Delta$ must entail the
outlives constraints in $\Delta_f$ on captured lifetimes. Moreover,
$q \sqsubseteq \Qual{\Gamma}$; that is,
a closure capturing an owned reference or a lender must be linear, while one capturing
only mutable borrows may be affine.

Rule \rref{T-App} instantiates the bound parameters $\bar p$ via 
lifetime substitution $\theta$, written $\theta\tau$ and $\theta\Delta_f$.
The substitution judgment $\theta \Vdash_{\Delta_2} \ArrT{p}{\Delta_f}{L}{q}{\tau_1}{\tau_2}$
requires $\theta$ to map $\bar p$ injectively to live lifetimes in $\Delta_2$
that are disjoint from the free lifetimes of the function type (a weaker
condition is possible, but injectivity and disjointness keep the rule simple).
Otherwise, $\theta$ could alias an ended lifetime with one that remains alive in the
return type $\theta\tau_2$, or alias the head and subtracted lifetimes of a reference,
breaking \rref{WF-Ref}.
Moreover, the caller context $\Delta_2$ must entail the instantiated constraints
$\theta\Delta_f$, and conversely $\theta\Delta_f$ must entail
$\Delta_2|_{\theta L}$, meaning that any outlives constraint in $\Delta_2$
between lifetimes in $\theta L$ also holds in $\theta\Delta_f$, so the callee
cannot end them in an order that violates constraints in $\Delta_2$. Finally, the call ends $\theta L$ one by one
via $\Delta_2 \LftEnd{\theta L} \Delta_3$ and returns $\theta\tau_2$. We omit
the exact definitions of these auxiliary judgments for brevity, and we invite
readers to check the appendix for details.

\section{Semantics and Metatheory}
\label{sec:semantics}

This section presents the semantics of \LBorrow{}, covering the operational
semantics (\Cref{sec:opsem}) and semantic model (\Cref{sec:semantic-typing}). We then
establish its type soundness and memory safety (\Cref{sec:soundness}).
We have mechanized \LBorrow{} and its metatheory
in the Rocq proof assistant~\cite{therocqdevelopmentteamRocqProver2025}
on top of Iris~\cite{jungIrisGroundModular2018}.

\subsection{Operational Semantics}
\label{sec:opsem}

The dynamic semantics of \LBorrow{} is given by a standard call-by-value
reduction relation $\sigma; e \longrightarrow \sigma'; e'$ (\Cref{app:opsem}),
which reads that under heap $\sigma$, expression $e$ reduces to $e'$, updating
the heap to $\sigma'$. Heap $\sigma$ maps
each allocated location $\ell$ to either a stored value $v$ or a tombstone $\bot$
marking a deallocated cell. Dropping a reference overwrites its location with $\bot$ so
that freed addresses are never reused, while all memory operations require
the accessed value $v$ to be live, i.e., $v \neq \bot$. Consequently, any invalid memory
access, such as use-after-free or double-free, gets stuck.
As in $\lambda_{\text{Rust}}$~\cite{jungRustBeltSecuringFoundations2017},
lifetimes are a purely compile-time concept with no runtime representation, and
the lifetime operations $\Kw{letlft}\ c :> \bar a;\, e$ and
$\Kw{endlft}\ a;\, e$ simply reduce to $e$ at runtime.

\subsection{Semantic Typing}
\label{sec:semantic-typing}

\newcommand{\wand}{\mathrel{-\!\!*}}
\newcommand{\bigsep}{\mathop{\raisebox{-0.05ex}{\normalsize$\ast$}}\limits}
\newcommand{\SemV}[3]{\mathcal{V}\llbracket #1 \rrbracket_{#2}^{#3}}
\newcommand{\SemD}[3]{\mathcal{D}_{#1}\llbracket #2 \rrbracket_{#3}}
\newcommand{\SemG}[3]{\mathcal{G}\llbracket #1 \rrbracket_{#2}(#3)}

To establish type soundness for \LBorrow{}, we define a logical relation in
Iris~\cite{jungIrisGroundModular2018} that interprets types and contexts as
separation-logic predicates. We present simplified definitions here to convey the key ideas and defer the full model
(including $\mathsf{LoanTable}$, $\mathsf{Shr}_\eta$, and $\mathcal{R}_\eta$) to \Cref{app:model}.

\paragraph{Context interpretations}
A semantic lifetime assignment $\eta : \mathit{LVar} \to \mathit{SemLft}$ maps
syntactic lifetime variables $a$ to semantic lifetimes $\kappa$. Inspired by
lifetime tokens of RustBelt~\cite{jungRustBeltSecuringFoundations2017}, our
model represents each semantic lifetime $\kappa$ by either an exclusive
\emph{lifetime token} $[\kappa]_1$ or persistent \emph{dead token} $\dagger\kappa$,
witnessing $\kappa$ to be alive or dead, respectively.
Ending a lifetime consumes $[\kappa]_1$ and generates $\dagger\kappa$, so
$[\kappa]_1 * \dagger\kappa \vDash \Kw{False}$. Following
$\lambda_{\text{Rust}}$~\cite{jungRustBeltSecuringFoundations2017}, the lifetime
context interpretation $\SemD{E, L}{\Delta}{\eta}$ partitions $\Dom\Delta$ into
\emph{external} lifetimes $E$ that must remain alive, and
\emph{local} lifetimes $L = \Dom\Delta \setminus E$ that the current expression
may end:
\begingroup
\setlength{\abovedisplayskip}{0pt}\setlength{\belowdisplayskip}{4pt}
\[
\SemD{E, L}{\Delta}{\eta} \triangleq
  \overbrace{\mathsf{LoanTable}}^{\text{active borrows}} *
  \overbrace{\big(\bigsep_{a \in L} [\eta(a)]_1\big)}^{\text{local lifetimes alive}} *
  \overbrace{\big(\bigsep_{(a <: b) \in \Delta} \eta(a) \sqsubseteq \eta(b)\big)}^{\text{outlives constraints}}
\]
The typing context interpretation $\SemG{\Gamma}{\eta}{\gamma}$ relates each
$x : \tau$ in $\Gamma$ to its runtime value $\gamma(x)$:
\[
\SemG{\Gamma}{\eta}{\gamma} \triangleq
  \bigsep_{(x : \tau) \in \Gamma} \exists v.\, (\gamma(x) = v) * \SemV{\tau}{\eta}{1}(v)
\]
\endgroup
\paragraph{Semantic typing}
The semantic typing is defined in terms of weakest
precondition $\mathsf{WP}$.
That is, given the context interpretations, $\gamma(e)$ executes safely and, if
it terminates at a value $v$, returns the updated lifetime context, the value
interpretation of $v$, and dead tokens for the local lifetimes ended by $e$,
where $\eta'$ extends $\eta$ with lifetimes introduced by $e$.
\begingroup
\setlength{\abovedisplayskip}{4pt}\setlength{\belowdisplayskip}{4pt}
\[
\begin{array}{r@{\;}c@{\;}l}
\Delta; \Gamma \vDash_E e : \tau \Dashv \Delta' & \triangleq &
  \forall \eta, \gamma.\;
  \SemD{E, L}{\Delta}{\eta} * \SemG{\Gamma}{\eta}{\gamma} \wand \\
  & & \quad \mathsf{WP}\ \gamma(e)\ \big\{ v.\, \exists \eta'.\, (\eta|_N = \eta'|_N) *
    \SemD{E, L'}{\Delta'}{\eta'} * \SemV{\tau}{\eta'}{1}(v) *
    \bigsep_{a \in L \setminus L'} \dagger\eta(a) \big\} \\
  & & \quad \text{where } L = \Dom\Delta \setminus E,\; L' = \Dom\Delta' \setminus E,\text{ and } N = \Dom\Delta \cup \dagger\Delta
\end{array}
\]
\endgroup

\begin{figure}[tb]
\vspace{-1.5ex}
\small
\[
\begin{array}{r@{\;}c@{\;}l}
\SemV{()}{\eta}{q}(v) & \triangleq & (v = ()) \qquad
\SemV{\Kw{int}}{\eta}{q}(v) \;\triangleq\; \exists n.\, (v = n) \qquad
\SemV{\Kw{bool}}{\eta}{q}(v) \;\triangleq\; \exists b.\, (v = b) \\[0.2em]
\SemV{\tau_1 \otimes \tau_2}{\eta}{q}(v) & \triangleq &
  \exists v_1, v_2.\, (v = (v_1, v_2)) * \SemV{\tau_1}{\eta}{q}(v_1) * \SemV{\tau_2}{\eta}{q}(v_2) \\[0.2em]
\SemV{\RefT^{\rho}\tau}{\eta}{q}(v) & \triangleq &
  \exists \ell.\, (v = \ell) *
  \begin{cases}
    \mathsf{Shr}_\eta(\eta(a), \tau, \ell) & \text{if } \Head(\rho) = a \\
    \mathcal{R}_\eta(\rho, \tau, q, \ell) & \text{if } \Head(\rho) \in \{U, a^{\Mut}\}
  \end{cases} \\[0.4em]
\SemV{\ArrT{p}{\Delta_f}{L}{q_f}{\tau_1}{\tau_2}}{\eta}{q}(v) & \triangleq &
  \exists x, e.\, (v = \lambda x.\, e) * \triangleright \Box_{q_f} \forall \theta, w.\,
    \SemD{E_f, L}{\Delta_f}{\eta_\theta} * \SemV{\tau_1}{\eta_\theta}{1}(w) \wand \\
  & & \quad \mathsf{WP}\ [w/x]e\ \big\{ r.\,
    \SemD{E_f, \emptyset}{\Delta_f \setminus L}{\eta_\theta} *
    \big(\bigsep_{b \in L} \dagger\eta_\theta(b)\big) *
    \SemV{\tau_2}{\eta_\theta}{1}(r) \big\} \\
  & & \quad \text{where } E_f = \Dom\Delta_f \setminus L,\;
    \eta_\theta = \eta[\bar p \mapsto \theta],\;
    \Box_{\Un} P = \Box P,\; \Box_{\Lin} P = \Box_{\Aff} P = P
\end{array}
\]
\vspace{-1ex}
\normalsize
\caption{Value interpretation}
\Description{Value interpretation of the logical relation.}
\label{fig:logical-relation}
\vspace{-4pt}
\end{figure}

\paragraph{Value interpretations}
The value interpretation $\SemV{\tau}{\eta}{q}(v)$, shown in
\Cref{fig:logical-relation}, is indexed by a fractional permission $q \in (0,
1]$~\cite{boylandCheckingInterferenceFractional2003}, which models shared
references and which we elaborate with reference interpretations below. The
interpretation is standard for base types and pairs $\tau_1 \otimes \tau_2$. For functions, it requires
$v$ to be a closure $\lambda x.\, e$ such that for any lifetime substitution
$\theta$ and argument $w$, combining the closure's
captured resources with the caller-provided resources
$\SemD{E_f, L}{\Delta_f}{\eta_\theta}$ and $\SemV{\tau_1}{\eta_\theta}{1}(w)$
yields the weakest precondition of $[w/x]e$, whose postcondition returns the
remaining context $\SemD{E_f, \emptyset}{\Delta_f \setminus L}{\eta_\theta}$ and
dead tokens $\dagger\eta_\theta(b)$ for all $b \in L$. When $q_f = \Un$, the
predicate is persistent $\Box$, so that unrestricted closures are freely
duplicable.

\paragraph{Reference interpretations}
\label{par:reference-interpretations}
Recall that in \LBorrow{}, splitting a reference $x : \RefT^{\rho}\tau$ produces a lender
$x : \RefT^{\rho - \alpha}\tau$ and a borrow $x : \RefT^{\alpha}\tau$ that alias
the same runtime location $\ell$,
and the lender's permission is restored via
$\tau\Rst{a}$ when $a$ ends. To model
this non-local restoration of permissions, borrowing deposits the lent
permission into $\mathsf{LoanTable}$ under $\eta(a)$, while the lender retains a
ghost inheritance token recording the loan so that it can reclaim the deposited permission
once $\eta(a)$ ends. Furthermore, when a reference is shared immutably as
$\RefT^{\rho - c}\tau$ and $\RefT^{c}\tau$, both the lender and the borrow hold
read permission to $\ell$ at the same time. Hence
$\SemV{\tau}{\eta}{q}(v)$ is indexed by the fractional permission $q$: sharing
can split a points-to assertion $\ell \mapsto_q w$ between
$\mathsf{LoanTable}$ and the lender.

We interpret a reference $\RefT^{\rho}\tau$ at location $\ell$ according to
$\Head(\rho)$.
When $\Head(\rho) = a$, the reference is an immutable borrow and is modeled by the \emph{sharing
predicate} $\mathsf{Shr}_\eta(\eta(a), \tau, \ell)$, which witnesses a shared
loan of $\ell$ in $\mathsf{LoanTable}$ at some lifetime outliving $\eta(a)$.
Notably, $\mathsf{Shr}_\eta$ is persistent, so that immutable borrows are freely
duplicable.
When $\Head(\rho) \in \{U, a^{\Mut}\}$, the reference is either an owned reference or
a mutable borrow and is modeled by the \emph{borrowability residual predicate}
$\mathcal{R}_\eta(\rho, \tau, q, \ell)$, which tracks the permission still
retained by the reference together with the inheritance tokens for the borrows
subtracted in $\rho$. Specifically, $\mathcal{R}_\eta$ is defined inductively
over the subtractions in $\rho$. In the base case with no subtractions,
$\mathcal{R}_\eta$ asserts direct heap ownership
$\exists w.\, \ell \mapsto_q w * \SemV{\tau}{\eta}{q}(w)$ for $\rho = U$, while
holding the loan token of fraction $q$ of the active loan in $\mathsf{LoanTable}$ for
$\rho = a^{\Mut}$. In the inductive step, each shared subtraction $\rho - c$
splits the underlying resource of fraction $q$, depositing a sub-fraction into
$\mathsf{LoanTable}$ and keeping the remaining fraction $q_1 \le q$ in
$\mathcal{R}_\eta(\rho, \tau, q_1, \ell)$ so that the lender retains read access
matching $\Erase{U - c} = (\Kw{r}, \bullet)$. A mutable subtraction
$\rho - c^{\Mut}$ from an unborrowed $\rho \in \{U, a^{\Mut}\}$ instead deposits the entire resource of fraction $q$ into
$\mathsf{LoanTable}$ and leaves the lender with only its inheritance token,
satisfying $\Erase{U - c^{\Mut}} = (\emptyset, \bullet)$.

\paragraph{Lazy reclamation}
A remaining question is when a lender reclaims the resource it deposited.
When a lifetime $a$ ends, \LBorrow{} restores types from $\tau$ to
$\tau\Rst{a}$, so one might reclaim the deposited resource for $a$ from
$\mathsf{LoanTable}$ at that point. However, the lender may have been framed
out or captured in a closure, where $\mathsf{LoanTable}$ cannot be updated.
Similar to lazy unblocking in RefinedRust~\cite{gaherRefinedRustTypeSystem2024},
we decouple type restoration from reclamation by generalizing
$\mathcal{R}_\eta(\rho, \tau, q, \ell)$ to allow subtractions at dead
lifetimes in $\rho$ and $\tau$. Thus, type restoration
$\dagger\eta(a) * \SemV{\tau}{\eta}{q}(v) \vdash \SemV{\tau\Rst{a}}{\eta}{q}(v)$
holds purely without requiring immediate reclamation of the deposited resource,
and the lender can reclaim its deposited permission
from $\mathsf{LoanTable}$ lazily next time it is accessed.

\subsection{Type Soundness}
\label{sec:soundness}

To show semantic type soundness, we first establish well-formedness preservation and
lifetime monotonicity, ensuring that result types remain valid in the output context
and that ended lifetimes are never reintroduced across sequential steps.

\begin{lemma}[Well-Formedness and Monotonicity\rocq]
\label{lem:wf-preservation}
If $\Delta; \Gamma \vdash e : \tau \dashv \Delta'$ and $\Delta \vDash \Gamma$,
then $\Delta' \vDash \tau$,
$\dagger\Delta \subseteq \dagger\Delta'$, and
$\Dom\Delta \cup \dagger\Delta \subseteq \Dom\Delta' \cup \dagger\Delta'$.
\end{lemma}

We write $\Delta; \Gamma \vDash e : \tau \Dashv \Delta'$ for top-level semantic
typing with $E = \emptyset$.
Then, we prove the fundamental property by proving compatibility lemmas for typing,
splitting, and restoration rules.

\begin{theorem}[Fundamental Property\rocq]
\label{thm:fundamental}
If $\Delta; \Gamma \vdash e : \tau \dashv \Delta'$ and $\Delta \vDash \Gamma$,
then $\Delta; \Gamma \vDash e : \tau \Dashv \Delta'$.
\end{theorem}

Combining \Cref{thm:fundamental} with Iris adequacy~\cite{jungIrisGroundModular2018},
we establish that well-typed closed programs never get stuck, which guarantees
both type soundness and memory safety, with no axiom assumed.

\begin{theorem}[Type Soundness\rocq]
\label{thm:soundness}
\label{thm:memory-safety}
If $\cdot; \cdot \vdash e : \tau \dashv \Delta'$ and $\emptyset; e \longrightarrow^* \sigma; e'$,
then either $e'$ is a value or there exist $\sigma'', e''$ such that
$\sigma; e' \longrightarrow \sigma''; e''$.
\end{theorem}

\section{Case Study}
\label{sec:case-study}

In this section, we evaluate how \LBorrow prevents real-world soundness bugs in
borrow checking.

\paragraph{Methodology}
We surveyed soundness issues in the Rust compiler \texttt{rustc}. We queried the
official \texttt{rust-lang/rust} issue
tracker\footnote{\url{https://github.com/rust-lang/rust/issues}} for all issues,
open or closed, labeled \texttt{I-unsound} and carrying at least one of eight area
labels related to borrowing and lifetimes, namely \texttt{A-borrow-checker},
\texttt{A-lifetimes}, \texttt{A-NLL}, \texttt{A-destructors}, \texttt{A-closures},
\texttt{A-implied-bounds},
\texttt{A-async-await}, and \texttt{A-variance}. The query returned 119 issues reported between
2014 and 2026.

Since \LBorrow is a core calculus and does not model the full Rust language, we exclude
issues outside its scope. Of the 119 issues, 92 are excluded for one of three reasons: they
rely on language features \LBorrow does not model, specifically the trait system and
associated-type projections, unstable nightly features, \texttt{unsafe} code, global \texttt{'static}
promotion, or \texttt{async} coroutine lowering; they concern unreleased or transitional
borrow-checker implementations such as Polonius~\cite{stjernaModellingRustsReference2020} and NLL migration mode; or they are
low-level representation, backend code generation, or standard library implementation bugs
unrelated to typing and borrow checking. Notably, about half of the excluded issues expose
the same kinds of lifetime-tracking gaps as the issues we keep, only triggered through
features outside our scope. A further 3 are excluded as duplicates.
This leaves 24 issues expressible with references, lifetimes, and functions alone, 6 of
them still open as of October 21, 2026.

\paragraph{Issue categories}
\Cref{tab:rust-bugs} groups the 24 issues into four categories, with the issues
still open at the snapshot date shown in bold; \Cref{app:case-study}
gives a per-issue breakdown along with a summary of the 95 excluded issues. Across all four
categories, most bugs in \texttt{rustc} arise because facts about active borrows and
lifetime bounds are split across separate compiler passes, such as type well-formedness
checking, MIR lowering, drop elaboration, and dataflow-based borrow checking.
Unsound programs slip through when one pass assumes an outlives bound, delays a mutable borrow, or rewrites a variable path that
another pass fails to track. \LBorrow prevents all 24 issues because borrowability and
lifetime constraints are tracked directly in types and contexts, and
\Cref{thm:soundness} proves that type checking applies sufficient checks.

\begin{table}[tb]
  \caption{Summary of in-scope \texttt{rustc} soundness issues. (open issues in \textcolor{ibmred}{\textbf{bold}}).}
  \label{tab:rust-bugs}
  \footnotesize
  \centering
  \setlength{\tabcolsep}{6pt}
  \newcommand{\rbug}[1]{\href{https://github.com/rust-lang/rust/issues/#1}{\texttt{\##1}}}
  \newcommand{\rbopen}[1]{\href{https://github.com/rust-lang/rust/issues/#1}{\textcolor{ibmred}{\textbf{\texttt{\##1}}}}}
  \newcommand{\rcat}[3]{\multicolumn{2}{@{}l@{}}{\textbf{#1} (#2): {\scriptsize #3}}}
  \begin{tabularx}{\linewidth}{@{}p{0.45\linewidth}X@{}}
    \toprule
    \textbf{Root Cause in \texttt{rustc}} & \textbf{Resolution in \LBorrow} \\
    \midrule
    \rcat{Unchecked lifetime bounds}{6, 1 open}{\rbopen{25860}, \rbug{29793}, \rbug{48803}, \rbug{52126}, \rbug{57170}, \rbug{57265}} \\
    Outlives bounds assumed on function parameters,
    return types, or mutated references are unchecked at call, coercion, or
    assignment sites. & Type well-formedness conditions, specifically \rref{WF-Ref} and
    \rref{WF-Arr}, require all implied outlives bounds to hold and are enforced at
    every call and reborrow via rules \rref{T-App} and \rref{S-Borrow}. \\
    \addlinespace[1pt]\cmidrule[0.3pt]{1-2}
    \rcat{Two-phase borrow reservation}{2, 2 open}{\rbopen{160598}, \rbopen{164163}} \\
    When activating a reserved mutable borrow,
    \texttt{rustc} misses conflicting borrows if evaluating the arguments
    reassigns the borrowed path or alters control flow. & Context splitting records
    active borrows in the lender's type (\rref{S-Borrow}), and restoration recovers the
    receiver's permission once argument borrows end (\rref{T-Pair}), eliminating the need
    for a reservation phase. \\
    \addlinespace[1pt]\cmidrule[0.3pt]{1-2}
    \rcat{Destructors on dead borrows}{6, 1 open}{\rbug{26656}, \rbug{31567}, \rbug{90838}, \rbug{115175}, \rbug{148854}, \rbopen{164250}} \\
    Drop checking incorrectly allows a borrow's
    lifetime to end prematurely before it is dropped. & Rule \rref{T-LftEnd}
    forbids ending a lifetime while any active borrow at that lifetime or closure capturing it remains in the context, and
    \rref{T-Drop} cannot free the lender until that lifetime has ended. \\
    \addlinespace[1pt]\cmidrule[0.3pt]{1-2}
    \rcat{Unchecked place accesses}{10, 2 open}{\rbug{27282}, \rbug{27868}, \rbug{29723}, \rbug{38899}, \rbug{45045}, \rbug{51117}, \rbug{51415}, \rbug{56797}, \rbopen{160525}, \rbopen{160599}} \\
    Borrow checking misses implicit reads or
    dereferences of a place or disagrees with lowering on subexpression order, allowing
    accesses that conflict with an active borrow. & Borrowing and memory accesses are
    tracked directly in types with no separate lowering or analysis pass, where context
    splitting (\rref{S-Borrow}) and \rref{T-Read} prevent conflicting accesses across
    subexpressions. \\
    \bottomrule
  \end{tabularx}
  \vspace{-8pt}
\end{table}

\paragraph{Walkthrough of Issue \texttt{\#164163}}
We illustrate how \LBorrow rules out these bugs using
\href{https://github.com/rust-lang/rust/issues/164163}{issue \texttt{\#164163}}
as an example. It is a
soundness bug in \textit{two-phase borrows} affecting stable Rust, reported in
October 2026 and still open. Two-phase borrows exist to accept nested method calls such as
\lstinline[language=Rust]|v.push(v.len())|. The call desugars to
\lstinline[language=Rust]|Vec::push(&mut v, v.len())|, and the receiver
\lstinline[language=Rust]|&mut v| is evaluated before the argument
\lstinline[language=Rust]|v.len()|, so under the ordinary borrow rules the call would be
rejected: the argument reads \lstinline[language=Rust]|v| while a mutable borrow of it
is already live. Two-phase borrows therefore give the receiver's mutable borrow two
phases. When evaluated, it is only \textit{reserved}, and \texttt{rustc} treats it like
a shared borrow, so the read in \lstinline[language=Rust]|v.len()| is allowed. When
\lstinline[language=Rust]|push| is entered, the borrow is \textit{activated} and becomes
a genuine mutable borrow. At that point \texttt{rustc} checks that no shared borrow
created during the reservation is still live, since any such borrow would now alias
the mutable one.

\begin{wrapfigure}{r}{0.55\textwidth}
\vspace{-\baselineskip}
\begin{taggedbox}[unbreakable]{Rust}
\begin{lstlisting}[language=Rust, aboveskip=0pt, belowskip=0pt]
let mut a = vec![1]; let mut b = vec![2];
let mut p: &mut Vec<i32> = &mut a;
let s: &Vec<i32>;
p.push({ s = &*p; p = &mut b; 0 });
println!("{:?}", s);  // UB: reads a via s
\end{lstlisting}
\end{taggedbox}
\vspace{-\baselineskip}
\end{wrapfigure}
Issue \texttt{\#164163} defeats this activation check by reassigning the receiver
variable inside the argument expression, as the following adapted reproducible
example shows.
Evaluating the receiver of \lstinline[language=Rust]|p.push(...)| reserves a
mutable reborrow of \lstinline[language=Rust]|*p|, which at this point is
\lstinline[language=Rust]|a|. Inside the argument block,
\lstinline[language=Rust]|s = &*p| takes a shared borrow of
\lstinline[language=Rust]|a| and stores it in \lstinline[language=Rust]|s|;
since the receiver's mutable reborrow is still only reserved, \texttt{rustc}
allows it. Then \lstinline[language=Rust]|p = &mut b| overwrites
\lstinline[language=Rust]|p| to point to \lstinline[language=Rust]|b|. The
borrow checker tracks borrows by syntactic paths and discards all borrows of
\lstinline[language=Rust]|*p| when \lstinline[language=Rust]|p| is reassigned,
so the assignment also discards the pending reservation. When
\lstinline[language=Rust]|push| finally runs on the receiver evaluated earlier,
namely \lstinline[language=Rust]|a|, the activation check finds nothing to
check, and \lstinline[language=Rust]|push| mutates \lstinline[language=Rust]|a|
while \lstinline[language=Rust]|s| still borrows it. Aliasing a shared and a
mutable reference this way is undefined behavior in safe Rust, which can be
confirmed by running the program under
Miri~\cite{jungMiriPracticalUndefined2026}; had \lstinline[language=Rust]|s|
borrowed an element instead, \lstinline[language=Rust]|s = &p[0]|, the push
would have to grow the full vector \lstinline[language=Rust]|a|, and the last
line would read a buffer that \lstinline[language=Rust]|push| has freed.

We now turn to \LBorrow. First, we note that \LBorrow accepts the nested call that motivated
two-phase borrows without any special mechanism. With
$\texttt{v} : \RefT^{a^\Mut}\,\Kw{Vec}$,
$\Kw{len} : \forall b.\, \RefT^{b}\,\Kw{Vec} \multimap^{b} \Kw{int}$, and
$\Kw{push} : \forall b.\, \RefT^{b^\Mut}\,\Kw{Vec} \otimes \Kw{int} \multimap^{b} ()$,
the program
\lstinline[language=Rust]|v.push(v.len())| is typed as follows:
\begin{taggedbox}{\LBorrow}
\begin{lstlisting}[language=calc, escapeinside={(*@}{@*)}, aboveskip=0pt, belowskip=0pt]
letlft c :> a;                   -- (*@$\texttt{v} : \RefT^{a^\Mut}\,\Kw{Vec}$@*)
push (v,                         -- (*@receiver $\texttt{v} : \RefT^{a^\Mut - c}\,\Kw{Vec}$@*)
  let n = len v in endlft c; n)  -- (*@argument $\texttt{v} : \RefT^{c}\,\Kw{Vec}$; ending $c$ restores $\texttt{v} : \RefT^{a^\Mut}\,\Kw{Vec}$@*)
\end{lstlisting}
\end{taggedbox}
\noindent
The context split for the pair (\rref{T-Pair}) gives the receiver $\texttt{v} : \RefT^{a^\Mut - c}\,\Kw{Vec}$
and the argument $\texttt{v} : \RefT^{c}\,\Kw{Vec}$. The argument reads the length
through the latter and ends $c$. The rule then restores the receiver's type by
the lifetimes the argument ended, to $\RefT^{a^\Mut}\,\Kw{Vec}$, as $\Kw{push}$
requires. Rust's reserved phase thus corresponds to the lender type
$\RefT^{a^\Mut - c}\,\Kw{Vec}$, and its activation to restoration.

Now let us see what happens when we try to type-check the buggy program in \LBorrow. We model
the reassignable variable \texttt{p} as an owned cell $\texttt{p} :
\RefT^U(\RefT^{a^\Mut}\,\Kw{Vec})$ holding a mutable borrow of \texttt{a}.
\begin{taggedbox}{\LBorrow}
\begin{lstlisting}[language=calc, escapeinside={(*@}{@*)}, aboveskip=0pt, belowskip=0pt]
let p = ref ma in               -- (*@$\texttt{ma}, \texttt{mb} : \RefT^{a^\Mut}\,\Kw{Vec},\quad \texttt{p} : \RefT^U(\RefT^{a^\Mut}\,\Kw{Vec})$@*)
letlft c :> a;
let r: (*@\lstann{15}{\RefT^{c^\Mut}(\RefT^{a^\Mut}\,\Kw{Vec})}@*) = p in   -- (*@$\texttt{p} : \RefT^{U - c^\Mut}(\RefT^{a^\Mut}\,\Kw{Vec}),\quad \texttt{r} : \RefT^{c^\Mut}(\RefT^{a^\Mut}\,\Kw{Vec})$@*)
let target = read r in          -- (*@consumes $\texttt{r}$; $\texttt{target} : \RefT^{c^\Mut}\,\Kw{Vec}$@*)
push (target,
  let r2: (*@\lstann{13}{\RefT^{c}(\RefT^{a^\Mut}\,\Kw{Vec})}@*) = p in  -- (*@REJECT: $\Erase{U - c^\Mut} = (\emptyset, \bullet)$ cannot split off $c$@*)
  let s = read r2 in
  let (p', old) = swap p mb in  -- (*@REJECT: $\texttt{p}$ is not writable while $c$ is alive@*)
  0);
endlft c
\end{lstlisting}
\end{taggedbox}
\noindent
To obtain the mutable receiver $\texttt{target} : \RefT^{c^\Mut}\,\Kw{Vec}$ for
$\Kw{push}$, line 3 splits a mutable borrow $\texttt{r}$ from $\texttt{p}$ at lifetime
$c$ via \rref{S-Borrow} and \rref{S-Ref}, and line 4 reads $\texttt{target}$ from
$\texttt{r}$ via \rref{T-Read} since $\Rview{c^\Mut}{\RefT^{a^\Mut}\,\Kw{Vec}} =
\RefT^{c^\Mut}\,\Kw{Vec}$. Crucially, this split immediately updates the lender's type in
the context to $\texttt{p} : \RefT^{U - c^\Mut}(\RefT^{a^\Mut}\,\Kw{Vec})$ that has
permission $\Erase{U - c^\Mut} = (\emptyset, \bullet)$. Because the receiver
$\texttt{target}$ is headed by $c$,
lifetime $c$ must remain alive inside the second component of the pair, and thus
both steps inside it are rejected. Line 6 cannot split an immutable borrow
$\texttt{r2}$ from $\texttt{p}$ because \rref{S-Borrow} requires splitting read
permission $(\Kw{r}, \circ)$ from $\Erase{U - c^\Mut} = (\emptyset, \bullet)$, which is
impossible. Likewise, line 8 cannot $\Kw{swap}$ a new borrow $\texttt{mb}$ into $\texttt{p}$:
$\Kw{swap}$ needs a mutable borrow of $\texttt{p}$, which $(\emptyset, \bullet)$ cannot
provide either. Even if $\texttt{target}$ were reborrowed directly from
$\texttt{ma} : \RefT^{a^\Mut}\,\Kw{Vec}$ without going through $\texttt{p}$, the
subtraction $- c^\Mut$ would be recorded in $\texttt{ma} : \RefT^{a^\Mut -
c^\Mut}\,\Kw{Vec}$, so reassigning $\texttt{p}$ to $\texttt{mb}$ could never erase the
active borrow on $\texttt{ma}$.

More broadly, \LBorrow shines by capturing the essence of Rust-style borrowing in the type
system from first principles. Giving lenders first-class types and enforcing borrowing and
lifetime rules directly during type checking avoids the need for separate static analysis
passes and special-case mechanisms such as two-phase borrow reservation, providing a
unified foundation that is easier to reason about and prevents these classes of bugs once
and for all.

\section{Related Work}
\label{sec:related}

\makeatletter
\let\origparagraph\paragraph
\renewcommand\paragraph{\def\@toclevel{4}%
  \@startsection{paragraph}{4}{\parindent}%
  {-.1\baselineskip \@plus -1\p@ \@minus -.2\p@}%
  {-3.5\p@}%
  {\ACM@NRadjust{\@parfont\@adddotafter}}}
\makeatother

\paragraph{Type systems for borrowing}
There are several recent attempts to bring Rust-style borrowing to substructural type systems.
\Cref{tbl:borrowing-foundations} compares \LBorrow{} with Rust and these calculi
in terms of how they handle borrowing. The baseline,
Rust, supports unrestricted shared borrows and non-lexical lifetimes without capability passing,
yet it checks borrowing outside of the type system.
Fractional uniqueness~\cite{marshallLinearityUniquenessEntente2022,marshallFunctionalOwnershipFractional2024}
supports borrows via a higher-order primitive that consumes a unique value and passes a
full-fraction borrow to a function, along with operations to split and rejoin fractions.
The function must thread borrows through every operation and explicitly rejoin
split fractions before returning. Therefore, it is still prone to the capability passing problem and shared borrows do not enjoy
weakening or contraction.
Similarly, BoCa~\cite{wagnerLinearityBorrowing2025} also introduces a higher-order
operation that consumes the lender and passes a fresh borrow to a function. In their system, shared
borrows are unrestricted and borrows can be discarded at the end of the call without being returned.
However, the operation returns the restored lender together with the result, so the
caller must still unpack the pair to rebind the lender (marked $\dagger$ in \Cref{tbl:borrowing-foundations}).
Another line of work avoids capability passing by restoring lenders in place, but freezes lenders in the typing context while borrowed.
Affe~\cite{radanneKindlyBentFree2020} splits a borrowed variable into a
suspended binding scoped by a region, and a restored binding for the continuation after the region.
OxCaml~\cite{lorenzenOxidizingOCamlModal2024,georgesDataRaceFreedom2025} attaches modes
to values, temporarily downgrading a lender to \verb|local| \verb|aliased| modes within a
borrowing expression and restoring its original mode in the continuation.
Both systems keep borrows lexically scoped to support type and mode inference without heavy
lifetime annotations.
Oxide~\cite{weissOxideEssenceRust2021} goes beyond lexical scope to support non-lexical lifetimes by recording active loan
sets per region in the typing context and clearing them once a region is no longer used.
Conversely, Pure Borrow~\cite{matsushitaPureBorrowLinear2026} models lenders as a special type and supports non-lexical lifetimes
by introducing lifetime tokens to track lifetime scopes in Linear Haskell~\cite{bernardyLinearHaskellPractical2017}. However, they split
access permissions between references and lifetime tokens.
As a result, using borrows and restoring lenders require threading these tokens through the program and explicitly reclaiming each lender, reintroducing capability passing.

\begin{table}[tb]
\caption{Related work comparison}
\label{tbl:borrowing-foundations}
\centering
\footnotesize
\renewcommand{\arraystretch}{1.15}
\newcommand{\featureyes}{\textcolor{green!40!black}{\ding{51}}}
\newcommand{\featureno}{\textcolor{red!70!black}{\ding{55}}}
\newcommand{\featurehalf}{\featureyes\rlap{\textsuperscript{$\dagger$}}}
\setlength{\tabcolsep}{2.2pt}
\begin{tabular}{@{}lcccccccc@{}}
\toprule
  & Rust~\cite{matsakisRustLanguage2014}
  & Fractional~\cite{marshallFunctionalOwnershipFractional2024}
  & BoCa~\cite{wagnerLinearityBorrowing2025}
  & Affe~\cite{radanneKindlyBentFree2020}
  & OxCaml~\cite{lorenzenOxidizingOCamlModal2024}
  & Oxide~\cite{weissOxideEssenceRust2021}
  & Pure Borrow~\cite{matsushitaPureBorrowLinear2026}
  & \LBorrow{} \\
\midrule
Type-system based           & \featureno  & \featureyes & \featureyes & \featureyes & \featureyes & \featureyes & \featureyes & \featureyes \\
Unrestricted sharing        & \featureyes & \featureno  & \featureyes & \featureyes & \featureyes & \featureyes & \featureyes & \featureyes \\
Capability-passing free     & \featureyes & \featureno  & \featurehalf & \featureyes & \featureyes & \featureyes & \featureno  & \featureyes \\
Non-lexical lifetimes       & \featureyes & \featureno  & \featureno  & \featureno  & \featureno  & \featureyes & \featureyes & \featureyes \\
First-class lenders         & \featureno  & \featureno  & \featureno  & \featureno  & \featureno  & \featureno  & \featureyes & \featureyes \\
\bottomrule
\end{tabular}
\vspace{-1.2ex}
\end{table}

\paragraph{Semantics and verification of Rust}
There is also a rich literature on the semantics and verification of Rust programs.
Aliasing models, such as Stacked Borrows~\cite{jungStackedBorrowsAliasing2019} and Tree
Borrows~\cite{villaniTreeBorrows2025}, define the memory model of Rust
to detect undefined aliasing behaviour in unsafe code and verify compiler optimizations.
On program logics, semantic type soundness models and verification frameworks, including
RustBelt~\cite{jungRustBeltSecuringFoundations2017},
RustHornBelt~\cite{matsushitaRustHornBeltSemanticFoundation2022}, and
RefinedRust~\cite{gaherRefinedRustTypeSystem2024}, reason about the safety and functional
correctness of Rust programs. In these systems, an active borrow attaches a lifetime-indexed
blocking marker to its lender's place or context entry, which is removed when the lifetime
ends. \LBorrow{}'s subtracted borrowability draws inspiration from this
mechanism.

\paragraph{Linear types, permissions, and regions}
More broadly, modern borrowing systems build on a long line of work that relaxes
strict linearity and tracks aliasing through permissions and regions~\cite{wadlerLinearTypesCan1990,
oderskyObserversLinearTypes1992,
swamySafeManualMemory2006,
nadenTypeSystemBorrowing2012}.
In program logics and permission systems, fractional and
counting permissions~\cite{boylandCheckingInterferenceFractional2003,bornatPermissionAccountingSeparation2005}
and temporary read-only permissions~\cite{chargueraudTemporaryReadOnlyPermissions2017}
support shared read access by splitting and recombining permissions, which we
also use in our semantic model in \Cref{sec:semantics}. Finally, region type
systems~\cite{tofteRegionBasedMemoryManagement1997,fluetLinearRegionsAre2006}
and capability
calculi~\cite{craryTypedMemoryManagement1999,smithAliasTypes2000,fahndrichAdoptionFocusPractical2002}
track captured pointers via region effects on function arrows and let
functions consume or temporarily revoke capabilities, which motivated
\LBorrow{}'s lifetime effects.

\paragraph{Future work} We see two main avenues for future work. First,
\LBorrow{} requires explicit lifetime annotations. We plan to develop an
algorithmic version with automatic lifetime
inference~\cite{brandonFullyAutomaticTypeInference2026,radanneKindlyBentFree2020,2094nllRustRFC,stjernaModellingRustsReference2020,tofteRegionInferenceAlgorithm1998},
for which \LBorrow{} can serve as the elaboration target.
Second, we would like to prove \textit{leak freedom}. \LBorrow{} statically
forbids discarding owned references, since any reference with a drop obligation
is linear and must eventually be consumed by $\Kw{drop}$. We have not, however,
proved that well-typed programs free all memory at runtime, because Iris is an affine
separation logic in which resources may be discarded. Prior Iris-based
soundness proofs for
borrowing~\cite{jungRustBeltSecuringFoundations2017,matsushitaRustHornBeltSemanticFoundation2022,gaherRefinedRustTypeSystem2024}
likewise establish memory safety but not leak freedom, while borrowing calculi
with leak freedom~\cite{radanneKindlyBentFree2020,wagnerLinearityBorrowing2025}
rely on custom linear models rather than Iris. We will study a linear variant of Iris such as
Iron~\cite{bizjakIronManagingObligations2019}.

\let\paragraph\origparagraph

\section*{AI Use}

The authors designed key proof strategies themselves and used Claude Code to
assist in mechanizing them in Rocq on top of Iris. The authors have carefully
reviewed the resulting Rocq definitions and statements and take full responsibility for their
correctness.

\bibliographystyle{ACM-Reference-Format}
\bibliography{references}

\newpage

\appendix
\section{Type System}
\label[appendix]{app:calculus}

\subsection{Syntax}

\[
\begin{array}{rcl@{\qquad}l}
a, b, c, p & & & \text{lifetime variable} \\
\alpha & ::= & a \mid a^{\Mut} & \text{borrowability atom} \\
\rho & ::= & U \mid \alpha \mid \rho - \alpha & \text{borrowability} \\
\beta & ::= & \Kw{w} \mid \Kw{r} \mid \emptyset & \text{access permission} \\
d & ::= & \bullet \mid \circ & \text{drop permission} \\
\pi & ::= & (\beta, d) & \text{permission} \\
\Delta & ::= & \cdot \mid \Delta, a \mid \Delta, a <: b \mid \Delta, \dagger a & \text{lifetime context} \\
q & ::= & \Lin \mid \Aff \mid \Un & \text{qualifier} \\[0.4em]
\tau & ::= & () \mid \Kw{int} \mid \Kw{bool} \mid \tau_1 \otimes \tau_2 \mid \RefT^{\rho}\tau
  \mid \ArrT{p}{\Delta_f}{L}{q}{\tau_1}{\tau_2} & \text{type} \\[0.4em]
e & ::= & x \mid \lambda x.\, e \mid e_1\, e_2 \mid \Kw{let}\ x = e_1\ \Kw{in}\ e_2 \mid e_1; e_2 & \text{term} \\
& \mid & () \mid n \mid \Kw{true} \mid \Kw{false} \mid (e_1, e_2) \mid \Kw{let}\ (x, y) = e_1\ \Kw{in}\ e_2 \\
& \mid & \RefT\ e \mid \Kw{read}\ e \mid \Kw{swap}\ e_1\ e_2 \mid \Kw{drop}\ e \\
& \mid & \Kw{if}\ e\ \Kw{then}\ e_1\ \Kw{else}\ e_2 \mid \Kw{loop}\ x = e_1\ \Kw{in}\ e_2 \\
& \mid & \Kw{letlft}\ c :> \bar a;\, e \mid \Kw{endlft}\ a;\, e \\[0.4em]
\Gamma & ::= & \cdot \mid \Gamma, x : \tau & \text{typing context}
\end{array}
\]

\subsection{Borrowabilities}

\begin{align*}
\Head(U) &= U & \Subs(U) = \Subs(\alpha) &= \emptyset \\
\Head(\alpha) &= \alpha & \Subs(\rho - \alpha) &= \Subs(\rho) \cup \Heads(\alpha) \\
\Head(\rho - \alpha) &= \Head(\rho) & \Heads(\rho) &= \{\, a \mid \Head(\rho) \in \{a, a^{\Mut}\} \,\}
\end{align*}
\begin{align*}
\Heads(()) = \Heads(\Kw{int}) = \Heads(\Kw{bool}) &= \emptyset \\
\Heads(\tau_1 \otimes \tau_2) &= \Heads(\tau_1) \cup \Heads(\tau_2) \\
\Heads(\RefT^{\rho}\tau) &= \Heads(\rho) \cup \Heads(\tau) \\
\Heads(\ArrT{p}{\Delta_f}{L}{q}{\tau_1}{\tau_2}) &= (\Dom\Delta_f \cup \Heads(\tau_1) \cup \Heads(\tau_2)) \setminus \bar p \\[0.3em]
\Subs(()) = \Subs(\Kw{int}) = \Subs(\Kw{bool}) &= \emptyset \\
\Subs(\tau_1 \otimes \tau_2) &= \Subs(\tau_1) \cup \Subs(\tau_2) \\
\Subs(\RefT^{\rho}\tau) &= \Subs(\rho) \cup \Subs(\tau) \\
\Subs(\ArrT{p}{\Delta_f}{L}{q}{\tau_1}{\tau_2}) &= (\Subs(\tau_1) \cup \Subs(\tau_2)) \setminus \bar p
\end{align*}
\begin{align*}
\Vars(\rho) &= \Heads(\rho) \cup \Subs(\rho) \\
\FL(\tau) &= \Heads(\tau) \cup \Subs(\tau) \\
\Vars(\Delta) &= \Dom\Delta \cup \{ a, b \mid (a <: b) \in \Delta \}
\end{align*}

\noindent\makebox[\linewidth]{%
  \(\boxed{\beta \leadsto \beta_1 \oplus \beta_2 \qquad d \leadsto d_1 \oplus d_2 \qquad \pi \leadsto \pi_1 \oplus \pi_2}\)\hfill\textit{(Permission Split)}}
\begin{mathpar}
\inferrule*[right=\textsc{Q-Write}]{ }{\Kw{w} \leadsto \Kw{r} \oplus \Kw{r}}
\and
\inferrule*[right=\textsc{Q-Read}]{ }{\Kw{r} \leadsto \Kw{r} \oplus \Kw{r}}
\and
\inferrule*[right=\textsc{Q-Empty}]{ }{\beta \leadsto \beta \oplus \emptyset}
\and
\inferrule*[right=\textsc{Q-Sym}]{\beta \leadsto \beta_2 \oplus \beta_1}{\beta \leadsto \beta_1 \oplus \beta_2}
\and
\inferrule*[right=\textsc{D-Keep}]{ }{d \leadsto d \oplus \circ}
\and
\inferrule*[right=\textsc{D-Sym}]{d \leadsto d_2 \oplus d_1}{d \leadsto d_1 \oplus d_2}
\and
\inferrule*[right=\textsc{P-Split}]{\beta \leadsto \beta_1 \oplus \beta_2 \\ d \leadsto d_1 \oplus d_2}{(\beta, d) \leadsto (\beta_1, d_1) \oplus (\beta_2, d_2)}
\end{mathpar}

\begin{align*}
\Erase{U} &= (\Kw{w}, \bullet) \\
\Erase{a} &= (\Kw{r}, \circ) \\
\Erase{a^{\Mut}} &= (\Kw{w}, \circ) \\
\Erase{\rho - \alpha} &= \max\{\, \pi \mid \Erase{\rho} \leadsto \pi \oplus \Erase{\alpha} \,\}
  && \emptyset < \Kw{r} < \Kw{w} \\[0.4em]
\Erase{\rho}_\beta &= \beta && \text{if } \Erase{\rho} = (\beta, d)
\end{align*}

\subsection{Splitting}

\noindent\makebox[\linewidth]{%
  \(\boxed{\Delta \vDash \rho \leadsto \rho_1 \oplus \rho_2}\)\hfill\textit{(Borrowability Split)}}
\begin{mathpar}
\inferrule*[right=\textsc{S-Borrow}]
{ \alpha \in \{a, a^{\Mut}\} \\
  a \in \Dom\Delta \\
  a \notin \Heads(\rho) \\
  \Erase{\rho} \leadsto \Erase{\rho - \alpha} \oplus \Erase{\alpha} \\
  \overline{\Delta \vDash a_0 <: a}^{\,a_0 \in \Heads(\rho)} }
{ \Delta \vDash \rho \leadsto (\rho - \alpha) \oplus \alpha }
\and
\inferrule*[right=\textsc{S-Sym}]
{ \Delta \vDash \rho \leadsto \rho_2 \oplus \rho_1 }
{ \Delta \vDash \rho \leadsto \rho_1 \oplus \rho_2 }
\end{mathpar}

\noindent\makebox[\linewidth]{%
  \(\boxed{\Delta \vDash \tau \leadsto \tau_1 \oplus \tau_2}\)\hfill\textit{(Type Split)}}
\begin{mathpar}
\inferrule*[right=\textsc{S-Ref}]
{ \Delta \vDash \rho \leadsto \rho_1 \oplus \rho_2 \\
  \Gpre{\Delta \vDash \RefT^{\rho_1}\tau} \\
  \Gpre{\Delta \vDash \RefT^{\rho_2}\tau} }
{ \Delta \vDash \RefT^{\rho}\tau \leadsto \RefT^{\rho_1}\tau \oplus \RefT^{\rho_2}\tau }
\and
\inferrule*[right=\textsc{S-Prod}]
{ \Delta \vDash \tau_1 \leadsto \tau'_1 \oplus \tau''_1 \\
  \Delta \vDash \tau_2 \leadsto \tau'_2 \oplus \tau''_2 }
{ \Delta \vDash \tau_1 \otimes \tau_2 \leadsto \tau'_1 \otimes \tau'_2 \oplus \tau''_1 \otimes \tau''_2 }
\and
\inferrule*[right=\textsc{S-Un}]
{ \Qual{\tau} = \Un }
{ \Delta \vDash \tau \leadsto \tau \oplus \tau }
\end{mathpar}

\noindent\makebox[\linewidth]{%
  \(\boxed{\Delta \vDash \Gamma \leadsto \Gamma_1 \oplus \Gamma_2}\)\hfill\textit{(Context Split)}}
\begin{mathpar}
\inferrule*[right=\textsc{C-Empty}]{ }{\Delta \vDash \cdot \leadsto \cdot \oplus \cdot}
\and
\inferrule*[right=\textsc{C-Left}]
{ \Delta \vDash \Gamma \leadsto \Gamma_1 \oplus \Gamma_2 }
{ \Delta \vDash \Gamma, x : \tau \leadsto (\Gamma_1, x : \tau) \oplus \Gamma_2 }
\and
\inferrule*[right=\textsc{C-Right}]
{ \Delta \vDash \Gamma \leadsto \Gamma_1 \oplus \Gamma_2 }
{ \Delta \vDash \Gamma, x : \tau \leadsto \Gamma_1 \oplus (\Gamma_2, x : \tau) }
\and
\inferrule*[right=\textsc{C-Split}]
{ \Delta \vDash \Gamma \leadsto \Gamma_1 \oplus \Gamma_2 \\
  \Delta \vDash \tau \leadsto \tau_1 \oplus \tau_2 }
{ \Delta \vDash \Gamma, x : \tau \leadsto (\Gamma_1, x : \tau_1) \oplus (\Gamma_2, x : \tau_2) }
\end{mathpar}

\subsection{Lifetime Contexts}

\begin{align*}
\Delta \vDash a <: b \ &:\Leftrightarrow\ a, b \in \Dom\Delta \quad (a, b) \in (<:_\Delta)^{*} \\
\Delta \vDash \Delta_0 \ &:\Leftrightarrow\ \Dom\Delta_0 \subseteq \Dom\Delta \quad
  \forall (a <: b) \in \Delta_0.\ \Delta \vDash a <: b \\
\Dom\Delta &= \{ a \mid a \in \Delta \} \\
\dagger\Delta &= \{ a \mid \dagger a \in \Delta \} \\
\Delta|_{S} &= \{ a \in \Delta \mid a \in S \} \cup \{ a <: b \in \Delta \mid a, b \in S \} \cup \{ \dagger a \in \Delta \} \\
\Delta \setminus L &= \{ a \in \Delta \mid a \notin L \} \cup \{ a <: b \in \Delta \mid a, b \notin L \} \cup \{ \dagger a \in \Delta \}
\end{align*}

\noindent\makebox[\linewidth]{%
  \(\boxed{\Delta \LftEnd{a} \Delta'}\)\hfill\textit{(Lifetime Ending)}}
\begin{mathpar}
\inferrule*[right=\textsc{L-End}]
{ a \in \Dom\Delta \\
  \overline{\Delta \nvDash a <: a'}^{\,a' \in \Dom\Delta \setminus \{a\}} }
{ \Delta \LftEnd{a} (\Delta \setminus \{a\}), \dagger a }
\and
\inferrule*[right=\textsc{L-Nil}]
{ }
{ \Delta \LftEnd{\emptyset} \Delta }
\and
\inferrule*[right=\textsc{L-Set}]
{ a \in S \\ \Delta \LftEnd{a} \Delta_1 \\ \Delta_1 \LftEnd{S \setminus \{a\}} \Delta' }
{ \Delta \LftEnd{S} \Delta' }
\end{mathpar}

\subsection{Restoration}

\begin{align*}
(\rho - \alpha)\Rst{\bar a} &= \rho\Rst{\bar a} && \alpha \in \{b, b^{\Mut}\},\ b \in \bar a \\
(\rho - \alpha)\Rst{\bar a} &= (\rho\Rst{\bar a}) - \alpha && \alpha \in \{b, b^{\Mut}\},\ b \notin \bar a \\
U\Rst{\bar a} = U, \quad \alpha\Rst{\bar a} &= \alpha && \alpha \in \{b, b^{\Mut}\},\ b \notin \bar a \\[0.3em]
(\RefT^{\rho}\tau)\Rst{\bar a} &= \RefT^{\rho\Rst{\bar a}}(\tau\Rst{\bar a}) \\
(\tau_1 \otimes \tau_2)\Rst{\bar a} &= \tau_1\Rst{\bar a} \otimes \tau_2\Rst{\bar a} \\
\tau\Rst{\bar a} &= \tau && \text{base type or function type } \tau \\[0.3em]
(\Gamma, x : \tau)\Rst{\bar a} &= \Gamma\Rst{\bar a}, x : \tau\Rst{\bar a}
\end{align*}

\subsection{Well-formedness}

\noindent\makebox[\linewidth]{%
  \(\boxed{\Delta \vDash \tau}\)\hfill\textit{(Type Well-formedness)}}
\begin{mathpar}
\inferrule*[right=\textsc{WF-Base}]{ }{\Delta \vDash ()} \and
\inferrule*[right=\textsc{WF-Int}]{ }{\Delta \vDash \Kw{int}} \and
\inferrule*[right=\textsc{WF-Bool}]{ }{\Delta \vDash \Kw{bool}}
\and
\inferrule*[right=\textsc{WF-Prod}]
{ \Delta \vDash \tau_1 \\ \Delta \vDash \tau_2 }
{ \Delta \vDash \tau_1 \otimes \tau_2 }
\and
\inferrule*[right=\textsc{WF-Ref}]
{ \Vars(\rho) \subseteq \Dom\Delta \\
  \Heads(\rho) \cap \Subs(\rho) = \emptyset \\
  \overline{\Delta \vDash b <: a}^{\,a \in \Heads(\rho),\ b \in \Heads(\tau)} \\
  \Delta \vDash \tau }
{ \Delta \vDash \RefT^{\rho}\tau }
\and
\inferrule*[right=\textsc{WF-Arr}]
{ (\Dom\Delta_f \cup \FL(\tau_1) \cup \FL(\tau_2)) \setminus \bar p \subseteq \Dom\Delta \\
  \Vars(\Delta_f) \cup L \subseteq \Dom\Delta_f \\
  L \cap \bar p \subseteq \FL(\tau_1) \\\\
  \Delta_f \vDash \tau_1 \\
  \Delta_f \setminus L \vDash \tau_2 }
{ \Delta \vDash \ArrT{p}{\Delta_f}{L}{q}{\tau_1}{\tau_2} }
\end{mathpar}

\[
\Delta \vDash \Gamma \ :\Leftrightarrow\ \forall (x : \tau) \in \Gamma.\ \Delta \vDash \tau
\]

\subsection{Qualifiers}

\[
\Lin \sqsubset \Aff \sqsubset \Un
\]
\begin{align*}
\Qual{()} = \Qual{\Kw{int}} = \Qual{\Kw{bool}} &= \Un \\
\Qual{\tau_1 \otimes \tau_2} &= \Qual{\tau_1} \sqcap \Qual{\tau_2} \\
\Qual{\RefT^{\rho}\tau} &= \Lin && \Erase{\Head(\rho)} = (\Kw{w}, \bullet) \\
\Qual{\RefT^{\rho}\tau} &= \Aff && \Erase{\Head(\rho)} = (\Kw{w}, \circ) \\
\Qual{\RefT^{\rho}\tau} &= \Un && \Erase{\Head(\rho)} = (\Kw{r}, \circ) \\
\Qual{\ArrT{p}{\Delta_f}{L}{q}{\tau_1}{\tau_2}} &= q \\
\Qual{\Gamma} &= \bigsqcap_{(x : \tau) \in \Gamma} \Qual{\tau}
\end{align*}

\subsection{Read View}

\begin{align*}
\Rview{\alpha}{\tau} &= \tau && \Qual{\tau} = \Un \\
\Rview{\alpha}{\tau_1 \otimes \tau_2} &= \Rview{\alpha}{\tau_1} \otimes \Rview{\alpha}{\tau_2} \\
\Rview{\alpha}{\RefT^{\rho}\tau} &= \RefT^{\alpha'}\tau
  && \Qual{\RefT^{\rho}\tau} \neq \Un \\
  &&& \alpha \in \{a, a^\Mut\}\; \alpha' \in \{a, a^\Mut\}\; \Erase{\alpha'}_\beta = \Erase{\rho}_\beta \sqcap \Erase{\alpha}_\beta \neq \emptyset
\end{align*}

\subsection{Lifetime Instantiation}

\begin{align*}
\theta \Vdash_{\Delta} \ArrT{p}{\Delta_f}{L}{q}{\tau_1}{\tau_2} \ :\Leftrightarrow\ \ 
  & \forall a \notin \bar p.\ \theta a = a; \quad \theta \text{ injective on } \bar p; \\
  & \theta(\bar p) \cap \big((\Dom\Delta_f \cup \FL(\tau_1) \cup \FL(\tau_2) \cup \Vars(\Delta_f) \cup L) \setminus \bar p\big) = \emptyset; \\
  & \theta(\bar p) \subseteq \Dom\Delta
\end{align*}

\subsection{Typing}

\noindent\makebox[\linewidth]{%
  \(\boxed{\Delta; \Gamma \vdash e : \tau \dashv \Delta'}\)\hfill\textit{(Typing)}\footnotemark}
\footnotetext{We write \(\Delta_1 \setminus \Delta_2\) as shorthand for \(\Dom\Delta_1 \setminus \Dom\Delta_2\) when there's no ambiguity.}
\[
\inferrule{P}{\Delta; \Gamma_1 \oplus \Gamma_2 \vdash e : \tau \dashv \Delta'}
\quad\text{abbreviates}\quad
\inferrule{\Delta \vDash \Gamma \leadsto \Gamma_1 \oplus \Gamma_2 \\ P}{\Delta; \Gamma \vdash e : \tau \dashv \Delta'}
\]
\begin{mathpar}
\inferrule*[right=\textsc{T-Var}]
{ }
{ \Delta; x : \tau \vdash x : \tau \dashv \Delta }
\and
\inferrule*[right=\textsc{T-Const}]
{ n \in \mathbb{Z} }
{ \Delta; \cdot \vdash () : () \dashv \Delta \quad
  \Delta; \cdot \vdash n : \Kw{int} \dashv \Delta \quad
  \Delta; \cdot \vdash \Kw{true} \mid \Kw{false} : \Kw{bool} \dashv \Delta }
\and
\inferrule*[right=\rdef{T-Weak}]
{ \Delta; \Gamma \vdash e : \tau \dashv \Delta' \\ \Aff \sqsubseteq \Qual{\tau_0} }
{ \Delta; \Gamma, x : \tau_0 \vdash e : \tau \dashv \Delta' }
\end{mathpar}

\begin{mathpar}
\inferrule*[right=\textsc{T-LftIntro}]
{ \Delta, c, \bar a <: c; \Gamma \vdash e : \tau \dashv \Delta'' \\
  c \notin \Dom\Delta \cup \dagger\Delta \\ \bar a \subseteq \Dom\Delta }
{ \Delta; \Gamma \vdash \Kw{letlft}\ c :> \bar a;\, e : \tau \dashv \Delta'' }
\and
\inferrule*[right=\textsc{T-LftEnd}]
{ \Delta \LftEnd{a} \Delta' \\
  \Delta' \vDash \Gamma\Rst{a} \\
  \Delta'; \Gamma\Rst{a} \vdash e : \tau \dashv \Delta'' }
{ \Delta; \Gamma \vdash \Kw{endlft}\ a;\, e : \tau \dashv \Delta'' }
\end{mathpar}

\begin{mathpar}
\inferrule*[right=\textsc{T-RefOwn}]
{ \Delta; \Gamma \vdash e : \tau \dashv \Delta' }
{ \Delta; \Gamma \vdash \RefT\ e : \RefT^{U}\tau \dashv \Delta' }
\and
\inferrule*[right=\textsc{T-Read}]
{ \Delta; \Gamma \vdash e : \RefT^{\alpha}\tau \dashv \Delta'}
{ \Delta; \Gamma \vdash \Kw{read}\ e : \Rview{\alpha}{\tau} \dashv \Delta' }
\and
\inferrule*[right=\textsc{T-Drop}]
{ \Delta; \Gamma \vdash e : \RefT^{U}\tau \dashv \Delta' }
{ \Delta; \Gamma \vdash \Kw{drop}\ e : \tau \dashv \Delta' }
\and
\inferrule*[right=\textsc{T-Swap}]
{ \Delta; \Gamma_1 \vdash e_1 : \RefT^{a^{\Mut}}\tau_0 \dashv \Delta' \\
  \Gpre{\Delta' \vDash \Gamma_2\Rst{(\Delta \setminus \Delta')}} \\
  \Delta'; \Gamma_2\Rst{(\Delta \setminus \Delta')} \vdash e_2 : \tau \dashv \Delta'' \\
  \tau = \tau_0\Rst{(\Delta' \setminus \Delta'')} \\
  \Delta'' \vDash \RefT^{a^{\Mut}}\tau }
{ \Delta; \Gamma_1 \oplus \Gamma_2 \vdash \Kw{swap}\ e_1\ e_2 : \RefT^{a^{\Mut}}\tau \otimes \tau \dashv \Delta'' }
\end{mathpar}

\begin{mathpar}
\inferrule*[right=\textsc{T-Seq}]
{ \Delta_1; \Gamma_1 \vdash e_1 : () \dashv \Delta_2 \\
  \Gpre{\Delta_2 \vDash \Gamma_2\Rst{(\Delta_1 \setminus \Delta_2)}} \\
  \Delta_2; \Gamma_2\Rst{(\Delta_1 \setminus \Delta_2)} \vdash e_2 : \tau \dashv \Delta_3 }
{ \Delta_1; \Gamma_1 \oplus \Gamma_2 \vdash e_1; e_2 : \tau \dashv \Delta_3 }
\and
\inferrule*[right=\textsc{T-Let}]
{ \Delta_1; \Gamma_1 \vdash e_1 : \tau_1 \dashv \Delta_2 \\
  \Gpre{\Delta_2 \vDash \Gamma_2\Rst{(\Delta_1 \setminus \Delta_2)}} \\
  \Delta_2; \Gamma_2\Rst{(\Delta_1 \setminus \Delta_2)}, x : \tau_1 \vdash e_2 : \tau_2 \dashv \Delta_3 }
{ \Delta_1; \Gamma_1 \oplus \Gamma_2 \vdash \Kw{let}\ x = e_1\ \Kw{in}\ e_2 : \tau_2 \dashv \Delta_3 }
\and
\inferrule*[right=\textsc{T-LetPair}]
{ \Delta_1; \Gamma_1 \vdash e_1 : \tau_1 \otimes \tau_2 \dashv \Delta_2 \\
  \Gpre{\Delta_2 \vDash \Gamma_2\Rst{(\Delta_1 \setminus \Delta_2)}} \\
  \Delta_2; \Gamma_2\Rst{(\Delta_1 \setminus \Delta_2)}, x : \tau_1, y : \tau_2 \vdash e_2 : \tau \dashv \Delta_3 }
{ \Delta_1; \Gamma_1 \oplus \Gamma_2 \vdash \Kw{let}\ (x, y) = e_1\ \Kw{in}\ e_2 : \tau \dashv \Delta_3 }
\and
\inferrule*[right=\textsc{T-Pair}]
{ \Delta_1; \Gamma_1 \vdash e_1 : \tau_1 \dashv \Delta_2 \\
  \Gpre{\Delta_2 \vDash \Gamma_2\Rst{(\Delta_1 \setminus \Delta_2)}} \\
  \Delta_2; \Gamma_2\Rst{(\Delta_1 \setminus \Delta_2)} \vdash e_2 : \tau_2 \dashv \Delta_3 \\
  \Delta_3 \vDash \tau_1\Rst{(\Delta_2 \setminus \Delta_3)} }
{ \Delta_1; \Gamma_1 \oplus \Gamma_2 \vdash (e_1, e_2) : \tau_1\Rst{(\Delta_2 \setminus \Delta_3)} \otimes \tau_2 \dashv \Delta_3 }
\and
\inferrule*[right=\textsc{T-If}]
{ \Delta_1; \Gamma_1 \vdash e : \Kw{bool} \dashv \Delta_2 \\
  \Gpre{\Delta_2 \vDash \Gamma_2\Rst{(\Delta_1 \setminus \Delta_2)}} \\
  \Delta_2; \Gamma_2\Rst{(\Delta_1 \setminus \Delta_2)} \vdash e_1 : \tau \dashv \Delta_3, \dagger\bar c_1 \\
  \Delta_2; \Gamma_2\Rst{(\Delta_1 \setminus \Delta_2)} \vdash e_2 : \tau \dashv \Delta_3, \dagger\bar c_2 }
{ \Delta_1; \Gamma_1 \oplus \Gamma_2 \vdash \Kw{if}\ e\ \Kw{then}\ e_1\ \Kw{else}\ e_2 : \tau
  \dashv \Delta_3, \dagger(\bar c_1 \cup \bar c_2) }
\and
\inferrule*[right=\textsc{T-Loop}]
{ \Delta; \Gamma \vdash e_1 : \tau \dashv \Delta_l \\
  \Delta_l; x : \tau \vdash e_2 : \tau \otimes \Kw{bool} \dashv \Delta_l, \dagger\bar c }
{ \Delta; \Gamma \vdash \Kw{loop}\ x = e_1\ \Kw{in}\ e_2 : \tau \dashv \Delta_l, \dagger\bar c }
\end{mathpar}

\begin{mathpar}
\inferrule*[right=\textsc{T-Lam}]
{ \bar p = \Dom\Delta_f \setminus \Dom\Delta \\
  \Gpre{\Delta_f \vDash \Gamma} \\
  \Delta_f; \Gamma, x : \tau_1 \vdash e : \tau_2 \dashv \Delta_f \setminus L, \dagger\bar c \\
  \Delta \vDash \ArrT{p}{\Delta_f}{L}{q}{\tau_1}{\tau_2} \\
  \Delta \vDash \Delta_f|_{\Dom\Delta} \\
  q \sqsubseteq \Qual{\Gamma} }
{ \Delta; \Gamma \vdash \lambda x.\, e : \ArrT{p}{\Delta_f}{L}{q}{\tau_1}{\tau_2} \dashv \Delta }
\and
\inferrule*[right=\textsc{T-App}]
{ \Delta; \Gamma_1 \vdash e_1 : \ArrT{p}{\Delta_f}{L}{q}{\tau_1}{\tau_2} \dashv \Delta_1 \\
  \Gpre{\Delta_1 \vDash \Gamma_2\Rst{(\Delta \setminus \Delta_1)}} \\
  \Delta_1; \Gamma_2\Rst{(\Delta \setminus \Delta_1)} \vdash e_2 : \theta\tau_1 \dashv \Delta_2 \\
  \theta \Vdash_{\Delta_2} \ArrT{p}{\Delta_f}{L}{q}{\tau_1}{\tau_2} \\
  \Delta_2 \vDash \ArrT{p}{\Delta_f}{L}{q}{\tau_1}{\tau_2} \\
  \Delta_2 \vDash \theta\Delta_f \\
  \theta\Delta_f \vDash \Delta_2|_{\theta L} \\
  \Delta_2 \LftEnd{\theta L} \Delta_3 }
{ \Delta; \Gamma_1 \oplus \Gamma_2 \vdash e_1\, e_2 : \theta\tau_2 \dashv \Delta_3 }
\end{mathpar}

\section{Operational Semantics}
\label[appendix]{app:opsem}

\subsection{Runtime Syntax}

\[
\begin{array}{rcl@{\qquad}l}
v & ::= & () \mid n \mid \Kw{true} \mid \Kw{false} \mid \ell \mid (v_1, v_2) \mid \lambda x.\, e & \text{value} \\
e & ::= & \cdots \mid \Kw{iter}\ x\ e_2\ e & \text{runtime term} \\
\sigma & \in & \mathit{Loc} \rightharpoonup \mathit{Val} \cup \{\bot\} & \text{heap} \\[0.4em]
K & ::= & [\,] \mid K\ e_2 \mid v\ K \mid \Kw{let}\ x = K\ \Kw{in}\ e_2 \mid K; e_2 & \text{evaluation context} \\
& \mid & (K, e_2) \mid (v, K) \mid \Kw{let}\ (x, y) = K\ \Kw{in}\ e_2 \\
& \mid & \RefT\ K \mid \Kw{read}\ K \mid \Kw{swap}\ K\ e_2 \mid \Kw{swap}\ v\ K \mid \Kw{drop}\ K \\
& \mid & \Kw{if}\ K\ \Kw{then}\ e_1\ \Kw{else}\ e_2 \mid \Kw{loop}\ x = K\ \Kw{in}\ e_2 \mid \Kw{iter}\ x\ e_2\ K
\end{array}
\]

\subsection{Reduction}

\noindent\makebox[\linewidth]{%
  \(\boxed{\sigma; e \longrightarrow \sigma'; e'}\)\hfill\textit{(Head Reduction)}}
\begin{mathpar}
\inferrule*[right=\rdef{E-Beta}]
{ }
{ \sigma; (\lambda x.\, e)\ v \longrightarrow \sigma; [v/x]e }
\and
\inferrule*[right=\rdef{E-Let}]
{ }
{ \sigma; \Kw{let}\ x = v\ \Kw{in}\ e \longrightarrow \sigma; [v/x]e }
\and
\inferrule*[right=\rdef{E-Seq}]
{ }
{ \sigma; (); e \longrightarrow \sigma; e }
\and
\inferrule*[right=\rdef{E-LetPair}]
{ }
{ \sigma; \Kw{let}\ (x, y) = (v_1, v_2)\ \Kw{in}\ e \longrightarrow \sigma; [v_2/y]([v_1/x]e) }
\and
\inferrule*[right=\rdef{E-IfTrue}]
{ }
{ \sigma; \Kw{if}\ \Kw{true}\ \Kw{then}\ e_1\ \Kw{else}\ e_2 \longrightarrow \sigma; e_1 }
\and
\inferrule*[right=\rdef{E-IfFalse}]
{ }
{ \sigma; \Kw{if}\ \Kw{false}\ \Kw{then}\ e_1\ \Kw{else}\ e_2 \longrightarrow \sigma; e_2 }
\end{mathpar}

\begin{mathpar}
\inferrule*[right=\rdef{E-Loop}]
{ }
{ \sigma; \Kw{loop}\ x = v\ \Kw{in}\ e_2 \longrightarrow \sigma; \Kw{iter}\ x\ e_2\ ([v/x]e_2) }
\and
\inferrule*[right=\rdef{E-IterTrue}]
{ }
{ \sigma; \Kw{iter}\ x\ e_2\ (v, \Kw{true}) \longrightarrow \sigma; \Kw{loop}\ x = v\ \Kw{in}\ e_2 }
\and
\inferrule*[right=\rdef{E-IterFalse}]
{ }
{ \sigma; \Kw{iter}\ x\ e_2\ (v, \Kw{false}) \longrightarrow \sigma; v }
\end{mathpar}

\begin{mathpar}
\inferrule*[right=\rdef{E-Ref}]
{ \ell \notin \Dom\sigma }
{ \sigma; \RefT\ v \longrightarrow \sigma[\ell \mapsto v]; \ell }
\and
\inferrule*[right=\rdef{E-Read}]
{ \sigma(\ell) = v }
{ \sigma; \Kw{read}\ \ell \longrightarrow \sigma; v }
\and
\inferrule*[right=\rdef{E-Swap}]
{ \sigma(\ell) = w }
{ \sigma; \Kw{swap}\ \ell\ v \longrightarrow \sigma[\ell \mapsto v]; (\ell, w) }
\and
\inferrule*[right=\rdef{E-Drop}]
{ \sigma(\ell) = v }
{ \sigma; \Kw{drop}\ \ell \longrightarrow \sigma[\ell \mapsto \bot]; v }
\end{mathpar}

\begin{mathpar}
\inferrule*[right=\rdef{E-LftIntro}]
{ }
{ \sigma; \Kw{letlft}\ c :> \bar a;\, e \longrightarrow \sigma; e }
\and
\inferrule*[right=\rdef{E-LftEnd}]
{ }
{ \sigma; \Kw{endlft}\ a;\, e \longrightarrow \sigma; e }
\end{mathpar}

\noindent\makebox[\linewidth]{%
  \(\boxed{\sigma; e \longrightarrow \sigma'; e'}\)\hfill\textit{(Contextual Closure)}}
\begin{mathpar}
\inferrule*[right=\rdef{E-Ctx}]
{ \sigma; e \longrightarrow \sigma'; e' }
{ \sigma; K[e] \longrightarrow \sigma'; K[e'] }
\end{mathpar}

\section{Semantic Model}
\label[appendix]{app:model}

\newcommand{\CV}[1]{\mathcal{V}\llbracket #1 \rrbracket}
\newcommand{\CVex}[1]{\mathcal{V}^{\mathsf{ex}}\llbracket #1 \rrbracket}
\newcommand{\CF}[1]{\mathcal{F}\llbracket #1 \rrbracket}
\newcommand{\CD}[4]{\mathcal{D}_{#1}\llbracket #2 \rrbracket_{#3}(#4)}
\newcommand{\Shr}{\mathsf{Shr}}
\newcommand{\Res}{\mathcal{R}}
\newcommand{\Idx}{\mathsf{Idx}}
\newcommand{\Inh}{\mathsf{Inh}}
\newcommand{\Tok}{\mathsf{Tok}}
\newcommand{\Cell}{\mathsf{cell}}
\newcommand{\Pay}{\mathsf{Pay}}
\newcommand{\Sups}{\mathsf{Sup}}
\newcommand{\LT}{\mathsf{LoanTable}}
\newcommand{\World}{\mathsf{World}}
\newcommand{\Core}{\mathrm{core}}
\newcommand{\Uncore}{\mathrm{uncore}}
\newcommand{\Sub}[1]{\mathrm{sub}_{#1}}
\newcommand{\TopS}{\mathrm{top}}
\newcommand{\Strip}[1]{\lfloor #1 \rfloor}
\newcommand{\Unshare}[1]{\mathrm{unshare}_{#1}}
\newcommand{\RTop}[1]{\mathrm{rtop}_{#1}}
\newcommand{\Lends}{\mathrm{lends}}
\newcommand{\Owned}{\mathrm{owned}}
\newcommand{\Subbed}[1]{\mathrm{subbed}_{#1}}
\newcommand{\MHeads}{\mathrm{mheads}}
\newcommand{\MVars}{\mathrm{mvars}}
\newcommand{\HSep}{\mathrm{hsep}}
\newcommand{\Lft}{\mathrm{lft}}
\newcommand{\Alive}{\mathrm{alive}}
\newcommand{\Shape}{\mathsf{shape}}
\newcommand{\teq}{\approx}
\newcommand{\teqk}[1]{\approx^{\top}_{#1}}
\newcommand{\teqs}{\approx^{\mathsf{shr}}}
\newcommand{\DeadSet}[1]{\textstyle\bigsep_{\kappa \in #1} \dagger\kappa}

\subsection{Ghost State}
\label[appendix]{app:model-ghost}

\[
\small
\begin{array}{rcl@{\qquad}l}
\kappa & \in & \mathit{SemLft} & \text{semantic lifetime} \\
M & \in & \mathit{SemLft} \rightharpoonup \{\mathsf{alive}, \mathsf{dead}\} & \text{lifetime status map } (\Alive(M) = \{\kappa \mid M(\kappa) = \mathsf{alive}\}) \\
\Sups & \in & \mathit{SemLft} \rightharpoonup \mathcal{P}_{\mathrm{fin}}(\mathit{SemLft}) & \text{superlifetime map} \\
\iota = (\kappa, s) & \in & \mathit{BorIdx} = \mathit{SemLft} \times \mathit{Name} & \text{borrow index } (\Lft(\iota) = \kappa) \\
o & \in & \mathit{BorIdx} \uplus \{\circ\} & \text{loan base } (\circ\text{: root},\ \iota_0\text{: reborrow}) \\
\mathit{ln} & \in & \mathit{Loan} & \text{loan descriptor:}
\end{array}
\]
\[
\small
\mathit{ln} \;=\; \langle \mathsf{mut} \in \mathbb{B},\;
  \mathsf{lender} \in \mathit{BorIdx} \uplus \{\circ\},\;
  \mathsf{depth} \in \mathbb{N},\;
  \mathsf{root} \in \mathit{Loc},\;
  \mathsf{dep} \in \mathit{CTy} \rangle
\]

\[
\small
\begin{array}{r@{\quad}l}
[\kappa]_q & \text{lifetime token } (q \in (0, 1]) \\
\dagger\kappa & \text{dead token (persistent)} \\
\kappa \sqsubseteq \kappa' & \text{semantic outlives ($\kappa$ outlives $\kappa'$), } \exists S.\; \Sups(\kappa') = S * \lceil \kappa \in S \rceil\ \text{(persistent)} \\[0.3em]
\Idx(\iota, \mathit{ln}) & \text{borrow index mapping } \iota \text{ to descriptor } \mathit{ln}\ \text{(persistent)} \\
\Inh^{w}(\iota, q) & \text{lender's inheritance claim for fraction } q \in (0,1] \text{ and exclusive-token flag } w \in \mathbb{B} \\
\Tok_q(\iota) & \text{borrower's loan token at fraction } q \in (0,1]
\end{array}
\]
\[
\small
\Cell^{\mathcal{F}}_q(\ell, \hat\tau) \;\triangleq\; \exists w.\; \ell \mapsto_q w * \CF{\hat\tau}^{q}(w)
\]
\[
\small
\World(M, \Sups) \;\triangleq\;
  \mathsf{Auth}_{\mathrm{world}}(M) * \mathsf{Auth}_{\mathrm{sup}}(\Sups) *
  \big(\bigsep_{M(\kappa) = \mathsf{dead}} \dagger\kappa\big) *
  \lceil \mathsf{WorldWF}(M, \Sups) \rceil
\]
where $\mathsf{Auth}_{\mathrm{world}}(M)$ and $\mathsf{Auth}_{\mathrm{sup}}(\Sups)$ are the authoritative ghost-map ownerships backing $[\kappa]_q$, $\dagger\kappa$, and the persistent fragment $\Sups(\kappa) = S$, and $\mathsf{WorldWF}(M, \Sups)$ states that $\Dom\Sups = \Dom M$, each $\Sups(\kappa) \subseteq \Dom M$ defines a reflexive, transitive, and antisymmetric superlifetime relation on $\Dom M$, and $\Alive(M)$ is closed under $\Sups$.

\subsection{Closed Types and Agreement}
\label[appendix]{app:model-closed}

\[
\small
\begin{array}{rcl@{\qquad}l}
\hat\alpha & ::= & \kappa \mid \kappa^{\Mut} & \text{closed atom } (\Lft(\kappa) = \Lft(\kappa^{\Mut}) = \kappa) \\
\hat\rho & ::= & U \mid \hat\alpha \mid \hat\rho - \hat\alpha & \text{closed borrowability} \\
\hat q & ::= & \Lin \mid \Un \mid \Shape & \text{closed qualifier} \\
\hat\tau & ::= & () \mid \Kw{int} \mid \Kw{bool} \mid \hat\tau_1 \otimes \hat\tau_2 \mid \RefT^{\hat\rho}\hat\tau
  \mid (\ArrT{p}{\Delta_f}{L}{\hat q}{\tau_1}{\tau_2})_{\eta} & \text{closed type } (\mathit{CTy})
\end{array}
\]

{\small
\begin{align*}
\eta(U) &= U &
  \eta(a^{\Mut}) &= \eta(a)^{\Mut} \\
\eta(\rho - \alpha) &= \eta(\rho) - \eta(\alpha) &
  \eta(\tau) &= \tau \quad (\tau \in \{(), \Kw{int}, \Kw{bool}\}) \\
\eta(\tau_1 \otimes \tau_2) &= \eta(\tau_1) \otimes \eta(\tau_2) &
  \eta(\RefT^{\rho}\tau) &= \RefT^{\eta(\rho)}\eta(\tau) \\
\eta(\ArrT{p}{\Delta_f}{L}{q}{\tau_1}{\tau_2}) &= (\ArrT{p}{\Delta_f}{L}{\hat q}{\tau_1}{\tau_2})_{\eta|_{\FL}}
\end{align*}}%
where $\eta : \mathit{LVar} \to \mathit{SemLft}$ maps lifetime variables $a$ to semantic lifetimes $\eta(a)$, $\hat q = \Un$ if $q = \Un$, and $\hat q = \Lin$ if $q \in \{\Lin, \Aff\}$ (since standard Iris is affine, $\Lin$ and $\Aff$ closures have the same interpretation).
The qualifier $\Shape$ is used by $\Core(\hat\tau)$ when a cell containing a $\Lin$/$\Aff$ closure is borrowed immutably, stripping its captured resources ($\Box_{\Shape} P = \Kw{True}$) while keeping them with the lender.

{\small
\begin{align*}
\Lends(\hat\rho) &:\Leftrightarrow \Head(\hat\rho) \in \{U, \kappa^{\Mut}\} &
\Owned(\hat\rho) &:\Leftrightarrow \Head(\hat\rho) = U \\
\TopS(\hat\rho_0) &= \hat\rho_0 \quad (\hat\rho_0 \in \{U, \hat\alpha\}) &
\Unshare{\kappa}(\hat\rho_0) &= \hat\rho_0 \quad (\hat\rho_0 \in \{U, \hat\alpha\}) \\
\TopS(\hat\rho - \kappa) &= \TopS(\hat\rho) &
\TopS(\hat\rho - \kappa^{\Mut}) &= \TopS(\hat\rho) - \kappa^{\Mut} \\
\Unshare{\kappa}(\hat\rho - \kappa) &= \Unshare{\kappa}(\hat\rho) &
\Unshare{\kappa}(\hat\rho - \hat\alpha) &= \Unshare{\kappa}(\hat\rho) - \hat\alpha \quad (\hat\alpha \neq \kappa) \\
\Strip{\RefT^{\hat\rho}\hat\tau} &= \RefT^{\TopS(\hat\rho)}\hat\tau &
\Sub{\kappa}(\RefT^{\hat\rho}\hat\tau) &= \RefT^{\hat\rho - \kappa}\hat\tau \quad (\Lends(\hat\rho)) \\
\RTop{\kappa}(\RefT^{\hat\rho}\hat\tau) &= \RefT^{\Unshare{\kappa}(\hat\rho)}\hat\tau \quad (\Lends(\hat\rho)) &
\MVars(\hat\rho) &= \{\, \kappa \mid \kappa^{\Mut} \text{ subtracted in } \hat\rho \,\} \\
\Core(\RefT^{\hat\rho}\hat\tau) &= \RefT^{\hat\rho}\Core(\hat\tau) \quad (\Owned(\hat\rho)) &
\Uncore(\RefT^{\hat\rho}\hat\tau) &= \RefT^{\hat\rho}\Uncore(\hat\tau) \quad (\Owned(\hat\rho))
\end{align*}
\[
\Core((\ArrT{p}{\Delta_f}{L}{\Lin}{\tau_1}{\tau_2})_\eta) = (\ArrT{p}{\Delta_f}{L}{\Shape}{\tau_1}{\tau_2})_\eta \qquad
\Uncore((\ArrT{p}{\Delta_f}{L}{\Shape}{\tau_1}{\tau_2})_\eta) = (\ArrT{p}{\Delta_f}{L}{\Lin}{\tau_1}{\tau_2})_\eta
\]}%
Above, $\Strip{\cdot}$, $\Sub{\kappa}$, $\RTop{\kappa}$, $\Core$, and $\Uncore$ recurse through $\hat\tau_1 \otimes \hat\tau_2$ and are no-op on other types.
On the spine of $\hat\tau$ (reached through products and owned/mutable references):
$\Subbed{\kappa}(\hat\tau)$ holds when every top-level $\RefT^{\hat\rho}\hat\tau_0$ with $\Lends(\hat\rho)$ has $\hat\rho = \hat\rho' - \kappa$;
$\MHeads(\hat\tau)$ is the set of mutable head lifetimes $\kappa$ on the spine; and
$\HSep(\hat\tau)$ holds when no nested reference $\RefT^{\hat\rho}\hat\tau_0$ with $\Head(\hat\rho) = \kappa^{\Mut}$ strictly below the top-level references on the spine has $\kappa \in \Subs(\hat\rho)$.

\smallskip
\noindent\makebox[\linewidth][l]{\textbf{Agreement Up to Dead Lifetimes}}
\[
\small
\begin{array}{r@{\;}c@{\;}l}
\hat\tau \teq \hat\tau' & \triangleq & \exists D.\; \lceil \hat\tau\Rst{D} = \hat\tau'\Rst{D} \rceil * \DeadSet{D} \\[0.2em]
\hat\rho \teq \hat\rho' & \triangleq & \exists D.\; \lceil \hat\rho\Rst{D} = \hat\rho'\Rst{D} \rceil * \DeadSet{D} \\[0.2em]
\hat\tau \teqk{\kappa} \hat\tau' & \triangleq & \exists D.\; \lceil \RTop{\kappa}(\hat\tau)\Rst{D} = \RTop{\kappa}(\hat\tau')\Rst{D} \rceil * \DeadSet{D} \\[0.2em]
\hat\tau \teqs \hat\tau' & \triangleq & \exists D.\; \lceil \Strip{\Core(\hat\tau)}\Rst{D} = \Strip{\Core(\hat\tau')}\Rst{D} \rceil * \DeadSet{D}
\end{array}
\]

\subsection{Value, Sharing, and Residual Interpretations}
\label[appendix]{app:model-value}

\noindent\makebox[\linewidth][l]{\textbf{Borrowability Residual Predicate}}
\[
\small
\begin{array}{r@{\;}c@{\;}l}
\mathsf{Loan}^{o}_{\hat\alpha}(\hat\tau, \ell, s, \mathit{ln}, q_s) & \triangleq &
  \exists w.\; \lceil \mathit{ln}.\mathsf{root} = \ell \wedge \mathit{ln}.\mathsf{lender} = o \rceil *
  \Idx((\kappa_{\hat\alpha}, s), \mathit{ln}) * \Inh^{w}((\kappa_{\hat\alpha}, s), q_s) * {} \\
  & & \mathsf{LoanTy}_{\hat\alpha}(\hat\tau) \teq^{o}_{\kappa_{\hat\alpha}} \mathit{ln}.\mathsf{dep} *
  \mathsf{Kind}_{\hat\alpha}(\mathit{ln}) * {} \\
  & & \lceil o = \circ \wedge \hat\alpha = \kappa^{\Mut} \wedge \Core(\hat\tau) \neq \hat\tau \Rightarrow w \rceil \\[0.25em]
  & & \text{where } \kappa_{\hat\alpha} = \Lft(\hat\alpha), \quad
  \mathsf{LoanTy}_{\kappa}(\hat\tau) = \Core(\hat\tau), \quad
  \mathsf{LoanTy}_{\kappa^{\Mut}}(\hat\tau) = \Uncore(\hat\tau), \\
  & & \phantom{\text{where }}
  (\teq^{\circ}_{\kappa}) = (\teqk{\kappa}), \quad
  (\teq^{\iota_0}_{\kappa}) = (\teqs), \quad
  \mathsf{Kind}_{\kappa^{\Mut}}(\mathit{ln}) = \lceil \mathit{ln}.\mathsf{mut} \rceil, \\
  & & \phantom{\text{where }}
  \mathsf{Kind}_{\kappa}(\mathit{ln}) = \lceil \neg\mathit{ln}.\mathsf{mut} \wedge \Subbed{\kappa}(\mathit{ln}.\mathsf{dep}) \rceil \\[0.5em]
\Res^{\mathcal{F},o}_{\mathrm{sub}}(\hat\rho_0, \hat\tau, q, \ell) & \triangleq &
  \begin{cases}
    \Cell^{\mathcal{F}}_q(\ell, \hat\tau) & o = \circ \\
    \Tok_q(\iota_0) & o = \iota_0
  \end{cases} \qquad (\hat\rho_0 \in \{U, \hat\alpha\}) \\[0.8em]
\Res^{\mathcal{F},o}_{\mathrm{sub}}(\hat\rho - \hat\alpha, \hat\tau, q, \ell) & \triangleq &
  \exists R, q_r.\;
  \lceil q = q_r + \textstyle\sum_{s \mapsto (\mathit{ln}, q_s) \in R} q_s \rceil * {} \\
  & & \big(\bigsep_{s \mapsto (\mathit{ln}, q_s) \in R} \mathsf{Loan}^{o}_{\hat\alpha}(\hat\tau, \ell, s, \mathit{ln}, q_s)\big) *
  \big(\lceil R \neq \emptyset \rceil \vee \dagger\kappa_{\hat\alpha} \vee \lceil \hat\alpha = \kappa \rceil\big) * {} \\
  & & \big(\lceil R \neq \emptyset \wedge \hat\alpha = \kappa \wedge \Erase{\hat\rho}_\beta = \emptyset \rceil \Rightarrow \DeadSet{\MVars(\hat\rho)}\big) * {} \\
  & & \Big( \lceil \hat\alpha = \kappa^{\Mut} \wedge \Subs(\hat\rho) = \emptyset \wedge q_r = \bot \rceil
      \;\vee\; {} \\
  & & \phantom{\Big(}
      \exists q_1.\; \lceil q_r = q_1 \wedge \hat\alpha = \kappa \rceil * \Res^{\mathcal{F},o}_{\mathrm{sub}}(\hat\rho, \hat\tau, q_1, \ell) \Big) \\[0.5em]
\Res^{\mathcal{F}}(\hat\rho, \hat\tau, q, \ell) & \triangleq &
  \begin{cases}
    \Res^{\mathcal{F},\circ}_{\mathrm{sub}}(\hat\rho, \hat\tau, q, \ell) & \Head(\hat\rho) = U \\[0.3em]
    \begin{array}[t]{@{}l@{}}
    \exists \iota_0, \mathit{ln}_0.\;
      \lceil \Lft(\iota_0) = \kappa \wedge \mathit{ln}_0.\mathsf{mut} \wedge \mathit{ln}_0.\mathsf{root} = \ell \rceil * {} \\
    \quad \Idx(\iota_0, \mathit{ln}_0) * \hat\tau \teq \mathit{ln}_0.\mathsf{dep} *
      \Res^{\mathcal{F},\iota_0}_{\mathrm{sub}}(\hat\rho, \hat\tau, q, \ell)
    \end{array} & \Head(\hat\rho) = \kappa^{\Mut} \\[0.3em]
    \Kw{False} & \Head(\hat\rho) = \kappa
  \end{cases}
\end{array}
\]

\smallskip
\noindent\makebox[\linewidth][l]{\textbf{Sharing Predicate}}
\[
\small
\begin{array}{r@{\;}c@{\;}l}
\Shr(\kappa, \hat\tau, \ell) & \triangleq &
  \exists \iota, \mathit{ln}.\;
  \Idx(\iota, \mathit{ln}) *
  \lceil \neg\mathit{ln}.\mathsf{mut} \wedge \Subbed{\Lft(\iota)}(\mathit{ln}.\mathsf{dep}) \wedge \mathit{ln}.\mathsf{root} = \ell \rceil * {} \\
  & & \Lft(\iota) \sqsubseteq \kappa *
  \hat\tau \teqs \mathit{ln}.\mathsf{dep} *
  \big(\exists S.\; \Sups(\Lft(\iota)) = S * \lceil \MHeads(\mathit{ln}.\mathsf{dep}) \subseteq S \rceil\big)
\end{array}
\]

\smallskip
\noindent\makebox[\linewidth][l]{\textbf{Exact and Restored Value Interpretations}}
\[
\small
\begin{array}{r@{\;}c@{\;}l}
\CF{()}^{q}(v) & \triangleq & \lceil v = () \rceil \qquad
\CF{\Kw{int}}^{q}(v) \;\triangleq\; \exists n.\, \lceil v = n \rceil \qquad
\CF{\Kw{bool}}^{q}(v) \;\triangleq\; \exists b.\, \lceil v = b \rceil \\[0.2em]
\CF{\hat\tau_1 \otimes \hat\tau_2}^{q}(v) & \triangleq &
  \exists v_1, v_2.\, \lceil v = (v_1, v_2) \rceil * \CF{\hat\tau_1}^{q}(v_1) * \CF{\hat\tau_2}^{q}(v_2) \\[0.2em]
\CVex{\RefT^{\hat\rho}\hat\tau}^{q}(v) & \triangleq &
  \exists \ell.\, \lceil v = \ell \rceil *
  \begin{cases}
    \Shr(\kappa, \hat\tau, \ell) & \Head(\hat\rho) = \kappa \\
    \Res^{\mathcal{V}^{\mathsf{ex}}}(\hat\rho, \hat\tau, q, \ell) & \Head(\hat\rho) \in \{U, \kappa^{\Mut}\}
  \end{cases} \\[0.9em]
\CV{\RefT^{\hat\rho}\hat\tau}^{q}(v) & \triangleq &
  \exists \ell.\, \lceil v = \ell \rceil *
  \begin{cases}
    \Shr(\kappa, \hat\tau, \ell) & \Head(\hat\rho) = \kappa \\
    \exists \hat\rho'.\; \hat\rho \teq \hat\rho' * \Res^{\mathcal{V}}(\hat\rho', \hat\tau, q, \ell) & \Head(\hat\rho) \in \{U, \kappa^{\Mut}\}
  \end{cases} \\[0.9em]
\multicolumn{3}{l}{%
  \CF{(\ArrT{p}{\Delta_f}{L}{\hat q}{\tau_1}{\tau_2})_{\eta}}^{q}(v) \;\triangleq\;
  \exists x, e.\, \lceil v = \lambda x.\, e \rceil * \triangleright \Box_{\hat q} \forall \theta, A, w.\;
    \CD{E_f, L}{\Delta_f}{\eta_\theta}{A} * \SemV{\tau_1}{\eta_\theta}{1}(w) \wand} \\
  \multicolumn{3}{l}{%
    \qquad \mathsf{WP}\ [w/x]e\ \big\{ r.\,
    \CD{E_f, \emptyset}{\Delta_f \setminus L}{\eta_\theta}{A} *
    \big(\bigsep_{b \in L} \dagger\eta_\theta(b)\big) *
    \SemV{\tau_2}{\eta_\theta}{1}(r) \big\}} \\
  \multicolumn{3}{l}{%
    \qquad \text{where } E_f = \Dom\Delta_f \setminus L,\;
    \eta_\theta = \eta[\bar p \mapsto \theta],\;
    \Box_{\Un} P = \Box P,\; \Box_{\Lin} P = P,\; \Box_{\Shape} P = \Kw{True}}
\end{array}
\]
where $\CF{\cdot} \in \{\CVex{\cdot}, \CV{\cdot}\}$, and the notations in \Cref{sec:semantic-typing} abbreviate:
{\small
\begin{align*}
\SemV{\tau}{\eta}{q}(v) &\triangleq \CV{\eta\tau}^{q}(v) &
\Shr_\eta(\kappa, \tau, \ell) &\triangleq \Shr(\kappa, \eta\tau, \ell) \\
\Res_\eta(\rho, \tau, q, \ell) &\triangleq \exists \hat\rho'.\; \eta\rho \teq \hat\rho' * \Res^{\mathcal{V}}(\hat\rho', \eta\tau, q, \ell)
\end{align*}}

\subsection{Loan Table}
\label[appendix]{app:model-table}

In $\LT(M, \Sups, p, d, R)$, directory $d : \mathit{BorIdx} \rightharpoonup \mathit{Loan}$ records all loans created so far, $p : \mathit{BorIdx} \rightharpoonup (0,1]$ records the fraction currently deposited in the table for each un-reclaimed loan, $\mathit{tc} : \mathit{BorIdx} \rightharpoonup \mathit{CTy}$ records the current cell type of each root loan, $W \subseteq \mathit{BorIdx}$ records mutable root loans whose lender holds an exclusive inheritance token $\Inh^{\top}$, and $R \subseteq \mathit{BorIdx}$ records retired mutable loans whose token $\Tok_1(\iota)$ has been returned to the table.
Here $\mathsf{Auth}_{\mathrm{amt}}(p, W)$, $\mathsf{Auth}_{\mathrm{dir}}(d)$, and $\mathsf{Auth}_{\mathrm{tok}}(\Dom d)$ are the authoritative ghost-map ownerships backing $\Inh^{w}(\iota, q)$, $\Idx(\iota, \mathit{ln})$, and $\Tok_q(\iota)$.
\[
\small
\begin{array}{r@{\;}c@{\;}l}
\Pay_{\mathit{tc}}(\iota, \mathit{ln}, q) & \triangleq &
  \begin{cases}
    \Cell^{\mathcal{V}^{\mathsf{ex}}}_q(\mathit{ln}.\mathsf{root}, \mathit{tc}(\iota)) & \mathit{ln}.\mathsf{lender} = \circ \\
    \Tok_q(\iota') & \mathit{ln}.\mathsf{lender} = \iota'
  \end{cases} \\[0.8em]
\LT(M, \Sups, p, d, R) & \triangleq &
  \exists \mathit{tc}, W.\;
  \World(M, \Sups) * \mathsf{Auth}_{\mathrm{amt}}(p, W) * \mathsf{Auth}_{\mathrm{dir}}(d) * {} \\
  & & \mathsf{Auth}_{\mathrm{tok}}(\Dom d) *
  \big(\bigsep_{\iota \mapsto q \in p} \exists \mathit{ln}.\; \lceil d(\iota) = \mathit{ln} \rceil * \Pay_{\mathit{tc}}(\iota, \mathit{ln}, q)\big) * {} \\
  & & \big(\bigsep_{\iota \in R} \Tok_1(\iota)\big) *
  \lceil \mathsf{WF}(M, \Sups, p, d, \mathit{tc}, W, R) \rceil
\end{array}
\]

The invariant $\mathsf{WF}(M, \Sups, p, d, \mathit{tc}, W, R)$ requires that:
\begin{enumerate}
  \item every loan at a live lifetime is present in $p$.
  \item root loans have depth $0$ and each reborrow points to a mutable parent loan at the same root location and one depth lower that outlives it.
  \item live mutable loans hold full fraction $1$ and are unique per root and depth.
  \item each root loan's current type $\mathit{tc}(\iota)$ agrees with its initial deposited type $\mathit{ln}.\mathsf{dep}$ up to lifetimes that are dead or blocked by an active reborrow or retired loan.
  \item $W \cap \Dom p$ contains $\iota$ iff $\iota$ is a deposited mutable root loan whose current type is not equal to its core.
\end{enumerate}
Formally, for all $d(\iota) = \mathit{ln}$:
\[
\small
\begin{array}{@{}l@{\quad}r@{}}
\begin{array}[t]{@{}l@{}}
\Lft(\iota) \in \Dom M \;\wedge\; \Dom p \subseteq \Dom d \\
  \quad {} \wedge (\Lft(\iota) \in \Alive(M) \Rightarrow \iota \in \Dom p)
\end{array}
  & \text{(liveness)} \\[0.35em]
\mathit{ln}.\mathsf{lender} = \circ \;\Rightarrow\; \mathit{ln}.\mathsf{depth} = 0
  & \text{(root depth)} \\[0.35em]
\begin{array}[t]{@{}l@{}}
\mathit{ln}.\mathsf{lender} = \iota' \;\Rightarrow\;
  \exists \mathit{ln}'.\; d(\iota') = \mathit{ln}' \wedge \mathit{ln}.\mathsf{depth} = \mathit{ln}'.\mathsf{depth} + 1 \wedge \mathit{ln}'.\mathsf{root} = \mathit{ln}.\mathsf{root} \\
  \quad {} \wedge \mathit{ln}'.\mathsf{mut} \wedge \Lft(\iota') \in \Sups(\Lft(\iota)) \wedge (\mathit{ln}.\mathsf{mut} \Rightarrow \mathit{ln}.\mathsf{dep} = \mathit{ln}'.\mathsf{dep})
\end{array}
  & \text{(reborrow)} \\[0.35em]
\begin{array}[t]{@{}l@{}}
\mathit{ln}.\mathsf{mut} \wedge \Lft(\iota) \in \Alive(M) \;\Rightarrow\; p(\iota) = 1 \;\wedge \\
  \quad (\forall \iota', \mathit{ln}'.\; d(\iota') = \mathit{ln}' \wedge \mathit{ln}'.\mathsf{mut} \wedge \Lft(\iota') \in \Alive(M) \\
  \quad\phantom{(\forall \iota', \mathit{ln}'.\;} {} \wedge \mathit{ln}'.\mathsf{root} = \mathit{ln}.\mathsf{root} \wedge \mathit{ln}'.\mathsf{depth} = \mathit{ln}.\mathsf{depth} \Rightarrow \iota' = \iota)
\end{array}
  & \text{(exclusivity)} \\[0.35em]
\mathit{ln}.\mathsf{lender} = \circ \wedge \neg\mathit{ln}.\mathsf{mut} \;\Rightarrow\; \mathit{tc}(\iota) = \mathit{ln}.\mathsf{dep} = \Core(\mathit{ln}.\mathsf{dep})
  & \text{(shared root)} \\[0.35em]
\begin{array}[t]{@{}l@{}}
\mathit{ln}.\mathsf{lender} = \circ \wedge \mathit{ln}.\mathsf{mut} \;\Rightarrow\;
  \mathsf{T1}(\iota, \mathit{ln}, \mathit{tc}(\iota)) \\
  \quad {} \vee (M(\Lft(\iota)) = \mathsf{dead} \wedge \exists \hat\tau_1.\; \mathit{tc}(\iota) = \Core(\hat\tau_1) \wedge \mathsf{T1}(\iota, \mathit{ln}, \hat\tau_1))
\end{array}
  & \text{(mutable root)} \\[0.35em]
\begin{array}[t]{@{}l@{}}
\mathit{ln}.\mathsf{lender} = \circ \wedge \mathit{ln}.\mathsf{mut} \wedge \Lft(\iota) \in \Alive(M) \wedge \mathsf{ShRebor}(\iota_w, \mathit{ln}_w, \mathit{ln}.\mathsf{root}) \;\Rightarrow\; \\
  \quad \HSep(\mathit{ln}_w.\mathsf{dep}) \wedge \mathit{tc}(\iota) = \Sub{\kappa_1}(\cdots\Sub{\kappa_n}(\mathit{ln}_w.\mathsf{dep})) \\
  \quad {} \wedge (\exists D.\; (\forall c \in D.\; M(c) = \mathsf{dead}) \wedge \Strip{\mathit{ln}_w.\mathsf{dep}}\Rst{D} = \Strip{\mathit{ln}.\mathsf{dep}}\Rst{D})
\end{array}
  & \text{(active reader)} \\[0.35em]
\begin{array}[t]{@{}l@{}}
W \cap \Dom p \subseteq \{\, \iota_0 \in \Dom p \mid d(\iota_0).\mathsf{mut} \wedge d(\iota_0).\mathsf{lender} = \circ \,\} \;\wedge \\
  \quad (\iota \in \Dom p \wedge \mathit{ln}.\mathsf{mut} \wedge \mathit{ln}.\mathsf{lender} = \circ \;\Rightarrow\; (\iota \notin W \iff \Core(\mathit{tc}(\iota)) = \mathit{tc}(\iota)))
\end{array}
  & \text{(excl.\ token)}
\end{array}
\]
where:
\[
\small
\begin{array}{r@{\;}c@{\;}l}
\mathsf{ShRebor}(\iota_w, \mathit{ln}_w, \ell) & :\Leftrightarrow &
  d(\iota_w) = \mathit{ln}_w \wedge \neg\mathit{ln}_w.\mathsf{mut} \wedge \mathit{ln}_w.\mathsf{lender} \neq \circ \\
  & & {} \wedge \Lft(\iota_w) \in \Alive(M) \wedge \mathit{ln}_w.\mathsf{root} = \ell \\[0.25em]
\mathsf{Blocked}(\ell, c) & :\Leftrightarrow &
  \exists \iota_w, \mathit{ln}_w.\; d(\iota_w) = \mathit{ln}_w \wedge \Lft(\iota_w) = c \wedge \mathit{ln}_w.\mathsf{root} = \ell \\
  & & {} \wedge ((\mathit{ln}_w.\mathsf{mut} \wedge \iota_w \in R) \vee (\neg\mathit{ln}_w.\mathsf{mut} \wedge \mathit{ln}_w.\mathsf{lender} \neq \circ)) \\[0.25em]
\mathsf{T1}(\iota, \mathit{ln}, \hat\tau) & :\Leftrightarrow &
  \exists S.\; \hat\tau\Rst{S} = \mathit{ln}.\mathsf{dep}\Rst{S} \\
  & & {} \wedge \forall c \in S.\; M(c) = \mathsf{dead} \vee (\Lft(\iota) \in \Sups(c) \wedge \mathsf{Blocked}(\mathit{ln}.\mathsf{root}, c))
\end{array}
\]

\subsection{Context Interpretations and Semantic Typing}
\label[appendix]{app:model-ctx}

\noindent\makebox[\linewidth][l]{\textbf{Context Interpretations}}
\[
\small
\begin{array}{r@{\;}c@{\;}l}
\CD{E, L}{\Delta}{\eta}{A} & \triangleq &
  \exists M, \Sups, p, d, R.\;
  \LT(M, \Sups, p, d, R) * \mathsf{Snap}(\Sups) * {} \\
  & & \big(\bigsep_{a \in L} [\eta(a)]_1\big) *
  \big(\bigsep_{a \in E \cup L} \Sups(\eta(a)) = \Sups(\eta(a))\big) * {} \\
  & & \lceil \eta \text{ injective on } E \cup L \rceil *
  \lceil \eta(E) \subseteq A \rceil *
  \lceil \Alive(M) = A \uplus \eta(L) \rceil * {} \\
  & & \lceil \forall \kappa \in \eta(L), \kappa' \in \Alive(M).\; \kappa \in \Sups(\kappa') \Rightarrow \kappa' \in \eta(L) \rceil * {} \\
  & & \lceil \forall (a <: b) \in \Delta.\; a, b \in E \cup L \Rightarrow \eta(a) \in \Sups(\eta(b)) \rceil * {} \\
  & & \lceil \forall a \in L, b \in E \cup L.\; \eta(a) \in \Sups(\eta(b)) \Rightarrow a = b \vee \Delta \vDash a <: b \rceil \\[0.4em]
\SemG{\Gamma}{\eta}{\gamma} & \triangleq &
  \bigsep_{(x : \tau) \in \Gamma} \exists v.\, \lceil \gamma(x) = v \rceil * \SemV{\tau}{\eta}{1}(v)
\end{array}
\]
where $\mathsf{Snap}(\Sups) \triangleq \bigsep_{\kappa \mapsto S \in \Sups} (\Sups(\kappa) = S)$ is the persistent snapshot of $\Sups$.

\smallskip
\noindent\makebox[\linewidth][l]{\textbf{Semantic Typing}}
\[
\small
\begin{array}{r@{\;}c@{\;}l}
\Delta; \Gamma \vDash_E e : \tau \Dashv \Delta' & \triangleq &
  \forall \eta, A, \gamma.\;
  \CD{E, L}{\Delta}{\eta}{A} * \SemG{\Gamma}{\eta}{\gamma} \wand \\
  & & \quad \mathsf{WP}\ \gamma(e)\ \big\{ v.\, \exists \eta'.\, \lceil \eta|_N = \eta'|_N \rceil *
    \CD{E, L'}{\Delta'}{\eta'}{A} *
    \SemV{\tau}{\eta'}{1}(v) *
    \bigsep_{a \in L \setminus L'} \dagger\eta(a) \big\} \\
  & & \quad \text{where } L = \Dom\Delta \setminus E,\; L' = \Dom\Delta' \setminus E,\;
    N = \Dom\Delta \cup \dagger\Delta
\end{array}
\]

\section{Case Study Details}
\label[appendix]{app:case-study}

Here we provide the full breakdown of 119 \texttt{rustc} soundness issues
returned by the query described in \Cref{sec:case-study} as of October 21, 2026.

{\scriptsize
\setlength{\tabcolsep}{4pt}
\renewcommand{\arraystretch}{1.12}
\newcommand{\rbug}[1]{\href{https://github.com/rust-lang/rust/issues/#1}{\texttt{\##1}}}
\newcommand{\rbopen}[1]{\href{https://github.com/rust-lang/rust/issues/#1}{\textcolor{ibmred}{\textbf{\texttt{\##1}}}}}
\begin{longtable}{@{}lp{0.43\linewidth}p{0.46\linewidth}@{}}
  \caption{Analysis of in-scope \texttt{rustc}
    soundness issues. (open issues in \textcolor{ibmred}{\textbf{bold}}).}
  \label{tab:rust-bugs-detail} \\
  \toprule
  \textbf{Issue} & \textbf{Root Cause in \texttt{rustc}} & \textbf{Resolution in \LBorrow} \\
  \midrule
  \endfirsthead
  \toprule
  \textbf{Issue} & \textbf{Root Cause in \texttt{rustc}} & \textbf{Resolution in \LBorrow} \\
  \midrule
  \endhead
  \bottomrule
  \endfoot
  \bottomrule
  \endlastfoot
    \multicolumn{3}{@{}l}{\textbf{Unchecked lifetime bounds (6 issues, 1 open)}} \\*[1pt]
    \rbopen{25860} & Coercing a function to a function pointer drops the outlives bound implied by a nested reference parameter \texttt{\&'a \&'b ()}. & Rules \rref{WF-Ref} and \rref{WF-Arr} require all implied outlives bounds to hold in $\Delta_f$, and \rref{T-App} checks $\Delta_2 \vDash \theta\Delta_f$ at every call. \\
    \rbug{29793} & Closure signatures are not checked for well-formedness, allowing a returned inner closure to keep borrows of the outer closure's parameters after the outer call returns. & Rule \rref{WF-Arr} requires parameter and return types to be well-formed under $\Delta_f$, and \rref{T-LftEnd} forbids ending a lifetime while an inner closure still captures it. \\
    \rbug{48803} & Early NLL region inference fails to propagate the outlives bound from a nested reference parameter \texttt{\&'a \&'b T} to the returned borrow \texttt{\&'a T}. & Rule \rref{WF-Ref} requires the inner lifetime to outlive the outer lifetime, and \rref{L-End} forbids ending the inner lifetime while the outer borrow is alive. \\
    \rbug{52126} & Compound assignment \texttt{a += b} allows the mutated left-hand reference \texttt{a} to shorten its lifetime, storing a short-lived borrow into a longer-lived variable. & Updating a reference via \rref{T-Swap} preserves its referent type, and reborrowing via \rref{S-Borrow} requires the source borrow to outlive the target lifetime. \\
    \rbug{57170} & Borrowing a function as \texttt{\&f} promotes the reference to a constant, and NLL drops the function's explicit \texttt{where 'a:\ 'min} bound at calls through that reference. & Function types carry their lifetime constraints $\Delta_f$ explicitly under \rref{WF-Arr}, and allocating or reading a reference to a function preserves $\Delta_f$ at every call via \rref{T-App}. \\
    \rbug{57265} & Borrow checking omits well-formedness checks on function return types, ignoring outlives bounds implied by the return type. & Rule \rref{WF-Arr} checks $\Delta_f \setminus L \vDash \tau_2$, requiring all implied outlives bounds in the return type to hold. \\[3pt]
    \midrule
    \multicolumn{3}{@{}l}{\textbf{Two-phase borrow reservation (2 issues, 2 open)}} \\*[1pt]
    \rbopen{160598} & A \texttt{break} branch inside an argument expression causes the borrow checker to treat a two-phase mutable borrow as still reserved even after the call activates it. & Rule \rref{S-Borrow} splits permissions immediately without a reservation phase, leaving the mutable lender with empty access permission $\emptyset$ while the borrow is alive. \\
    \rbopen{164163} & Reassigning the receiver variable during argument evaluation causes the borrow checker to discard the pending reservation and miss a conflicting shared borrow. & Rule \rref{S-Borrow} records the active mutable borrow in the lender's type with access permission $\emptyset$, blocking both shared reborrows and \rref{T-Swap} during argument evaluation. \\[3pt]
    \midrule
    \multicolumn{3}{@{}l}{\textbf{Destructors on dead borrows (6 issues, 1 open)}} \\*[1pt]
    \rbug{26656} & Parametric drop checking assumes a generic type parameter cannot access borrowed data on drop, even when the destructor reaches the borrowed data through a trait object field. & Rule \rref{T-LftEnd} requires $\Delta' \vDash \Gamma\Rst{a}$, forbidding lifetime $a$ from ending while any active borrow at $a$ or closure capturing $a$ remains in the context. \\
    \rbug{31567} & Consuming a wrapper around a mutable borrow runs the wrapper's destructor at scope exit while a reborrow returned from the wrapper is still active. & Reborrowing via \rref{S-Borrow} records the sub-borrow in the lender's type, and \rref{T-LftEnd} forbids ending the parent borrow's lifetime until the reborrow has ended. \\
    \rbug{90838} & A \texttt{Drop} implementation declares a stricter \texttt{'static} lifetime bound than the struct itself, and drop checking assumes the bound holds without verifying it at drop sites. & Deallocation via \rref{T-Drop} and cleanup calls via \rref{T-App} are type-checked under the current lifetime context, enforcing all required lifetime constraints at the use site. \\
    \rbug{115175} & NLL omits well-formedness checks on local variable type annotations, allowing an unchecked outlives bound on a local variable to be assumed by later passes. & Context splitting and let-binding rules require all reference types in the context to satisfy \rref{WF-Ref} under the current lifetime context $\Delta$. \\
    \rbug{148854} & An unchecked \texttt{'static} bound on a local variable type is trusted by drop checking, allowing a destructor to read a borrow after its lifetime has ended. & Rules \rref{WF-Ref} and \rref{T-LftEnd} enforce well-formedness on all bindings in $\Gamma$ and forbid ending a lifetime while any active borrow at that lifetime remains. \\
    \rbopen{164250} & Assigning a short-lived borrow into a struct field bypasses the \texttt{'static} bound assumed by the struct's destructor, allowing the destructor to observe a dead borrow at scope exit. & Rule \rref{T-Swap} preserves the cell's referent type, and \rref{T-LftEnd} prevents the borrow's lifetime from ending while the cell holding it remains in the context. \\[3pt]
    \midrule
    \multicolumn{3}{@{}l}{\textbf{Unchecked place accesses (10 issues, 2 open)}} \\*[1pt]
    \rbug{27282} & A closure inside a match guard mutates the matched variable through a mutable pattern binding, invalidating the binding read by a subsequent match arm. & Context splitting in elimination forms partitions permissions across subexpressions via \rref{S-Borrow}, preventing a condition and a branch from holding conflicting permissions. \\
    \rbug{27868} & Borrow checking and code generation assume opposite operand evaluation orders for compound assignment \texttt{a += b}, missing that the left-hand side moves a value that the right-hand side then reads. & Typing rules such as \rref{T-Pair}, \rref{T-App}, and \rref{T-Swap} propagate ended lifetimes and restore types across subexpressions in a single, consistent order. \\
    \rbug{29723} & Moving a value inside a match guard is not tracked across fallthrough to subsequent match arms, allowing the moved value to be used twice. & Owned references have linear qualifier $\Lin$, and context splitting in \rref{T-If} and \rref{T-Let} prevents consuming a linear resource more than once. \\
    \rbug{38899} & Taking a shared reborrow of a nested reference field \texttt{*block.current} is permitted while the outer struct \texttt{block} is already mutably borrowed. & A mutable lender $\RefT^{\rho - a^\Mut}\tau$ has empty access permission $\emptyset$, so neither \rref{S-Borrow} nor \rref{T-Read} can access its inner reference while $a$ is alive. \\
    \rbug{45045} & Matching on an enum variable with fieldless variant patterns does not record a read of its discriminant, allowing inspection of a mutably borrowed variable. & Inspecting a reference requires \rref{T-Read} on an active borrow, which cannot be split from a mutable lender with empty access permission $\emptyset$. \\
    \rbug{51117} & Pattern matching implicitly dereferences a mutable reference without recording a borrow of its referent, allowing the referent to be mutated while bound in the match arm. & Dereferencing requires an explicit \rref{T-Read} on a borrow split via \rref{S-Borrow}, which records the active borrow in the lender's type until the borrow's lifetime ends. \\
    \rbug{51415} & A reference pattern in a closure parameter moves an owned value out from behind a shared reference without borrow checking rejecting the move. & Rule \rref{T-Read} returns the read view $\Rview{\alpha}{\tau}$, which downgrades any nested owned reference $\RefT^U\sigma$ to a borrow $\RefT^\alpha\sigma$ and never yields ownership. \\
    \rbug{56797} & NLL borrow checking treats an enum discriminant read as disjoint from the enum's variant fields, permitting a match on an enum while a variant field is mutably borrowed. & Inspecting a cell's contents requires \rref{T-Read}, which demands at least read permission $\Kw{r}$ and is blocked on a mutable lender. \\
    \rbopen{160525} & Borrow checking tracks constant array indices and variable indices with incompatible place representations, missing a move of an array inside an index expression while borrowing an element. & Context splitting across subexpressions via \rref{S-Borrow} tracks permissions on the reference itself, so borrowing an element prevents moving or mutating the container in another subexpression. \\
    \rbopen{160599} & Reassigning the matched variable inside a match guard causes a subsequent arm to bind from a new value whose discriminant was never tested. & While a borrow at lifetime $a$ is alive, the lender's subtracted borrowability $\rho - a$ lacks write permission $\Kw{w}$, so \rref{T-Swap} cannot overwrite the lender until $a$ ends. \\
\end{longtable}
}

{\footnotesize
\setlength{\tabcolsep}{4pt}
\renewcommand{\arraystretch}{1.15}
\newcommand{\rbug}[1]{\href{https://github.com/rust-lang/rust/issues/#1}{\texttt{\##1}}}
\newcommand{\rbopen}[1]{\href{https://github.com/rust-lang/rust/issues/#1}{\textcolor{ibmred}{\textbf{\texttt{\##1}}}}}
\begin{longtable}{@{}>{\raggedright\arraybackslash}p{0.20\linewidth}cp{0.36\linewidth}p{0.37\linewidth}@{}}
  \caption{Excluded \texttt{rustc} soundness issues (open issues in \textcolor{ibmred}{\textbf{bold}}).}
  \label{tab:rust-bugs-excluded} \\
  \toprule
  \textbf{Exclusion Reason} & \textbf{\#} & \textbf{Description} & \textbf{Issues} \\
  \midrule
  \endfirsthead
  \toprule
  \textbf{Exclusion Reason} & \textbf{\#} & \textbf{Description} & \textbf{Issues} \\
  \midrule
  \endhead
  \bottomrule
  \endfoot
  \bottomrule
  \endlastfoot
    Trait system & 37 & Traits, associated-type projections, trait objects, opaque types, specialization, or higher-ranked trait bounds. &
    {\scriptsize \rbug{45982}, \rbug{47470}, \rbug{58299}, \rbug{71550}, \rbug{80176}, \rbug{82633}, \rbug{84305}, \rbopen{84366}, \rbopen{84533}, \rbopen{84591}, \rbug{84657}, \rbug{85863}, \rbug{85873}, \rbug{88901}, \rbug{91068}, \rbug{97104}, \rbug{98117}, \rbug{98543}, \rbopen{100051}, \rbug{104005}, \rbopen{112905}, \rbug{118876}, \rbopen{129005}, \rbopen{141713}, \rbopen{144312}, \rbopen{151637}, \rbopen{152489}, \rbopen{153140}, \rbug{153596}, \rbopen{153597}, \rbug{155520}, \rbopen{161864}, \rbopen{162066}, \rbopen{163886}, \rbopen{164416}, \rbopen{164421}, \rbopen{164506}} \\[3pt]
    Unstable/unsafe features & 12 & Unsafe union field accesses or nightly features such as inline assembly, slice patterns, \texttt{async\_drop}, \texttt{arbitrary\_self\_types}, \texttt{min\_specialization}, or custom \texttt{Reborrow} traits. &
    {\scriptsize \rbug{41126}, \rbug{45157}, \rbug{45695}, \rbug{52786}, \rbug{66502}, \rbug{120217}, \rbug{142572}, \rbopen{144442}, \rbopen{164306}, \rbopen{164338}, \rbopen{164510}, \rbopen{164614}} \\[3pt]
    Globals and promotion & 8 & Global \texttt{static} items, thread-local storage, or temporary \texttt{'static} promotion. &
    {\scriptsize \rbug{17954}, \rbug{38520}, \rbug{45704}, \rbug{46557}, \rbopen{49682}, \rbug{49955}, \rbug{55223}, \rbug{69114}} \\[3pt]
    Coroutines & 7 & Lowering of \texttt{async} blocks and generators. &
    {\scriptsize \rbug{47516}, \rbug{59972}, \rbug{62312}, \rbug{119564}, \rbug{144155}, \rbug{152735}, \rbopen{161026}} \\[3pt]
    Polonius & 8 & Experimental Polonius borrow-checker implementation on nightly. &
    {\scriptsize \rbug{53142}, \rbug{160669}, \rbug{160670}, \rbopen{161038}, \rbopen{161471}, \rbopen{164337}, \rbopen{164420}, \rbug{164546}} \\[3pt]
    Transitional modes & 7 & Pre-release 2017 MIR borrow checker, NLL migration mode downgrading errors to warnings, or beta-only NLL regressions. &
    {\scriptsize \rbug{44833}, \rbug{44834}, \rbug{46023}, \rbug{53026}, \rbug{53432}, \rbug{58776}, \rbug{98693}} \\[3pt]
    Representation and backend & 8 & Packed struct layout, panic unwinding, \texttt{extern} FFI declarations, or LLVM codegen. &
    {\scriptsize \rbug{30380}, \rbug{35851}, \rbopen{47949}, \rbopen{63818}, \rbopen{143411}, \rbug{157011}, \rbug{157615}, \rbopen{164561}} \\[3pt]
    Standard library & 5 & Implementation bugs in standard library unrelated to type and borrow checking. &
    {\scriptsize \rbug{83618}, \rbug{157089}, \rbug{159334}, \rbug{159368}, \rbug{159370}} \\[3pt]
    Duplicate reports & 3 & Duplicates or minor variants of other issues. &
    {\scriptsize \rbug{72315}, \rbug{99158}, \rbug{157316}} \\
\end{longtable}
}

\section{Rocq Mechanization}
\label[appendix]{app:mechanization}

\begin{table}[ht]
  \caption{Correspondence between claims in \Cref{sec:soundness} and Rocq theorems.}
  \label{tab:mech-theorems}
  \footnotesize
  \centering
  \begin{tabularx}{\textwidth}{lXl}
    \toprule
    \textbf{Claim} & \textbf{Description} & \textbf{In Rocq} \\
    \midrule
    \multirow{2}{*}{\Cref{lem:wf-preservation}}
      & Type well-formedness & \texttt{LBorrow.Theorems.WellFormedness.typed\_wf} \\
      & Lifetime context monotonicity & \texttt{LBorrow.Theorems.WellFormedness.typed\_mono} \\
    \midrule
    \Cref{thm:fundamental}
      & Fundamental property of logical relations & \texttt{LBorrow.Theorems.Fundamental.fundamental} \\
    \midrule
    \multirow{3}{*}{\Cref{thm:soundness}}
      & Type soundness & \texttt{LBorrow.Theorems.Soundness.type\_soundness} \\
      & Adequacy & \texttt{LBorrow.Theorems.Soundness.sem\_adequacy} \\
      & Memory safety & \texttt{LBorrow.Theorems.Soundness.memory\_safety} \\
    \bottomrule
  \end{tabularx}
\end{table}

\begin{table}[tb]
  \caption{Code statistics for the Rocq mechanization.}
  \label{tab:mech-stats}
  \footnotesize
  \centering
  \begin{tabularx}{\textwidth}{lXrrr}
    \toprule
    \textbf{Module} & \textbf{Description} & \textbf{LOC} & \textbf{\#Def} & \textbf{\#Thm} \\
    \midrule
    Syntax & Lifetimes, permissions, qualifiers, types, terms, and typing rules & 521 & 82 & 2 \\
    Semantics & Operational semantics and logical relation & 1\,386 & 184 & 37 \\
    Tactics & Iris proof mode and proof automation tactics & 978 & 8 & 23 \\
    Lemmas & Auxiliary lemmas and properties & 9\,193 & 7 & 722 \\
    Theorems & Well-formedness preservation, fundamental property, and type soundness & 464 & 1 & 13 \\
    Examples & Mechanized example programs and counterexamples & 1\,651 & 249 & 176 \\
    \midrule
    \textbf{Total} & & \textbf{14\,193} & \textbf{531} & \textbf{973} \\
    \bottomrule
  \end{tabularx}
\end{table}

\begin{table}[ht]
  \caption{Correspondence between paper definitions, lemmas, examples, and the Rocq mechanization.}
  \label{tab:mech-correspondence}
  \footnotesize
  \centering
  \begin{tabularx}{\textwidth}{Xll}
    \toprule
    \textbf{Item} & \textbf{In Paper} & \textbf{In Rocq} \\
    \midrule
    \multicolumn{3}{l}{\textbf{Syntax and Type System (\Cref{sec:calculus}, \Cref{app:calculus})}} \\
    \quad Borrowabilities & \Cref{fig:calculus-grammar} & \texttt{LBorrow.Syntax.Lifetimes.lftg} \\
    \quad Permission splitting & \Cref{fig:permission-split} & \texttt{LBorrow.Syntax.Lifetimes.perm\_split} \\
    \quad Types and qualifiers & \Cref{fig:calculus-qual} & \texttt{LBorrow.Syntax.Types.ty\_qual} \\
    \quad Contexts & \Cref{fig:calculus-grammar} & \texttt{LBorrow.Syntax.Types.lctx} \\
    \quad Well-formedness & \Cref{fig:calculus-wf} & \texttt{LBorrow.Syntax.Types.wf\_ty} \\
    \quad Read view & \Cref{fig:calculus-rview} & \texttt{LBorrow.Syntax.Types.read\_ty} \\
    \quad Splitting & \Cref{fig:ref-split,fig:calculus-split} & \texttt{LBorrow.Syntax.Types.ty\_split} \\
    \quad Lifetime ending & \Cref{fig:calculus-lft-end} & \texttt{LBorrow.Syntax.Restoration.lft\_end} \\
    \quad Permission restoration & \Cref{fig:restoration} & \texttt{LBorrow.Syntax.Restoration.ty\_restore} \\
    \quad Lifetime instantiation & \Cref{sec:calculus-typing} & \texttt{LBorrow.Syntax.Instantiation.subst\_ty} \\
    \quad Syntactic typing & \Cref{fig:calculus-typing} & \texttt{LBorrow.Syntax.Typing.typed} \\
    \midrule
    \multicolumn{3}{l}{\textbf{Operational Semantics and Semantic Model (\Cref{sec:semantics}, \Cref{app:opsem,app:model})}} \\
    \quad Operational semantics & \Cref{sec:opsem} & \texttt{LBorrow.Semantics.SmallStep.lborrow.base\_step} \\
    \quad Ghost state & \Cref{app:model-ghost} & \texttt{LBorrow.Semantics.World.world} \\
    \quad Closed types & \Cref{app:model-closed} & \texttt{LBorrow.Semantics.ClosedTypes.cty} \\
    \quad Loan table & \Cref{app:model-table} & \texttt{LBorrow.Semantics.LoanTable.ctable} \\
    \quad Value interpretation & \Cref{fig:logical-relation} & \texttt{LBorrow.Semantics.Interp.V} \\
    \quad Context interpretation & \Cref{app:model-ctx} & \texttt{LBorrow.Semantics.Interp.D} \\
    \quad Semantic typing & \Cref{app:model-ctx} & \texttt{LBorrow.Semantics.Judgment.sem\_typed\_lc} \\
    \midrule
    \multicolumn{3}{l}{\textbf{Key Lemmas (\Cref{sec:semantics}, \Cref{app:model})}} \\
    \quad Lazy type restoration & \Cref{sec:semantic-typing} & \texttt{LBorrow.Lemmas.Compat.Core.V\_restore} \\
    \quad Lazy reclamation & \Cref{sec:semantic-typing} & \texttt{LBorrow.Lemmas.Restore.ValueRestore.Vdv\_reclaim} \\
    \quad Semantic type splitting & \Cref{fig:ref-split,fig:calculus-split} & \texttt{LBorrow.Lemmas.Compat.LoanSplit.Vsplit\_ok} \\
    \quad Reference sharing & \Cref{app:model-value} & \texttt{LBorrow.Lemmas.Restore.Sharing.Vex\_share} \\
    \quad Lifetime instantiation & \Cref{sec:calculus-typing} & \texttt{LBorrow.Lemmas.Compat.Fn.V\_inst} \\
    \midrule
    \multicolumn{3}{l}{\textbf{Mechanized Examples (\Cref{sec:overview,sec:semantics,sec:case-study})}} \\
    \quad Borrows and reborrows & \Cref{sec:overview-mem-ops} & \texttt{LBorrow.Examples.Borrows.example\_safe} \\
    \quad Polymorphic functions & \Cref{sec:overview-functions} & \texttt{LBorrow.Examples.Calls.call\_end\_safe} \\
    \quad Closures & \Cref{sec:overview-functions} & \texttt{LBorrow.Examples.Calls.closure\_end\_safe} \\
    \quad Sequential restoration & \Cref{sec:case-study} & \texttt{LBorrow.Examples.Restoring.pair\_wait\_safe} \\
    \quad Lazy reclamation & \Cref{sec:semantic-typing} & \texttt{LBorrow.Examples.LazyReclamation.lazy\_reclaim\_safe} \\
    \quad Rejected closures & \Cref{tab:rust-bugs} & \texttt{LBorrow.Examples.ClosureRejections.boxed\_G\_not\_endable} \\
    \bottomrule
  \end{tabularx}
\end{table}

\end{document}